\documentclass[3p]{elsarticle}

\usepackage{a4wide}
\usepackage{float}
\usepackage{amsmath}
\usepackage{amssymb}
\usepackage{graphicx}
\usepackage[latin1]{inputenc} %Kann auch Umlaute etc. erkennen
\usepackage{rotating}
\usepackage{setspace}
\usepackage{subfigure}
\usepackage{sidecap}
\usepackage{upgreek}
\usepackage{microtype}
\usepackage{url}
\begin{document}
\begin{frontmatter}

%\pagenumbering{Roman}

%------------------------------------ Title ------------------------
\title{Charge losses in segmented silicon sensors at the Si-SiO$_2$ interface}

\renewcommand{\thefootnote}{\fnsymbol{footnote}}

\author[]{Thomas Poehlsen \corref{cor1}}
\author[]{Eckhart Fretwurst}
\author[]{Robert Klanner}
\author[]{Sergej Schuwalow}
\author[]{Jörn~Schwandt}
\author[]{Jiaguo~Zhang}
\cortext[cor1]{Corresponding author. Email address: thomas.poehlsen@desy.de. Telephone: +49 40 8998 4725.}

\address{Institute for Experimental Physics, University of Hamburg, Luruper Chaussee 149, 22761 Hamburg, Germany}

\begin{abstract}
Using multi-channel time-resolved current measurements (multi-TCT),
the charge collection of $p^+n$ silicon strip sensors for electron-hole
pairs produced close to the Si-SiO$_2$~interface by a focussed
sub-nanosecond laser with a wavelength of 660~nm has been studied.
Sensors before and after irradiation with 1~MGy of X-rays have been
investigated. The charge signals induced in the readout strips and the
rear electrode as a function of the position of the light spot are
described by a model which allows a quantitative determination of the
charge losses and of the widths of the electron-accumulation and
hole-inversion layers close to the Si-SiO$_2$ interface. Depending on
the applied bias voltage, biasing history and environmental conditions,
like humidity, incomplete electron or hole collection and different
widths of the accumulation layers are observed. In addition, the results
depend on the time after biasing the sensor, with time constants which
can be as long as days. The observations are qualitatively explained
with the help of detailed sensor simulations. Finally, their relevance
for the detection of X-ray photons and charged particles, and for the
stable operation of segmented $p^+n$ silicon sensors is discussed.
\end{abstract}

%\centerline{\large{abstract}}

%\newpage{}

%\end{titlepage}

%\date{}  %\today


\end{frontmatter}

%\maketitle

%-------------------Abstract-------------------------------------------------

% ----------------Contents ---------------------------------------------------
\tableofcontents % \chapterfin

\section{Introduction}

 In the last 30 years segmented silicon sensors underwent an impressive development and found many applications. Examples are precision tracking in particle physics, imaging in photon science and many applications
 in industry and medicine. These sensors rely on a high quality SiO$_2$ layer grown on
 the silicon
 \cite{Kemmer:1980, Kemmer:1984}
% \cite{Kemmer:1984}
 to separate the finely segmented electrodes fabricated by doping the silicon. It is well known
 \cite{Nicollian:1982} that the positive charges, which are always present in
 the SiO$_2$, induce an electron-accumulation layer at the interface between the SiO$_2$ and $n$-type silicon, and that ionizing radiation, like X-rays or charged particles passing through the SiO$_2$, further increases the density of positive oxide charges and produces traps at the Si-SiO$_2$ interface.

 For sensors with  $p^+$ electrodes fabricated on high-ohmic $n$-type silicon, which are investigated in this work, potential consequences of the accumulation layer are:

 \begin{itemize}
   \item an increase of the capacitance between the $p^+$~electrodes,
   \item a decrease of the resistance between the $p^+$~electrodes,
   \item an increase of the depletion voltage of the sensor,
   \item an increase of the electric field near the $p^+$~implants, possibly causing charge-carrier multiplication or breakdown,
   \item charge losses at the Si-SiO$_2$ interface, and
   \item an increase of  charge-collection times which may result in a ballistic deficit or pile-up.
 \end{itemize}

 The accumulation layer is also sensitive to charges on the surface of the passivation layer.
 They influence the electric boundary condition
 \cite{Longoni:1990, Richter:1996}
 which have to be known for the mathematical modeling. The density of charges close to the interface depends on many parameters, like the biasing history, the quality of the oxide, the properties of the passivation layer, the cleanliness of surfaces and the humidity of the environment. The time scale to reach stable conditions can be as long as days, with a corresponding time dependence of sensor performance.

 In this work the time resolved pulses produced by a sub-nanosecond laser focussed to an rms of 3~$\upmu$m in two different $p^+n$ strip sensors are measured as a function of the position of the light spot. The wavelength of the light is 660 nm, which corresponds to an absorption length in silicon of 3.5 $\upmu $m at room temperature. The strip sensors investigated have a pitch of 50 and 80 $\upmu$m, respectively. The measurements are made before and after irradiating the sensors with 12 keV X-rays to 1 MGy (SiO$_2$). Similar measurements are discussed in Refs. 
  \cite{Eremin:2003, Verbitskaya:2003}.
% In addition, the C-V and I-V characteristics between the strips and for the entire sensor are measured.

 Above measurements are performed at different voltages and for different humidities. It is found that the measured current transients and the fraction of the generated charges collected are different for irradiated and non-irradiated sensors. They also depend on the biasing history and change with time, with time constants which depend on humidity. The measurements can be described quantitatively by losses of electrons and/or holes in the region close to the Si-SiO$_2$ interface. The results provide insight into the influence of X-ray radiation damage, biasing history and humidity on the accumulation layer, as well as on the local electric fields close to the Si-SiO$_2$ interface of segmented silicon sensors.

 The paper first describes the sensors investigated and the measurement techniques used. A qualitative discussion of the integrated charge expected as a function of the position of the generated $eh$-pairs with and without charge losses follows, and a simple model is developed. Next, the expectations are compared to selected measurements which demonstrate situations of electron losses, hole losses and no losses. Then the model is fitted to the data and parameters like the numbers of electrons and holes collected and the width of the accumulation layer are determined. It is observed that the measured parameters change with time, with a time constant that depends on humidity.
% For the irradiated sensor, where electron losses are observed, the time required for the sensor to reach steady-state conditions in dry and humid conditions is measured.
% Next, the number of electrons which have to be accumulated until the situation of no electron losses is reached, and the time constant to return to the steady state are determined.
 Finally, an interpretation of the results is presented and their relevance for the understanding and use of sensors discussed.

 The work has been done within the AGIPD collaboration
 \cite{AGIPD}
 which is developing a large area pixel detector system for experimentation at the European X-ray Free-Electron Laser XFEL
 \cite{XFEL}
 and other X-ray sources.

\section{Measurement techniques and analysis}

\subsection{Sensors under investigation}

\begin{figure}
	\centering
	 \includegraphics[width=10cm]{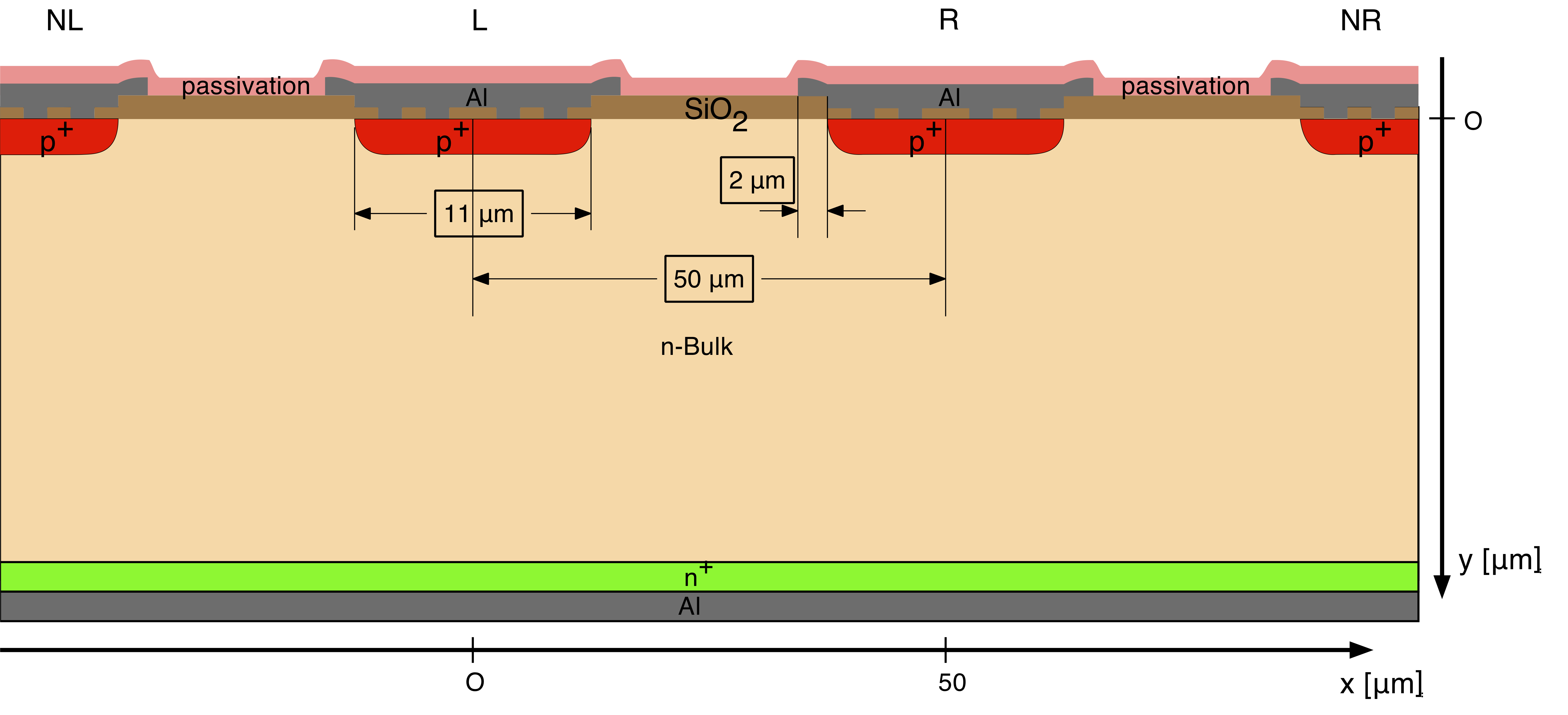}	
	\caption{Schematic layout of the strip region of the DC-coupled Hamamatsu $p^+n$ sensor, and definition
      of the $x$ and $y$ coordinates. The drawing is not to scale.}
	\label{fig:sensor}
\end{figure}

 Two different $p^+n$ strip sensors have been investigated: A DC-coupled sensor produced by Hamamatsu
 \cite{Hamamatsu},
 and an AC-coupled sensor produced by CiS
 \cite{CIS}.
 Sensors from two vendors have been studied to get an idea how much the observed effects depend on technology. Overall, the results found are similar. Finally we concentrated on the DC-coupled sensor because AGIPD
  \cite{AGIPD}
 will use a DC-coupled sensor, and detailed TCAD\footnote{TCAD = technology computer aided design} simulations of AC-coupled sensors are significantly more complex. The relevant parameters for both sensors are listed in Table~\ref{tab:sensors}. All measurements presented in this work refer to the Hamamatsu sensor whose cross-section is shown in Figure~\ref{fig:sensor}.
 The sensors are covered by a passivation layer with openings at the two ends of each strip for bonding. Both sensors were investigated as produced, and after irradiation with 12~keV photons to 1~MGy (SiO$_2$) followed by annealing for 60 minutes at 80$^\circ$C. The corresponding values for oxide charge density, $N_{ox}$, integrated interface trap density, $N_{it}$, and  surface current density, $I_{surf}$, shown in Table
  \ref{tab:irrad}
 have been derived from measurements on MOS capacitors and gate-controlled diodes from Hamamatsu
  \cite{Zhang:2011a, Zhang:2011b, Perrey:Thesis, Zhang:Thesis} and scaled to the measurement conditions of the sensor.
% produced on the same wafers as the sensors.

\begin{table}[b]
\centering
\begin{tabular}[b]{|c|c|c|}
\hline
producer 					& Hamamatsu 	& CiS \\ \hline
coupling 					& DC 			& AC\\ \hline
pitch 						& 50 $\upmu$m & 80 $\upmu$m 	\\ \hline
depletion voltage		    & $\sim$~155 V & $\sim$~63 V		\\ \hline
doping concentration		& $\sim$~10$^{12}$ cm$^{-3}$ & $\sim$~$8 \cdot 10^{11}$ cm$^{-3}$	\\ \hline
single strip capacitance 	& $\sim$~1.4 pF		& $\sim$~0.5 pF \\ \hline
rear side capacitance 		& $\sim$~12 pF		& $\sim$~24 pF \\ \hline
gap between $p^+$ implants	& 39 $\upmu$m 	& 60 $\upmu$m 	\\ \hline
width $p^+$ implant window	& 11 $\upmu$m	& 20 $\upmu$m	\\ \hline
depth $p^+$ implant  		& unknown 		& 1.2 $\upmu$m	\\ \hline
aluminium overhang 			& 2 $\upmu$m	& -2 $\upmu$m	\\ \hline
number of strips 			& 128        	& 98        	\\ \hline
strip length 				& 7.956 mm 		& 7.8 mm 		\\ \hline
sensor thickness			& 450 $\upmu$m	& 285 $\upmu$m	\\ \hline
thickness SiO$_2$			& 700 nm     	& 300 nm    	\\ \hline
thickness Si$_3$N$_4$	& none & 50 nm    	\\ \hline
passivation layer			& unknown & unknown    	\\ \hline
%thickness of Si$_3$N$_4$	& none	& 50 nm \\ \hline
crystal orientation 		& $\langle 1 1 1 \rangle$ & $\langle 1 0 0 \rangle$ \\ \hline
 \end{tabular}
   \caption{Parameters of the Hamamatsu sensor and of the CiS sensor.}
   \label{tab:sensors}
\end{table}

\begin {table}[b]
\centering
\begin{tabular}[b]{|c|c|c|c|}
\hline
   X-ray dose  & 0 Gy                          &  1 MGy (60 min. at 80$^\circ $C) \\ \hline
   $N_{ox}$    &  1.3$\cdot$10$^{11}$/cm$^{2}$ & 1.4$\cdot$10$^{12}$/cm$^{2}$ \\ \hline
   $N_{it}$    & 0.87$\cdot$10$^{10}$/cm$^{2}$ & 1.6$\cdot$10$^{12}$/cm$^{2}$ \\ \hline
   $I_{surf}$  & 9.8 nA/cm$^{2}$               & 2.2 $\upmu$A/cm$^{2}$        \\ \hline
 \end{tabular}
   \caption{Oxide charge density, $N_{ox}$, interface trap density integrated over the Si-band gap, $N_{it}$, and surface current density, $I_{surf}$, obtained from measurements on test structures (a MOS capacitor and a gate-controlled diode) produced by Hamamatsu. The values for a temperature of 22.9$^\circ$C before and after X-ray irradiation to 1~MGy and annealing for 60 minutes at 80$^\circ$C are presented. The actual measurements were taken at 21.8$^\circ$C and, for the irradiated structures after annealing for 10 minutes at 80$^\circ$C, and then scaled (scale factor $\sim $~0.7) to above values, which correspond to the measurement conditions of the sensor investigated.}
   \label{tab:irrad}
\end{table}

\subsection{TCT measurements} \label{sec:setup}

 To study the charge collection and charge transport in the sensors, the instantaneous currents which are induced in the electrodes by the moving charges are measured (Transient Current Technique - TCT
  \cite{Kraner:1993, Becker:2011, Becker:Thesis}).
% \cite{Becker:2010},
% \cite{Becker:Thesis}).
The multi-channel TCT setup described in detail in Ref. 
  \cite{Becker:Thesis}
 has been used for the measurements. Electron-hole pairs in the sensor close to its surface are generated by red light (660 nm) from a laser focussed to 3 $\upmu$m, which has an absorption length at room temperature of approximately 3.5 $\upmu$m. The number of generated $eh$-pairs is controlled by optical filters. Between 30 000 and few millions of $eh$-pairs are used for the measurements.
 The absolute number of $eh$-pairs produced has been obtained to an accuracy of $\sim $~5~\% from the integrated charge measured for non-irradiated sensors and the known gain of the amplifiers.
 The bias voltage (up to 500 V) is applied on the $n^+$-rear side of the sensor. The current signal is read out on the rear side and on two strips on the front side using Agilent 8496G attenuators, Femto HSA-X-2-40 current amplifiers with a rise time of 180~ps (10 to 90~\%) and a Tektronix digital oscilloscope with $2.5$~GHz bandwidth (DPO 7254). The readout strips are grounded through the DC-coupled amplifiers ($\sim$~50~$\Omega$ input impedance). The seven strips to right and the seven to the left of the readout strips are connected to ground by 50~$\Omega$ resistors.

 The collected charge per strip is calculated off-line by integrating the current signal. For most measurements the laser is used with a repetition rate of 1 kHz and the current is integrated over 25 ns. For longer integration times the noise increases and no significant change of the integrated charge is observed.

\subsection{Naive expectations from simulations and analysis method}

   In this section 2-D simulations using SYNOPSIS TCAD
 \cite{Synopsys, Schwandt:Thesis}
   are presented. They illustrate the distribution of the field and potential in the region of the strips and the Si-SiO$_2$ interface, and are used to calculate the weighting potentials
 \cite{Shockley:1938, Ramo:1939, Hamel:2008},
  which are required to estimate the expected signals, i.e. the charges induced in the readout strips and the rear electrode, as a function of the position of the injected light.

 The simulated electric potential in a strip sensor close to the Si-SiO$_2$ interface between the strip implants depends on the boundary conditions at the SiO$_2$ surface, the oxide charge density,  the density of charged interface states and the current distribution in the sensor.

\begin{figure}
	\centering
		\includegraphics[width=7cm]{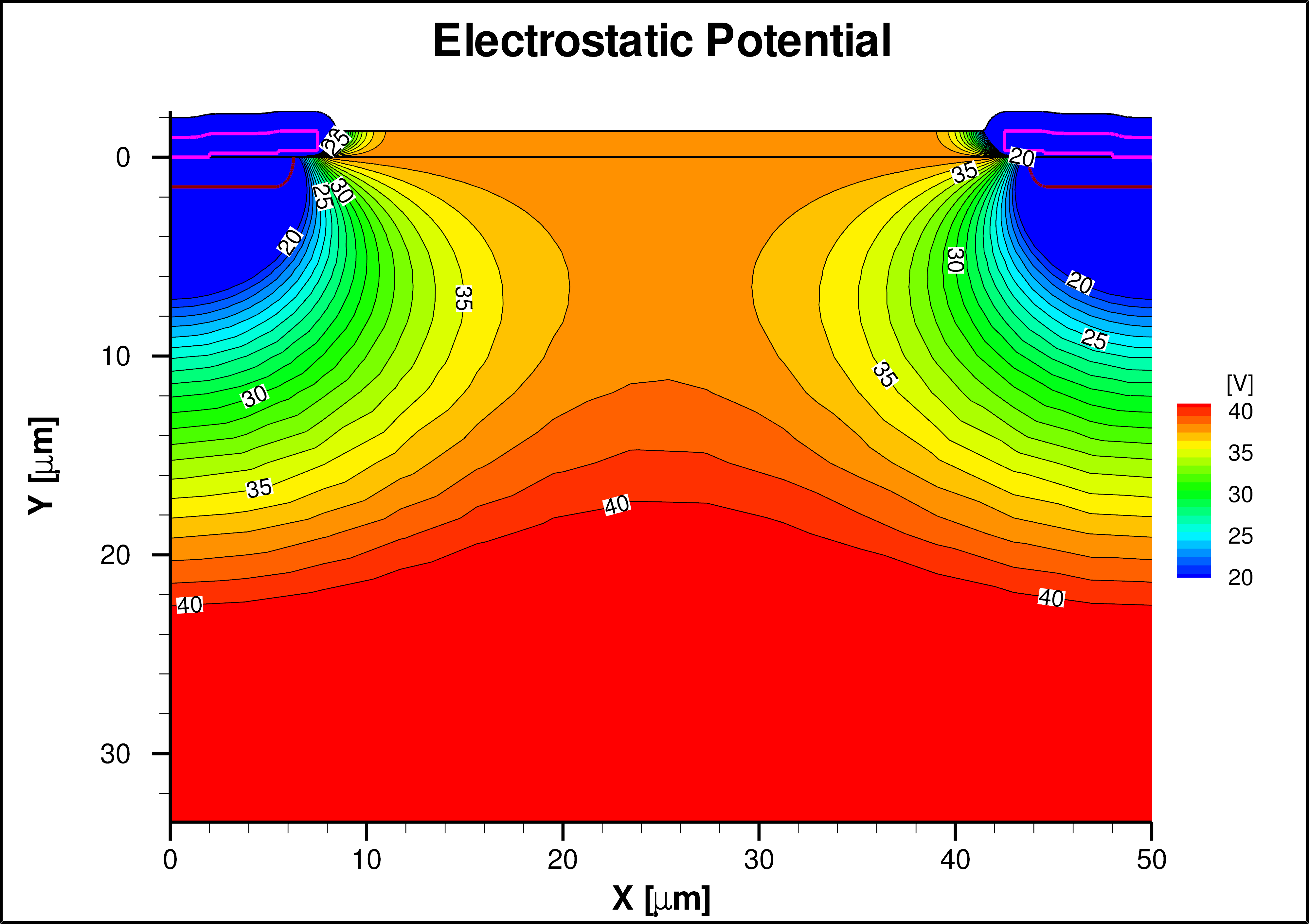}
		\includegraphics[width=7cm]{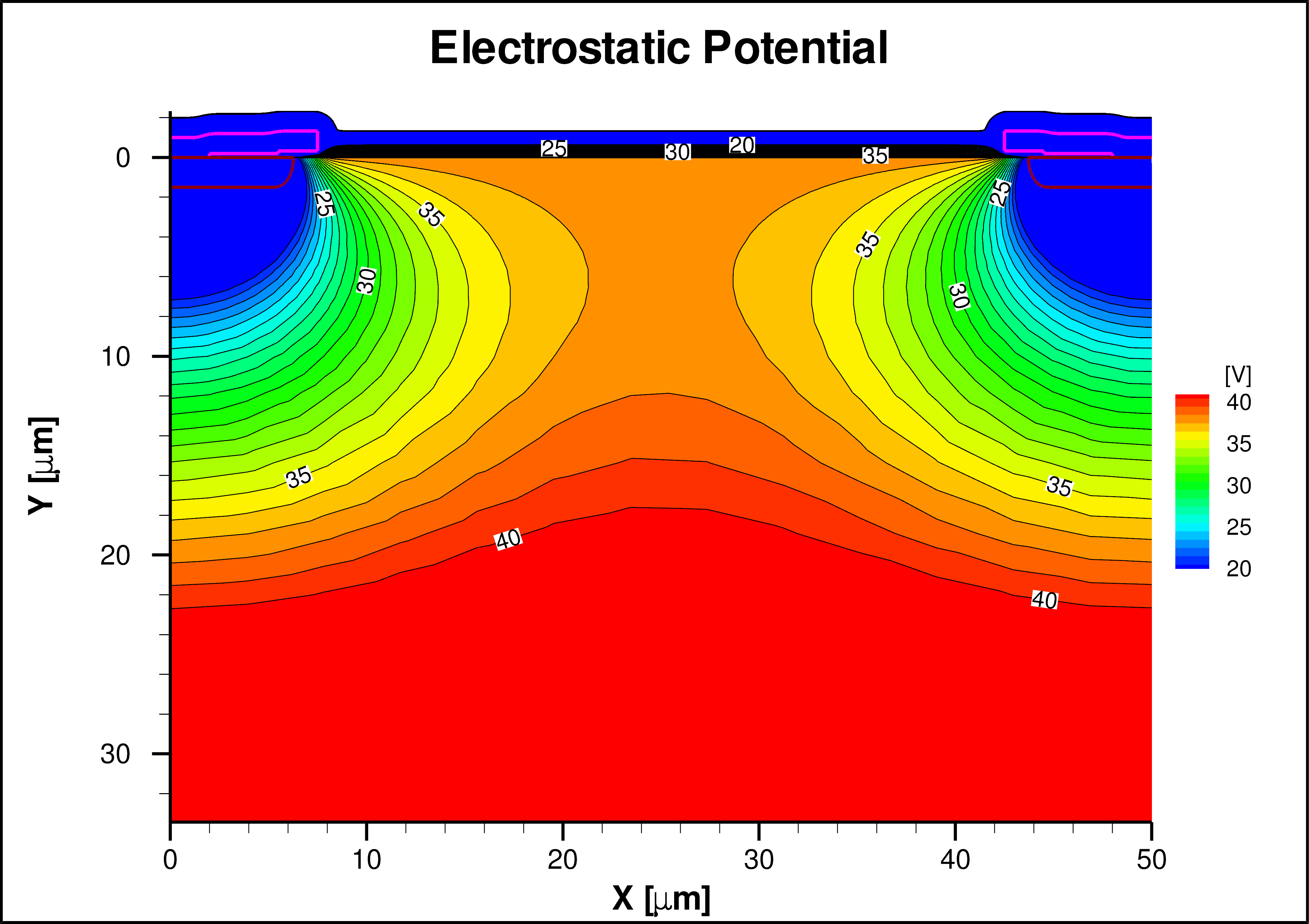}
	\caption{Electric potential for the Hamamatsu strip sensor calculated using SYNOPSIS  TCAD. Neumann boundary conditions (left) and Dirichlet boundary conditions (right) on the SiO$_2$ surface, a positive oxide charge density of $2\cdot 10^{12}$ cm$^{-2}$ and a surface current density of 8 $\upmu $A/cm$^2$ are assumed. The bias voltage, applied to the rear contact is 200~V. The sensor has a pitch of 50 $\upmu$m and a thickness of 450 $\upmu$m. Only the region 35 $\upmu$m from the strip surface is shown. The colour scale covers only the range between 20 and 40 V, and the distance between the equipotential lines is 1~V.}
    \label{fig:epot}
\end{figure}

   Figure \ref{fig:epot} shows for Neumann  (zero electric field perpendicular to the outer SiO$_2$ surface) and Dirichlet boundary conditions (zero potential at the outer SiO$_2$ surface) the electric potential for the Hamamatsu sensor with the strips at 0 V and the rear contact at $200$~V.

\begin{figure}
	\centering
	\includegraphics[width=7.4cm]{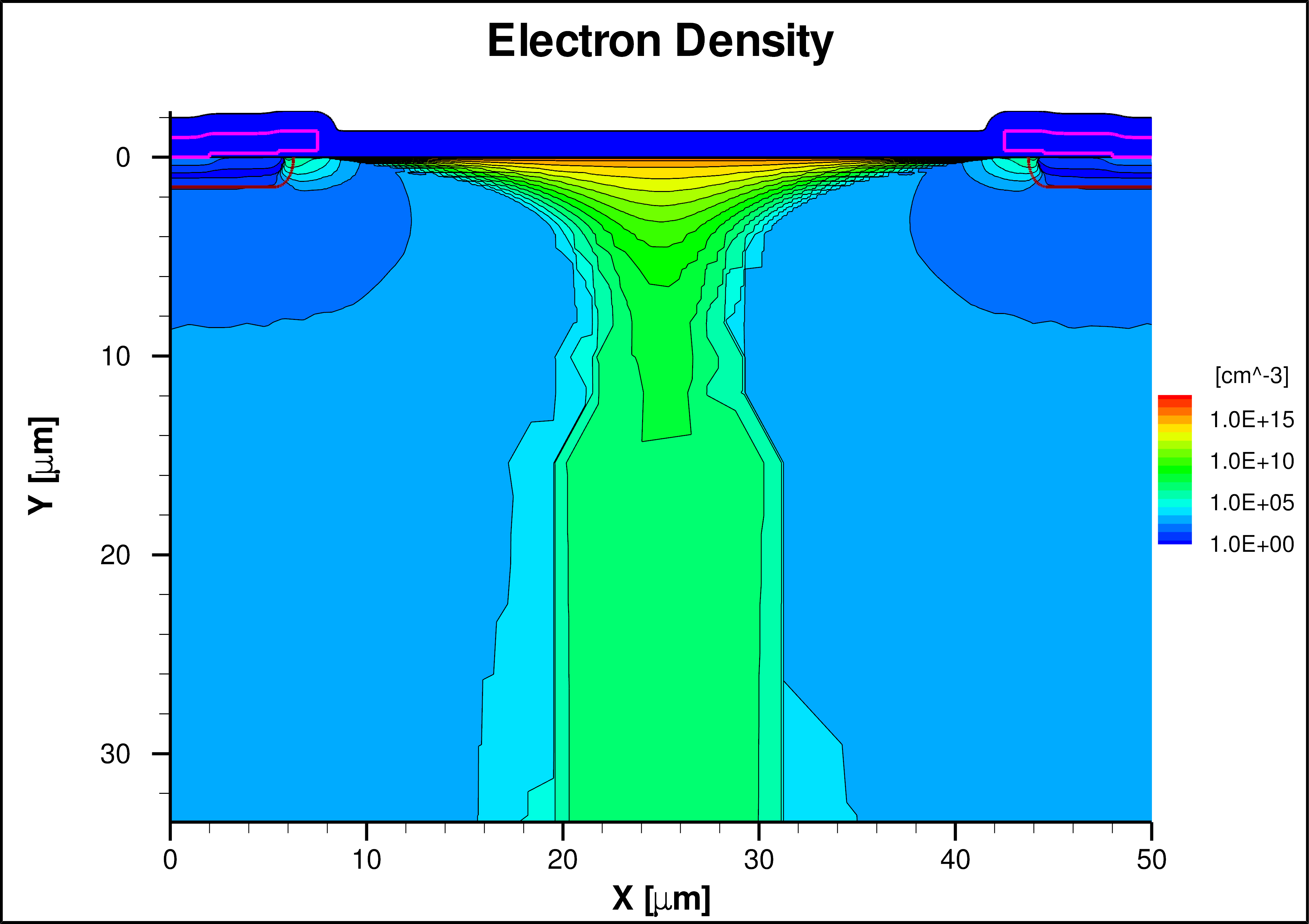}
	\includegraphics[width=7.4cm]{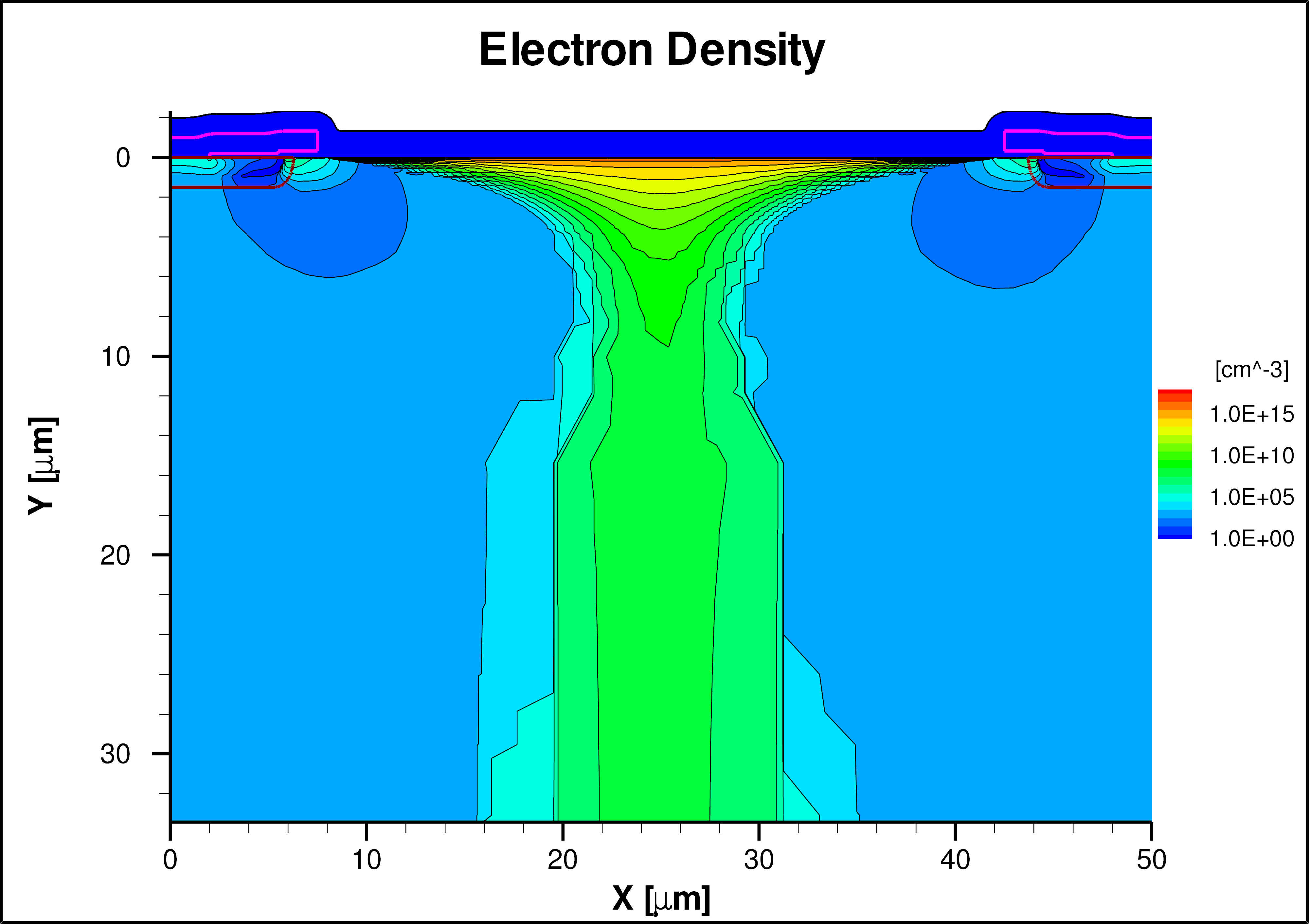}
	\caption{Free electron density for the Hamamatsu strip sensor.
     The conditions for the simulations are the same as for Figure \ref{fig:epot}.}
%  Only the region 30 $\upmu$m from the strip surface is shown.}

  \label{fig:edensity}
\end{figure}

 Figure~\ref{fig:edensity}
   shows the electron density for the same simulations.
  The high electron density of several 10$^{18}$ cm$^{-3}$ shows that an accumulation layer with a width of about 35~$\upmu $m has formed below the Si-SiO$_2$ interface. The dark current is mainly due to the surface generation current from the depleted Si-SiO$_2$ interface. The holes generated at the interface drift to the $p^+$ strips. The electrons drift along the field lines to the rear contact, as can be seen from the increased electron density at the symmetry plane between the strips for larger $y$ values.
  Between the readout strips, approximately 6~$\upmu $m below the Si-SiO$_2$ interface, the potential has a saddle point and the electric field points from the accumulation layer into the sensor. Thus, electrons produced close to the accumulation layer may not reach the rear contact during the integration time of the measurement.
  Simulations with an oxide  charge density of 10$^{11}$~cm$^{-2}$, which is typical for a non-irradiated sensor, also show an accumulation layer for  Neumann boundary conditions. For the Dirichlet boundary conditions however, the accumulation layer is absent. It should be noted that, given the uncertainties on the boundary conditions, the simulations presented only serve as an illustration.

\begin{figure}
	\centering
		\includegraphics[width=7.4cm]{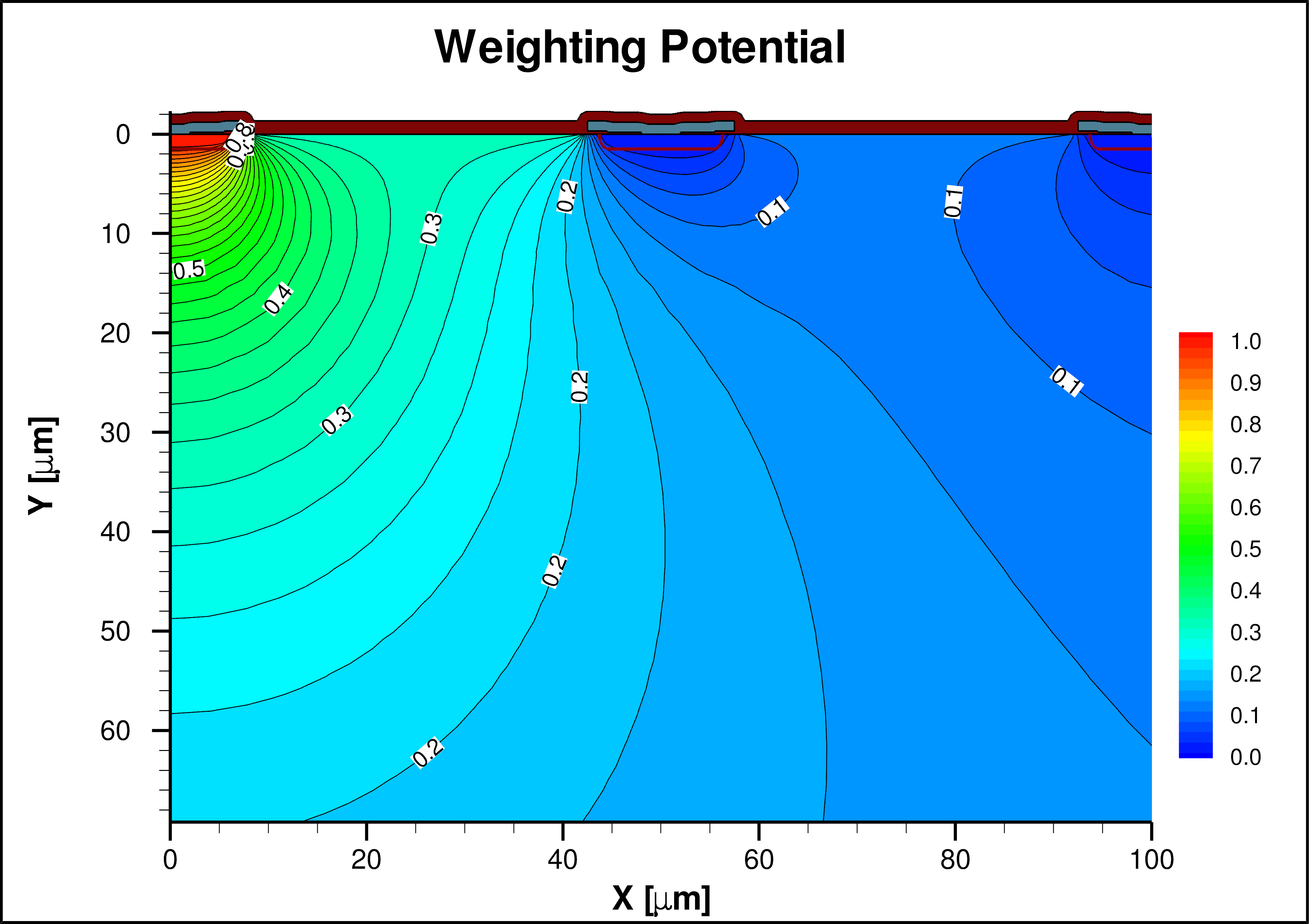}
		\includegraphics[width=7.4cm]{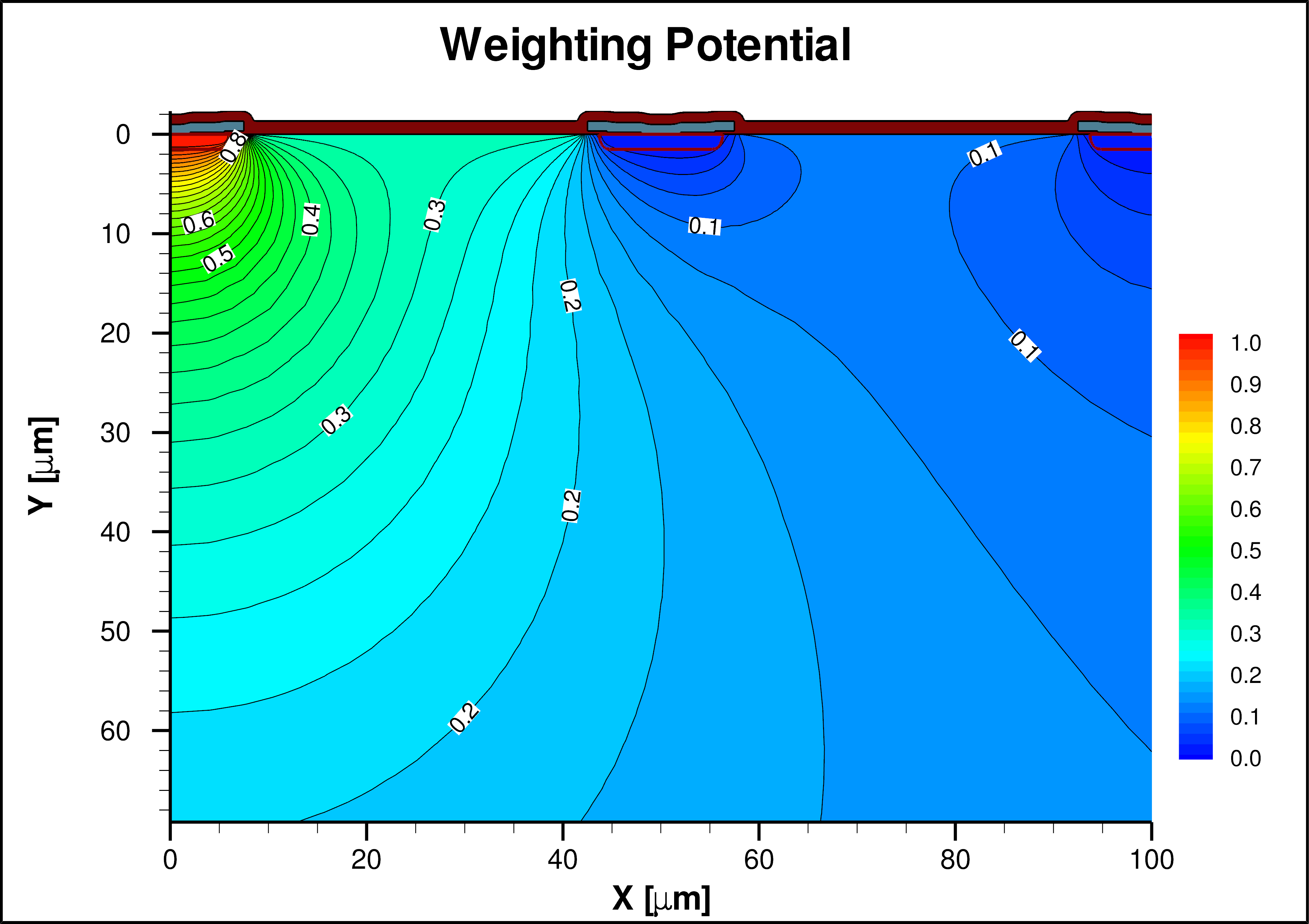}\\
		\includegraphics[width=7.4cm]{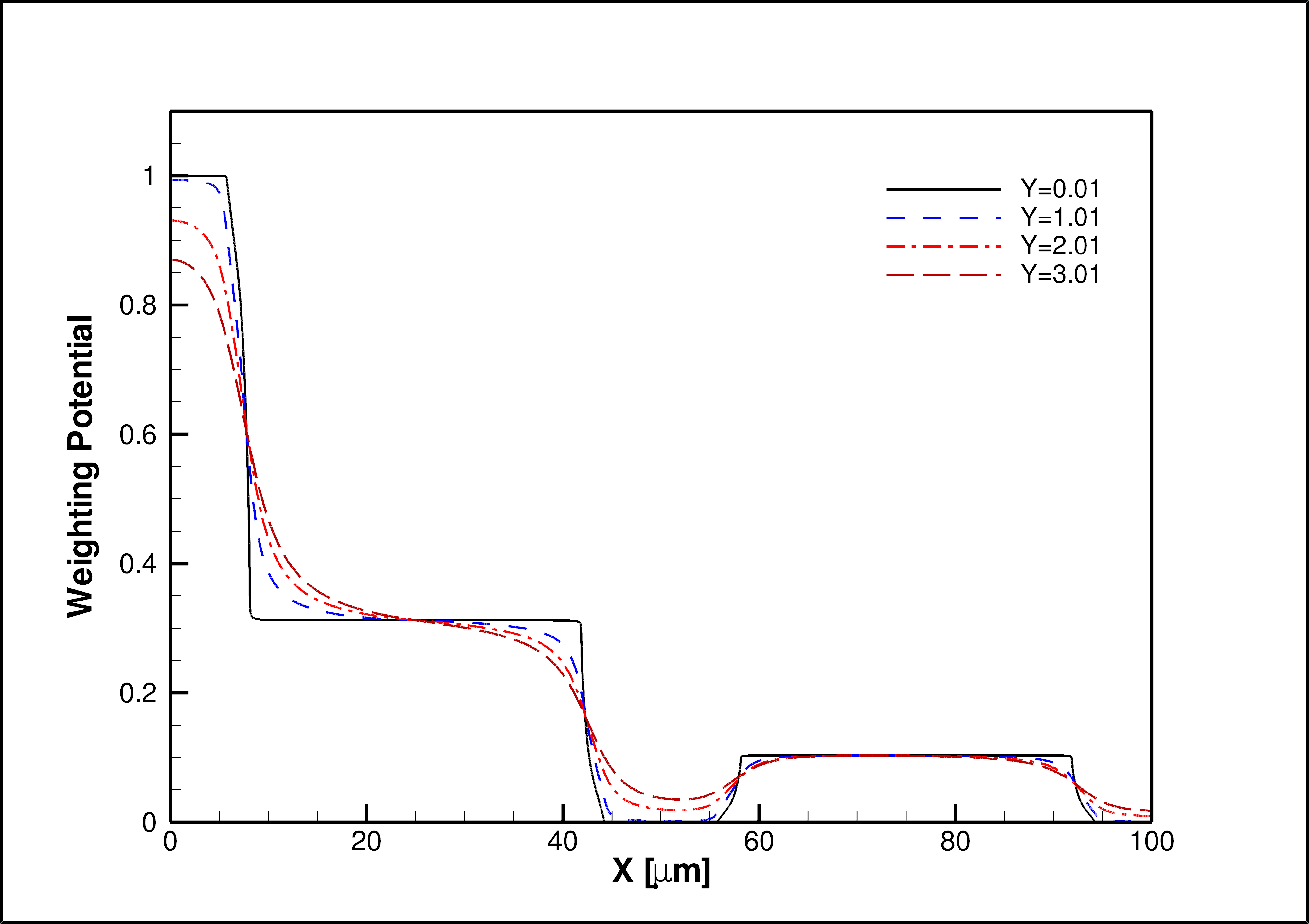}
		\includegraphics[width=7.4cm]{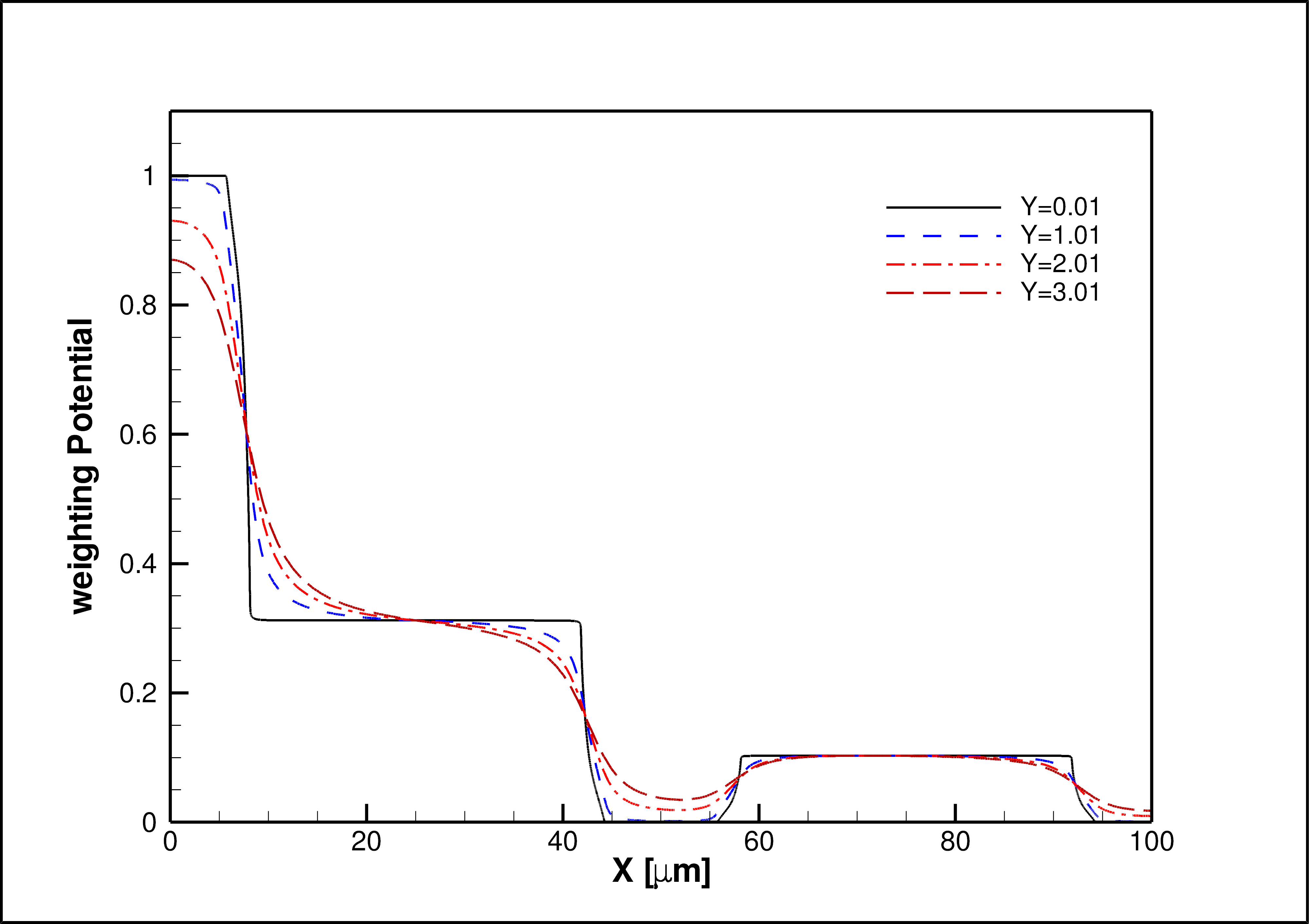}
		
	\caption{Weighting potential for strip L for the Hamamatsu strip sensor. The top row shows the two-dimensional distributions, the bottom row the one-dimensional distributions 0.01, 1.0, 2.0 and 3.0~$\upmu $m below the Si-SiO$_2$ interface. The parameters for the simulations are the same as for Figure \ref{fig:epot}.}
% Only 100 $\upmu$m depth is shown.

	\label{fig:wpot}
\end{figure}

 Figure
   \ref{fig:wpot}
 shows the weighting potential for strip L (centred at $x$ = 0), $\phi_{w,L}$,  for the above mentioned simulations of the Hamamatsu strip sensor. Both the two-dimensional distributions, as well as its $x$ dependence for  $y$ values of 0.01, 1, 2 and 3~$\upmu $m below the Si-SiO$_2$ interface are shown. $\phi_{w,L}$ is obtained from the difference of the potential calculated with strip L at 1~V and all other strips at 0~V minus the potential with all strips at 0~V. In both cases the backplane is biased to 200~V. In this way the effects of the mobile charge carriers in the accumulation layer are properly taken into account
   \cite{Hamel:2008}.
 One consequence is that $\phi_{w,L}$ is constant over the accumulation layer. We also note that, as expected, the weighting potential is hardly affected by the boundary conditions on the sensor surface.

%\include{chapters/analysis}
% \subsection{Analysis}

\begin{figure}
	\centering
		\includegraphics[width=7.4cm]{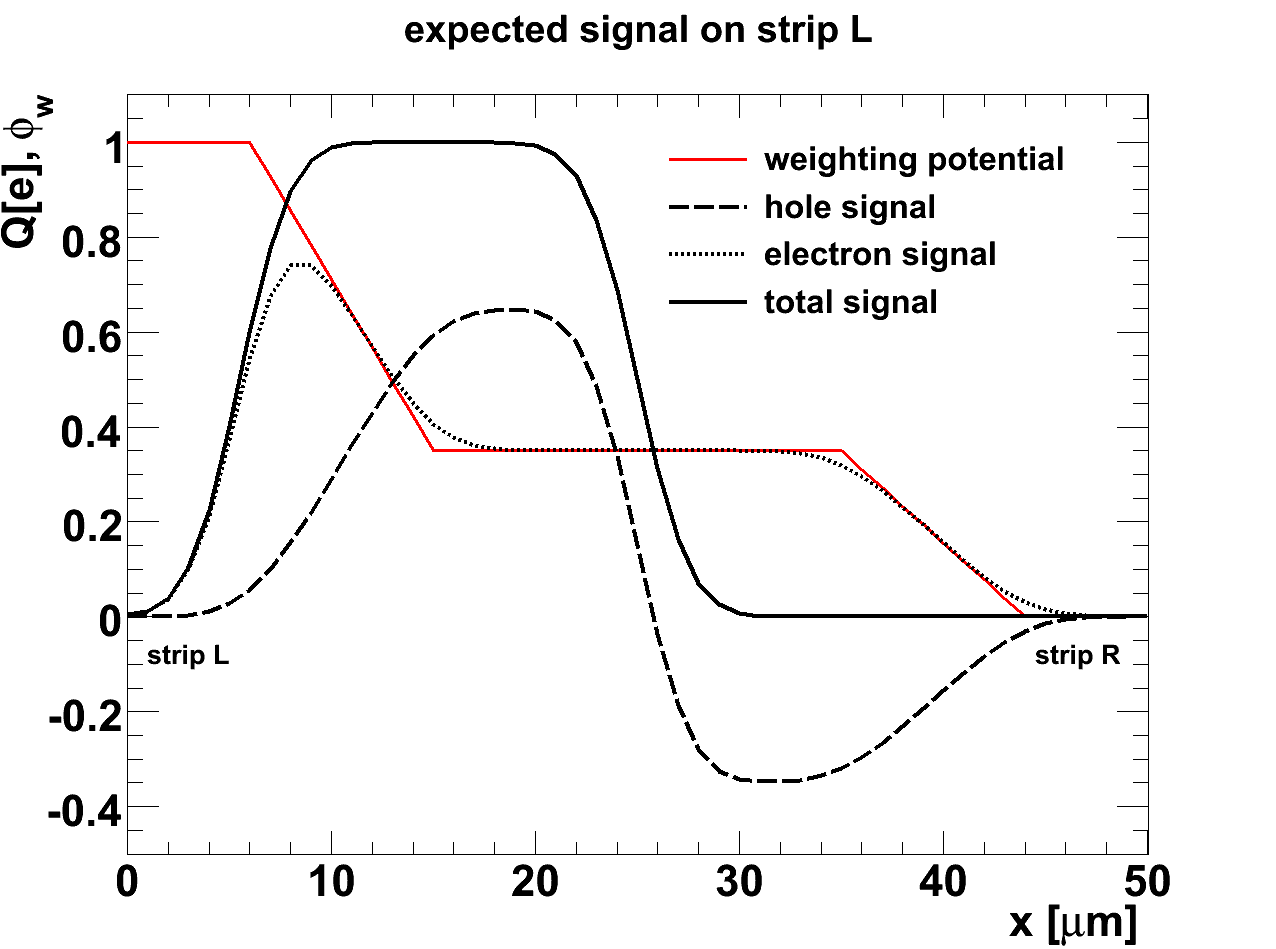}
		\includegraphics[width=7.4cm]{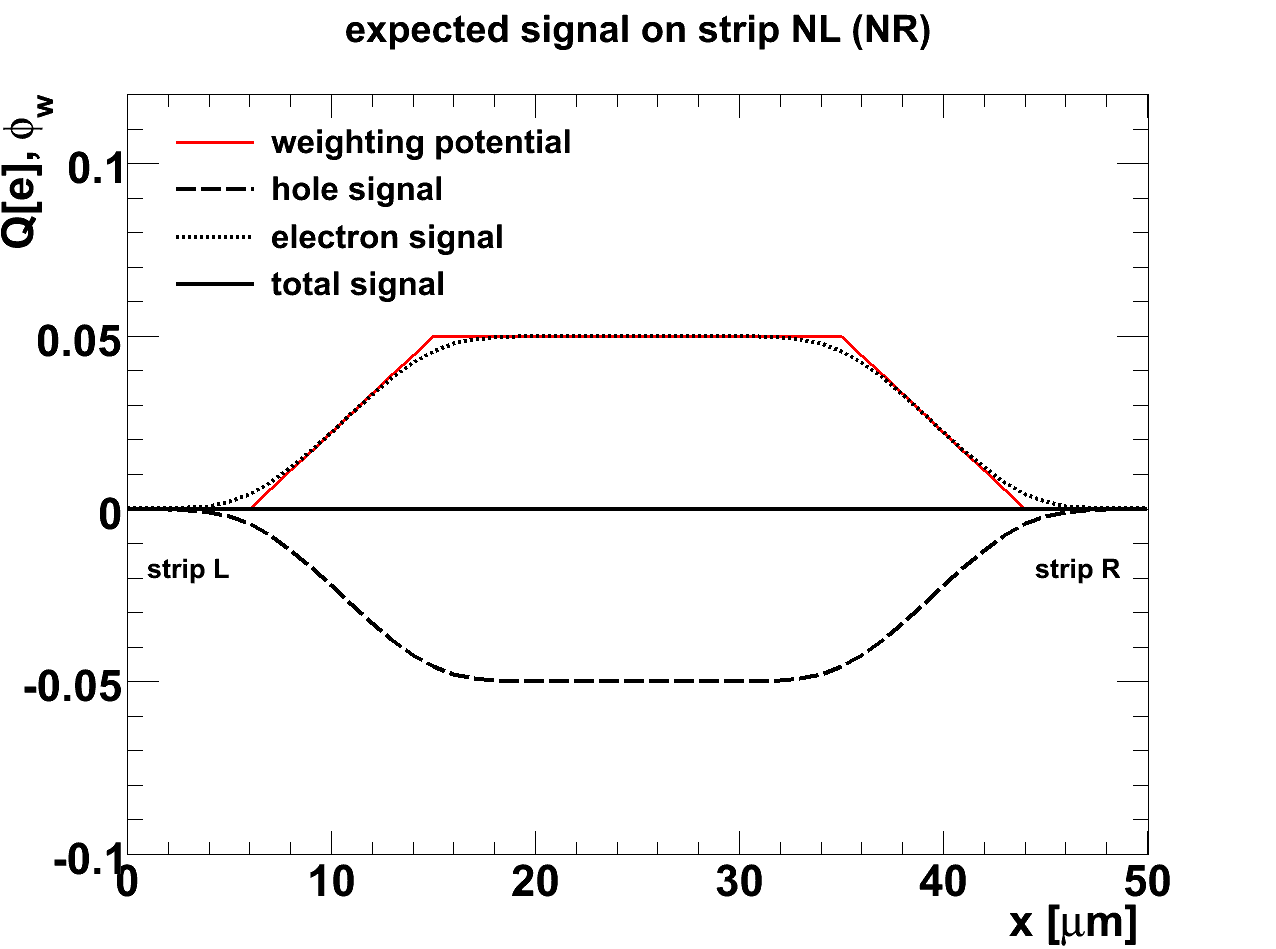}\\
		\includegraphics[width=7.4cm]{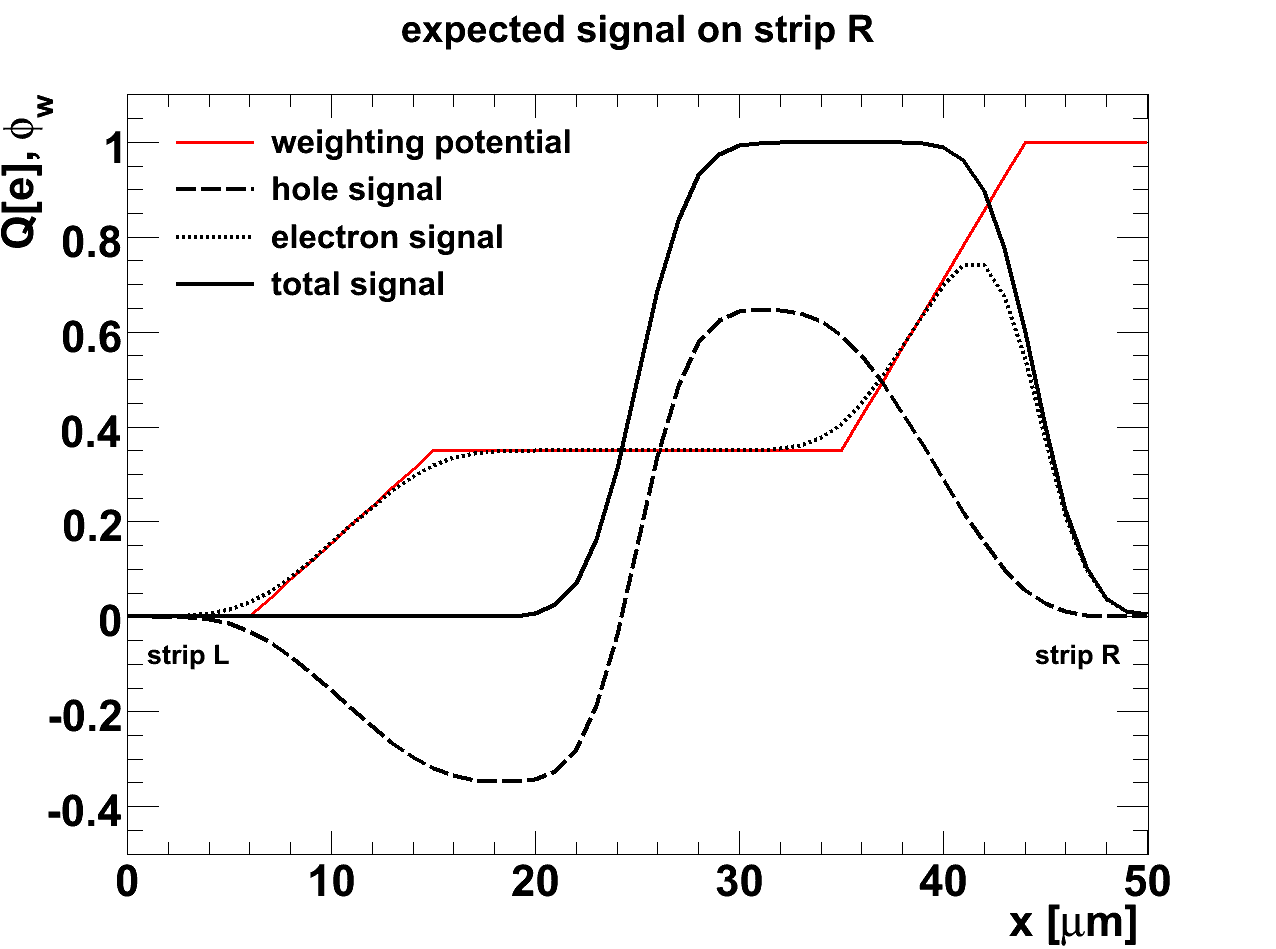}
		\includegraphics[width=7.4cm]{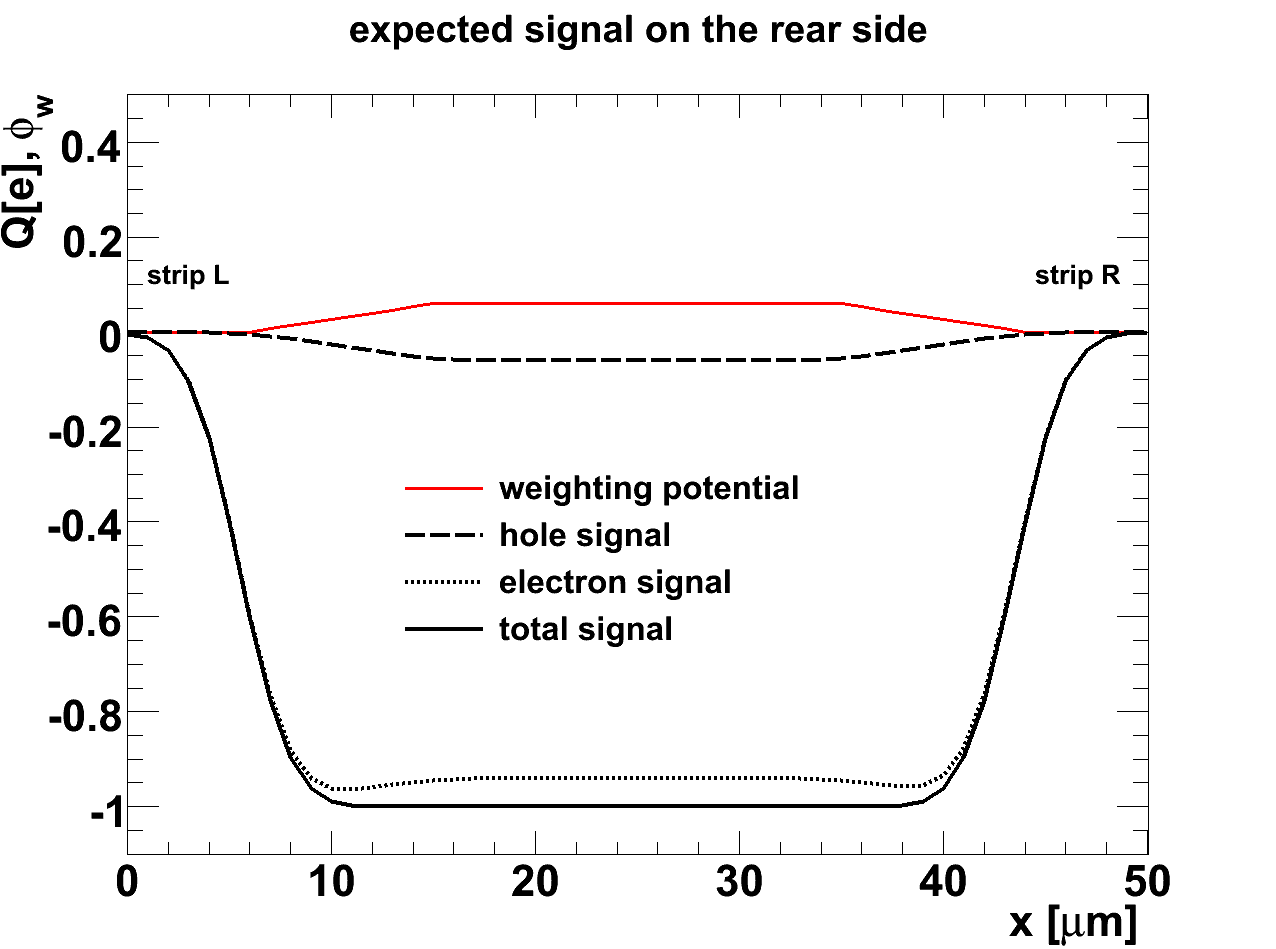}\\
		\includegraphics[width=7.4cm]{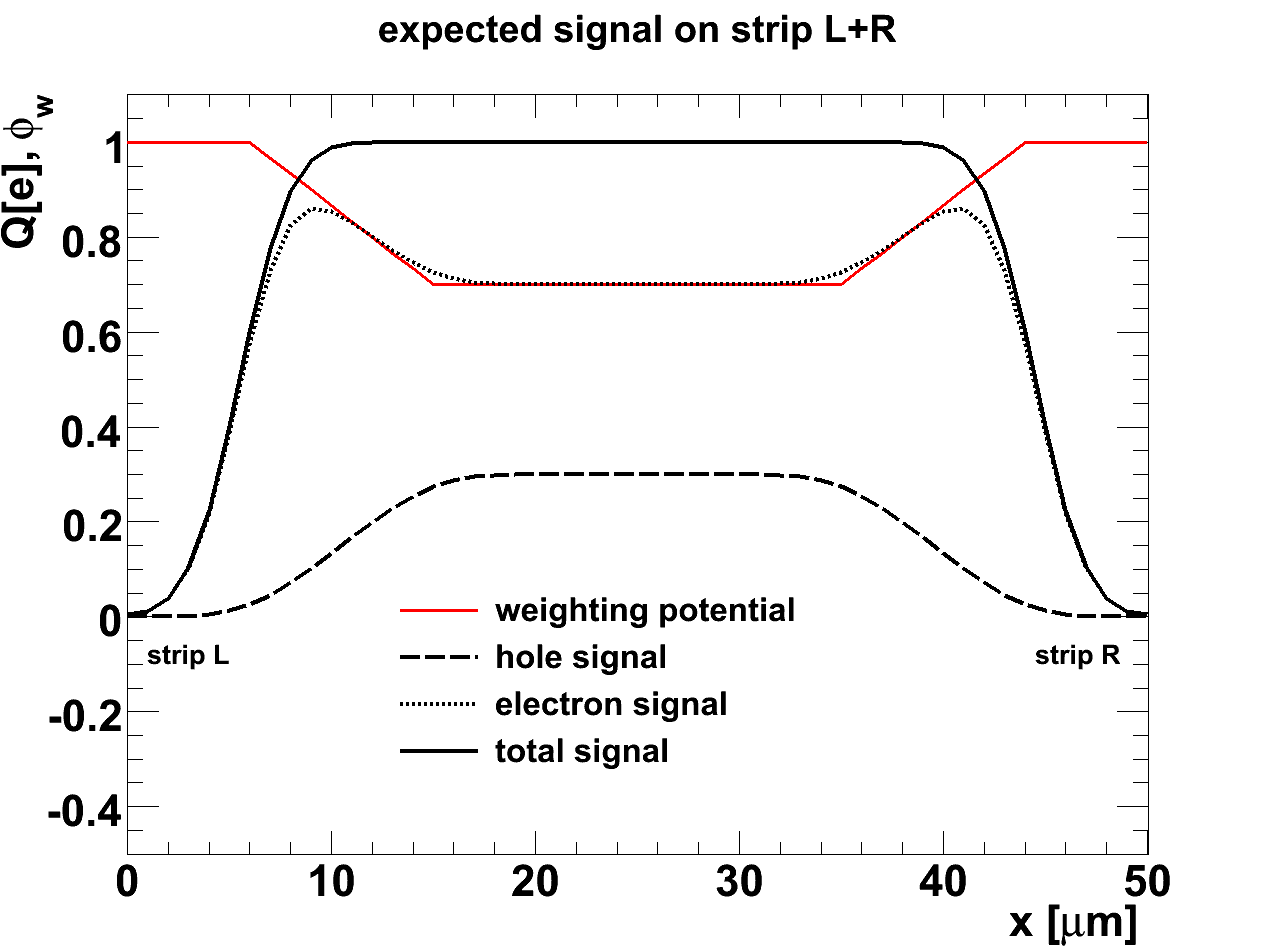}
		\includegraphics[width=7.4cm]{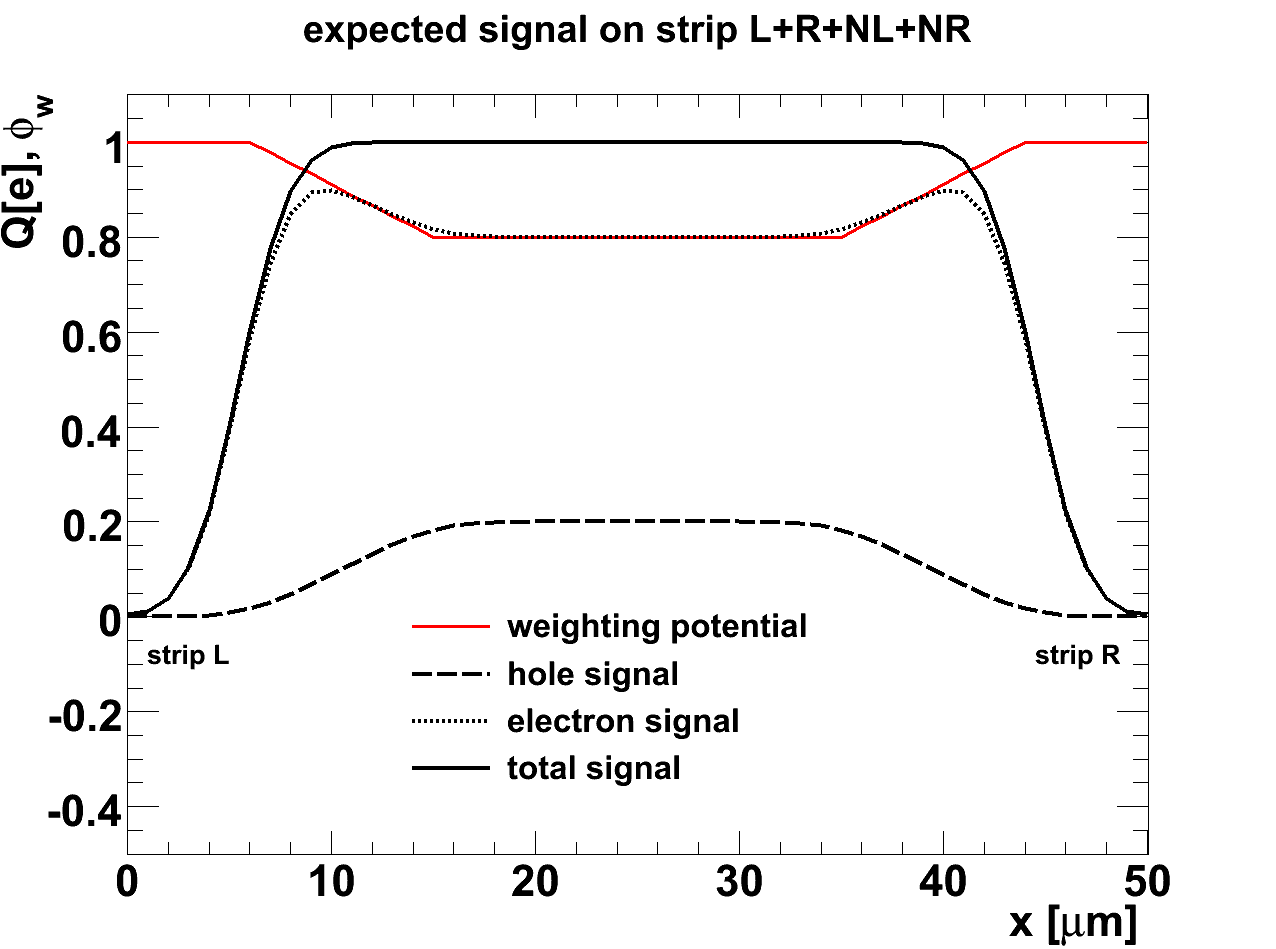}

  \caption{Weighting potential and expected integrated hole and electron signals as a function of the position $x$ of the light spot between the strips L and R for the situation of complete charge collection. The signals are normalised to one produced electron-hole pair. A Gaussian light profile with an rms width $\sigma=2$ $\upmu$m is used. For more details see text.}
	\label{fig:expected_charge}
\end{figure}

 In order to estimate the charge collected by the individual strips and by the rear contact separately for electrons and holes, the following assumptions for the spot size of the laser beam and for the weighting potential close to the Si-SiO$_2$ interface are made:

 \begin{itemize}
  \item
   The beam spot is assumed to have a Gaussian distribution with an rms width of 2~$\upmu$m. The light transmission is assumed to be zero for  aluminum and constant for SiO$_2$.
  \item
   The holes generated by the laser light, possibly only a fraction of them if hole losses occur, are collected by the closest strip. The electrons not trapped at the interface are collected by the rear electrode. Thus, diffusion effects are neglected.
  \item
   The charge $Q$ induced on electrode $i$ by a charge $q$ moving from position $x$ to electrode $j$ is given by $Q_i = q \cdot (\phi_{w}^i(j) - \phi_{w}^i(x))$, i.e. the product of the charge $q$ times the difference of the weighting potential at the electrode where the charge $q$ was collected minus the weighting potential at the position where it was generated. Note, that the sign of the induced signal $Q$ depends on the sign of the charge $q$. "Lost charges" are assumed to remain at the point where they have been generated, and therefore do not contribute to the signal.
  \item
   Inspired by the simulation shown in Figure \ref{fig:wpot}, we make following assumptions for the weighting potentials $\phi_{w}^i$. They are shown in Figure~\ref{fig:expected_charge}.
   By definition, $\phi_{w}^L$ is 1 at strip L, and 0 for all other strips and for the rear contact. Over the width of the accumulation layer (20~$\upmu$m in the Figure) $\phi_{w}^L$ is assumed to be constant, with a value of 0.35 at the  accumulation layers adjacent to L, and 0.05 at the next one. Between the edges of the Al strips and the edges of the accumulation layers weighting potentials linear in $x$ are assumed. The weighting potentials for the other strips considered are obtained by symmetry considerations.
  \item
   The weighting potential for the rear contact $\phi_{w}^{Rear}$ has a value of 1 at the rear contact, 0.06 at the accumulation layer, 0 at the strips, and a linear dependence in $x$ in-between.
 \end{itemize}

 For the situation where all electrons and holes generated are collected, Figure
   \ref{fig:expected_charge}
 shows separately for electrons, holes and for the sum of electrons and holes, the predictions for the charge collected by strips L and R, their sums L + R, the charge collected by the strips beyond the neighbours, NL and NR, the sum of  NL + L + R + NR, and the charge collected by the rear contact. A width of 20~$\upmu$m for the accumulation layer and the weighting potentials presented above, is assumed. The signals are normalised to one $eh$ pair generated and $q_0$ is the elementary charge. We note, that a hole produced close to the accumulation layer
 %to the left of the centre between the strips ($x<25$~$\upmu$m)
 and collected by strip L induces a positive signal in strip L and a negative signal in strip R. For the assumed weighting potential of 0.35 at the accumulation layer the hole signals are $Q_L^h = q_0 \cdot (1 - 0.35) =  0.65 \cdot q_0$ and $Q_R^h = q_0 \cdot (0 - 0.35) = - 0.35\cdot q_0$, respectively. An electron which drifts from the accumulation layer to the rear contact induces a positive signals $Q_L^e = Q_R^e = -q_0 \cdot (0 - 0.35) = 0.35\cdot q_0$. If for an $eh$ pair the hole is collected by strip L and the electron by the rear contact, the signals are $Q_L = Q_L^e + Q_L^h = q_0$ and $Q_R^e + Q_R^h = 0$. In the two strips NL and NR, positive electron and negative hole signals of $0.05\cdot q_0$ are induced, which cancel in the sum. In the rear contact a small negative hole signal ($- 0.06\cdot q_0$) and a much larger negative electron signal ($- 0.94\cdot q_0$) are induced, which again add to $q_0$.

 At the centre between the strips, at $x = 25$~$\upmu$m, the hole signal shows a rapid transition from positive to negative, whereas the electron signal remains constant.
 As expected for the situation without charge losses and no diffusion, the strip closest to the light spot and the rear contact record the total generated charge. However, current pulses are also induced in non-adjacent strips, as the collection time for the holes, which are produced close to the collecting strip, is much shorter than for the electrons, which drift through the entire detector. The integrals however are zero. In Section 3.1 measurements of such current signals will be shown.

\begin{figure}
	\centering
		\includegraphics[width=7.4cm]{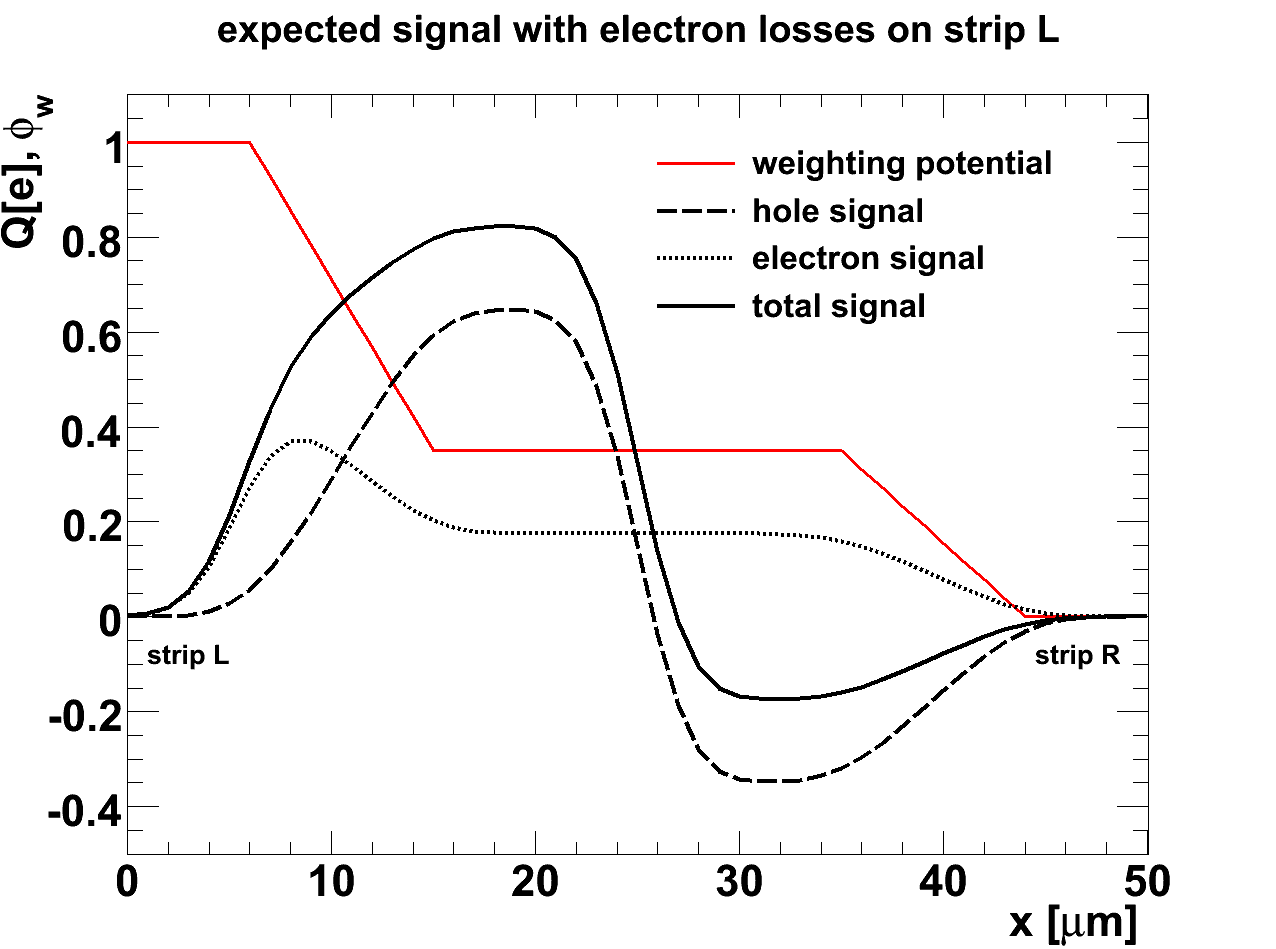}
		\includegraphics[width=7.4cm]{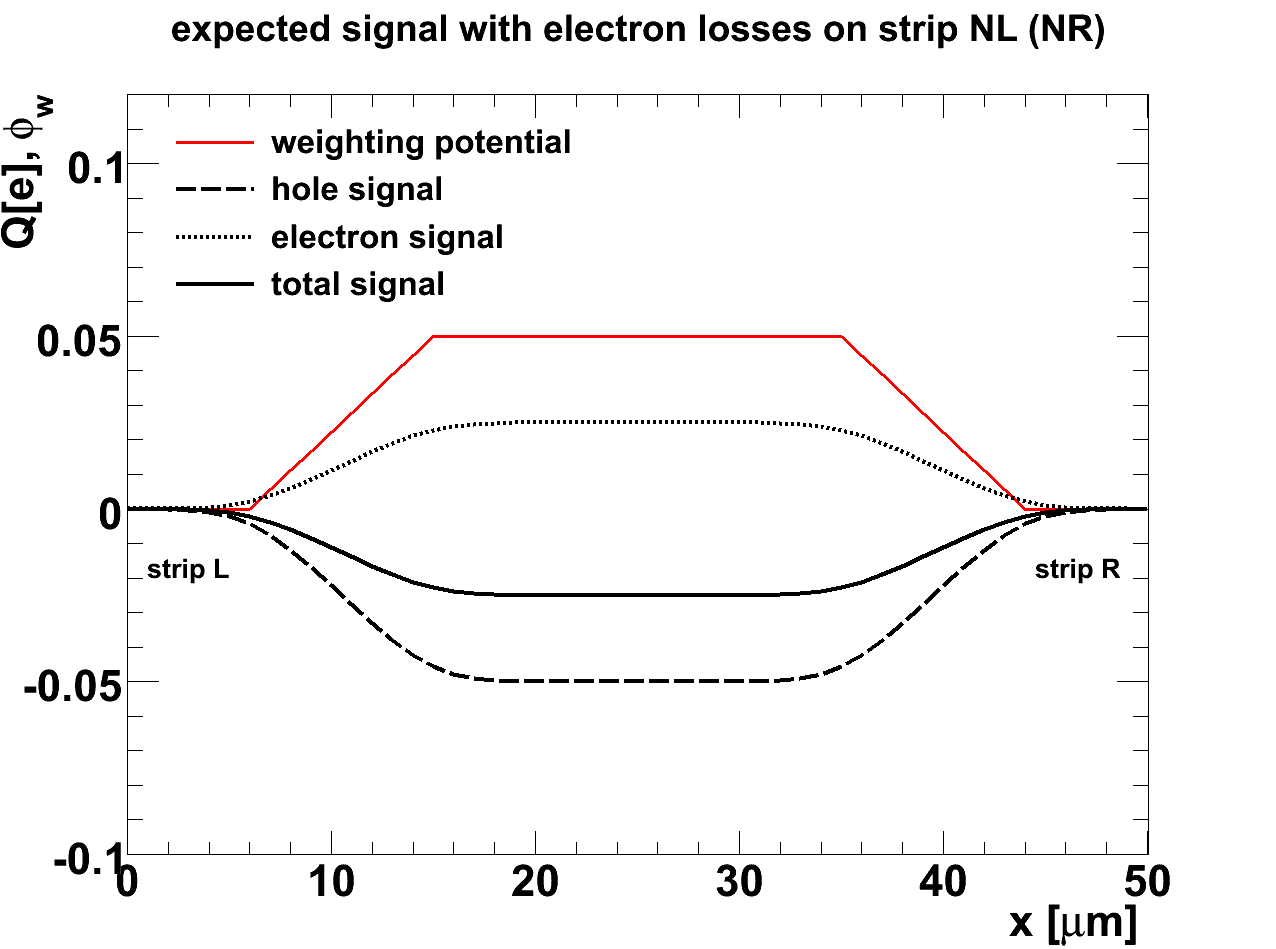}\\
		\includegraphics[width=7.4cm]{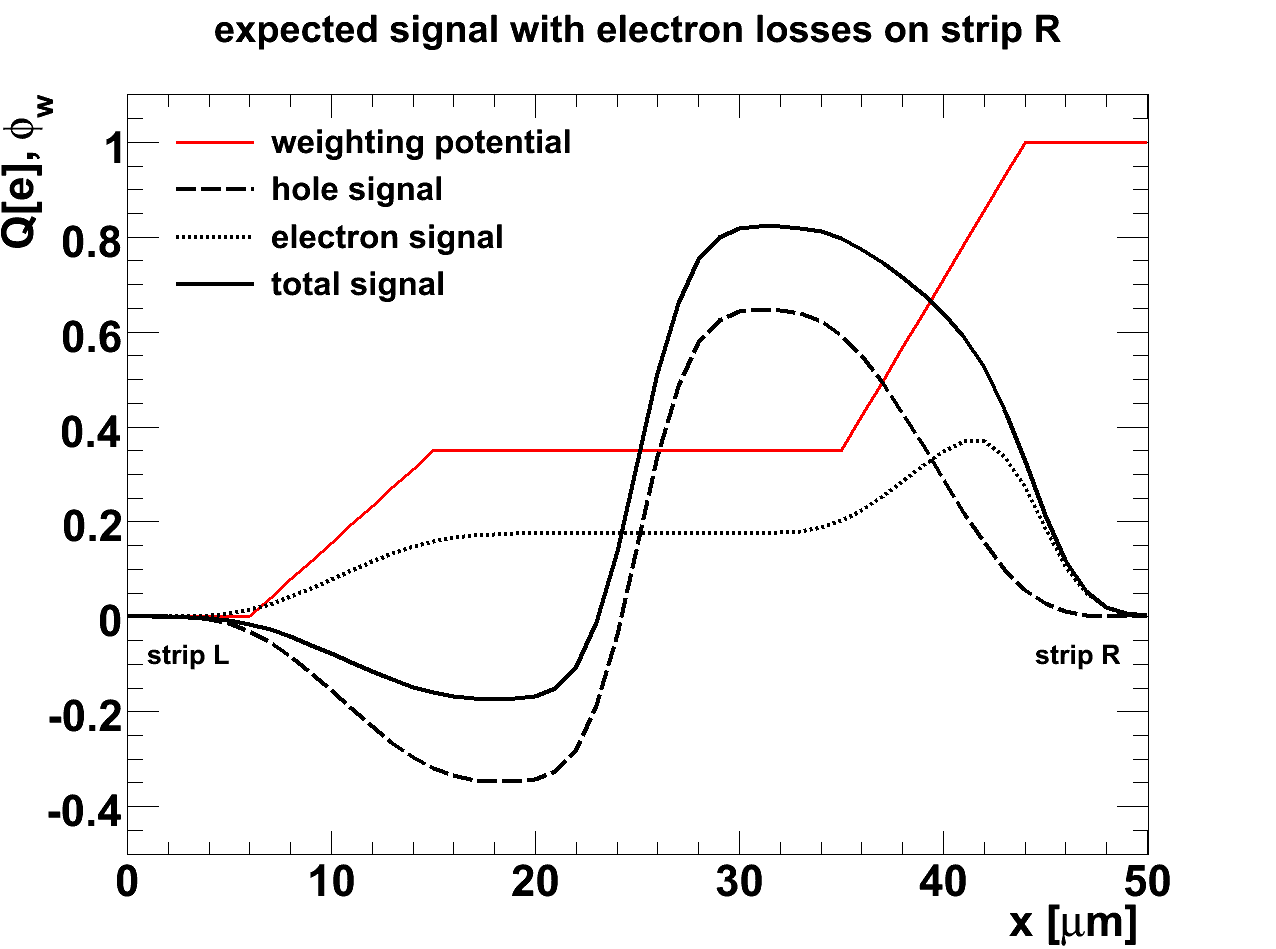}
		\includegraphics[width=7.4cm]{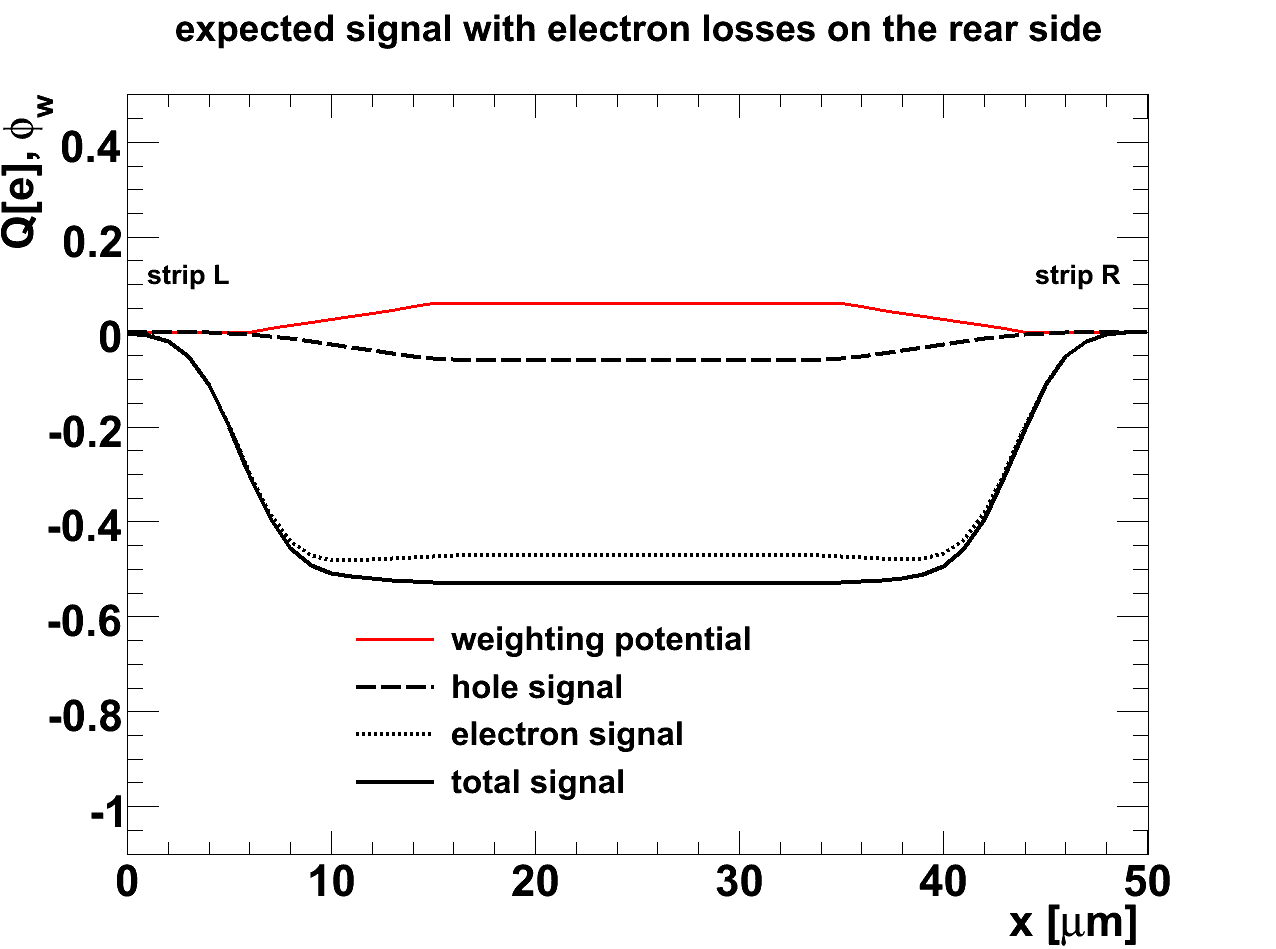}\\
		\includegraphics[width=7.4cm]{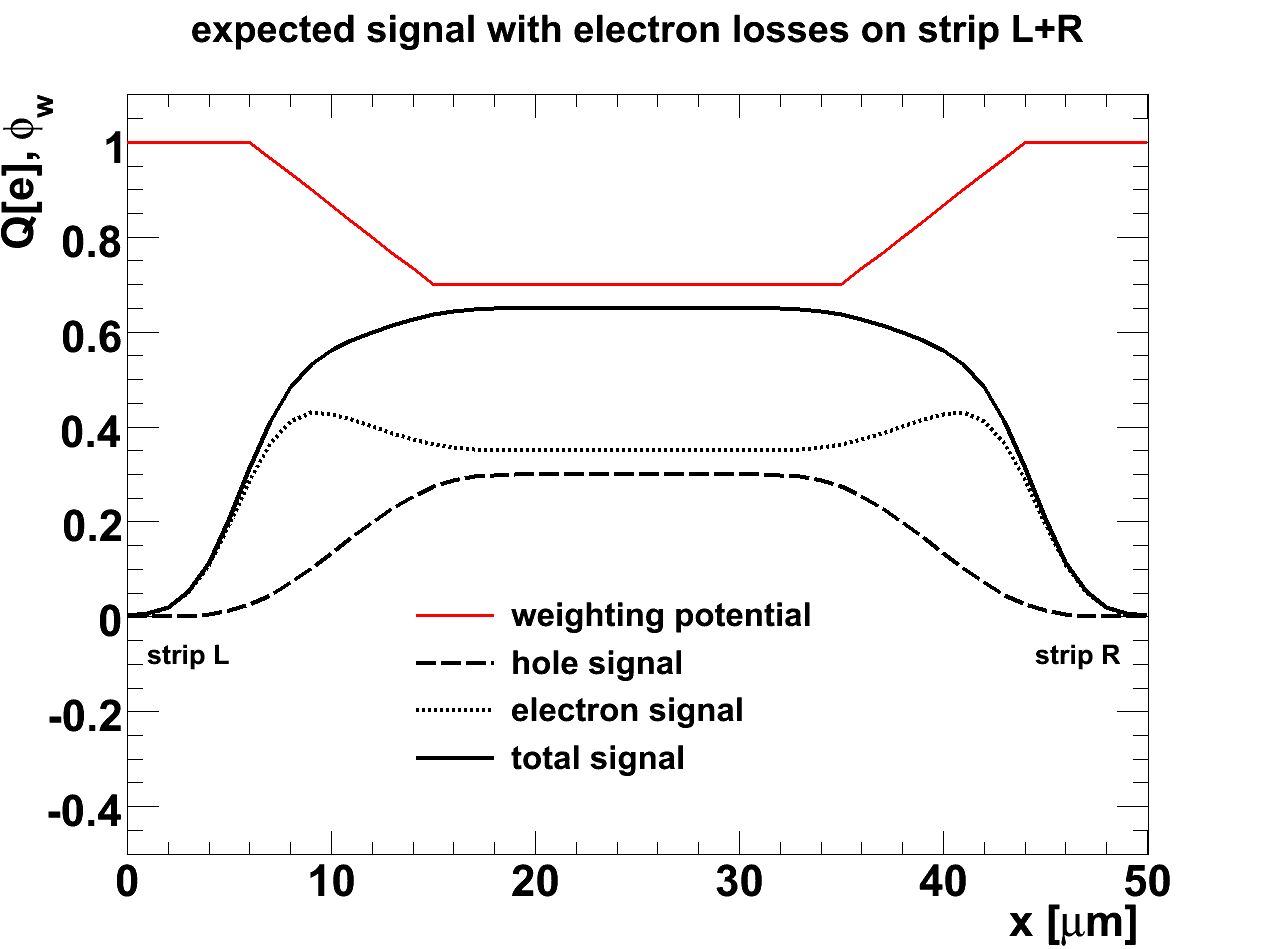}
		\includegraphics[width=7.4cm]{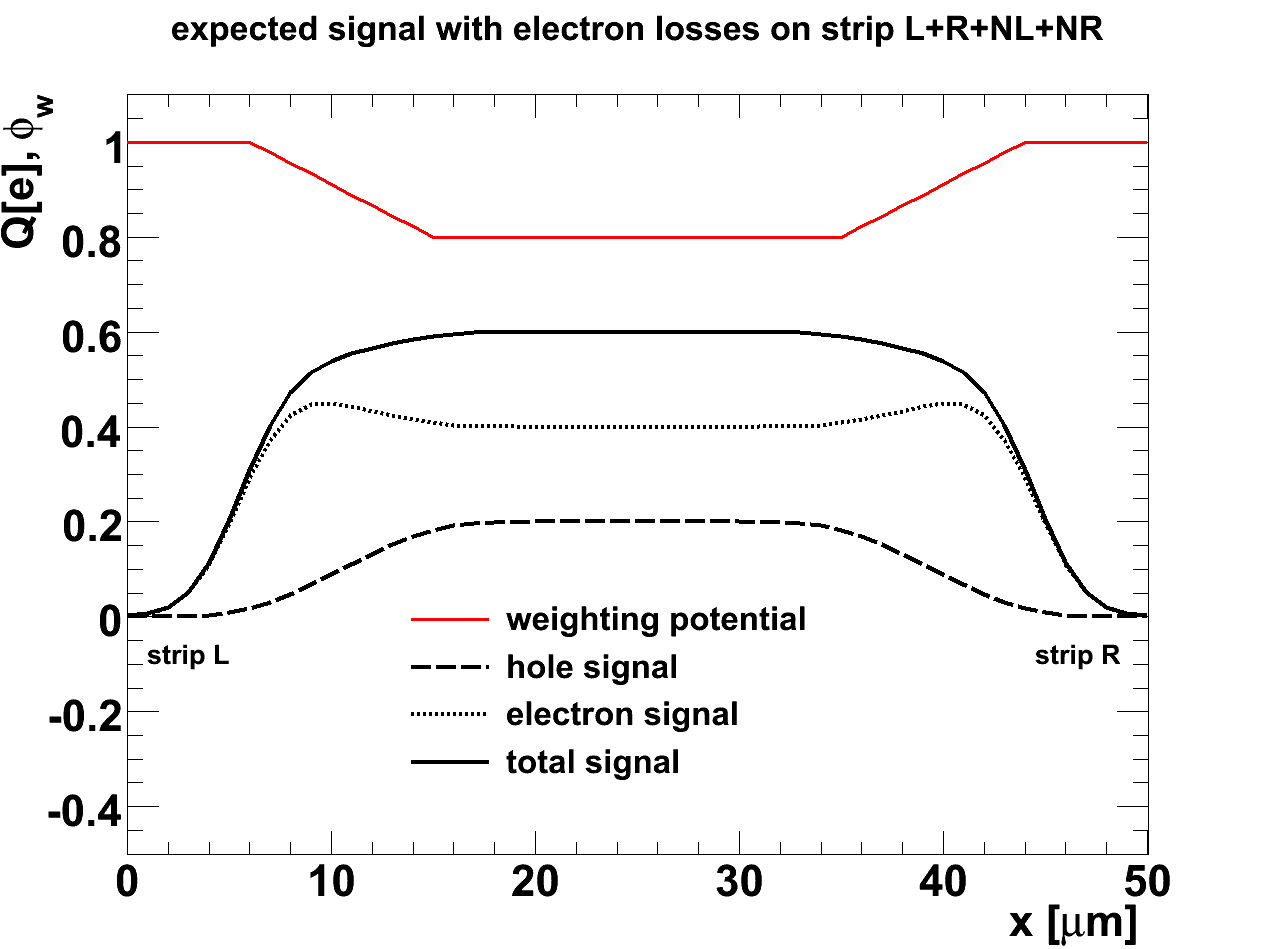}
  \caption{Weighting potential and expected hole and electron signals as a function of the position $x$ of the light spot between the strips L and R for the situation of 100~$\%$ hole collection and 50~$\%$ electron losses. The signals are normalised to one produced electron-hole pair. For more details see text.}

	\label{fig:expected_charge_electronloss}
\end{figure}

 Figure~\ref{fig:expected_charge_electronloss}
   shows the predicted charge distributions for complete hole collection and 50~$\%$ electron losses. Here the fraction of lost electrons is assumed to be independent of the position $x$ of illumination. It is assumed that the electrons are trapped at the $x$ position of the Si-SiO$_2$ interface where they were produced for a time longer than the integration time used in the analysis (25~ns). The weighting potential at this position is used to calculate the charge losses. A reduction of the sum of the electron plus hole signals from the neighbouring strips L and R to about 65~$\%$ is predicted. The signal on the rear side is reduced to about 50~$\%$. In addition, negative signals of about 3~$\%$ appear at the next to neighbouring strips NL and NR. In a similar way the signals for hole losses can be obtained by multiplying the signals due to the holes in
 Figure~\ref{fig:expected_charge}
   with the fraction of holes collected. In this case a positive signal is predicted for strips NL and NR.

   When comparing the measurements, which are shown in Chapter 3, to the predictions discussed above, differences are found, and some of the assumptions have to be changed for a quantitative description of the data. We note that these changes have been first derived from the experimental data and later, at least to some extent, explained by the simulations presented in Chapter 4.

\begin{figure}
	\centering
		\includegraphics[width=7.4cm]{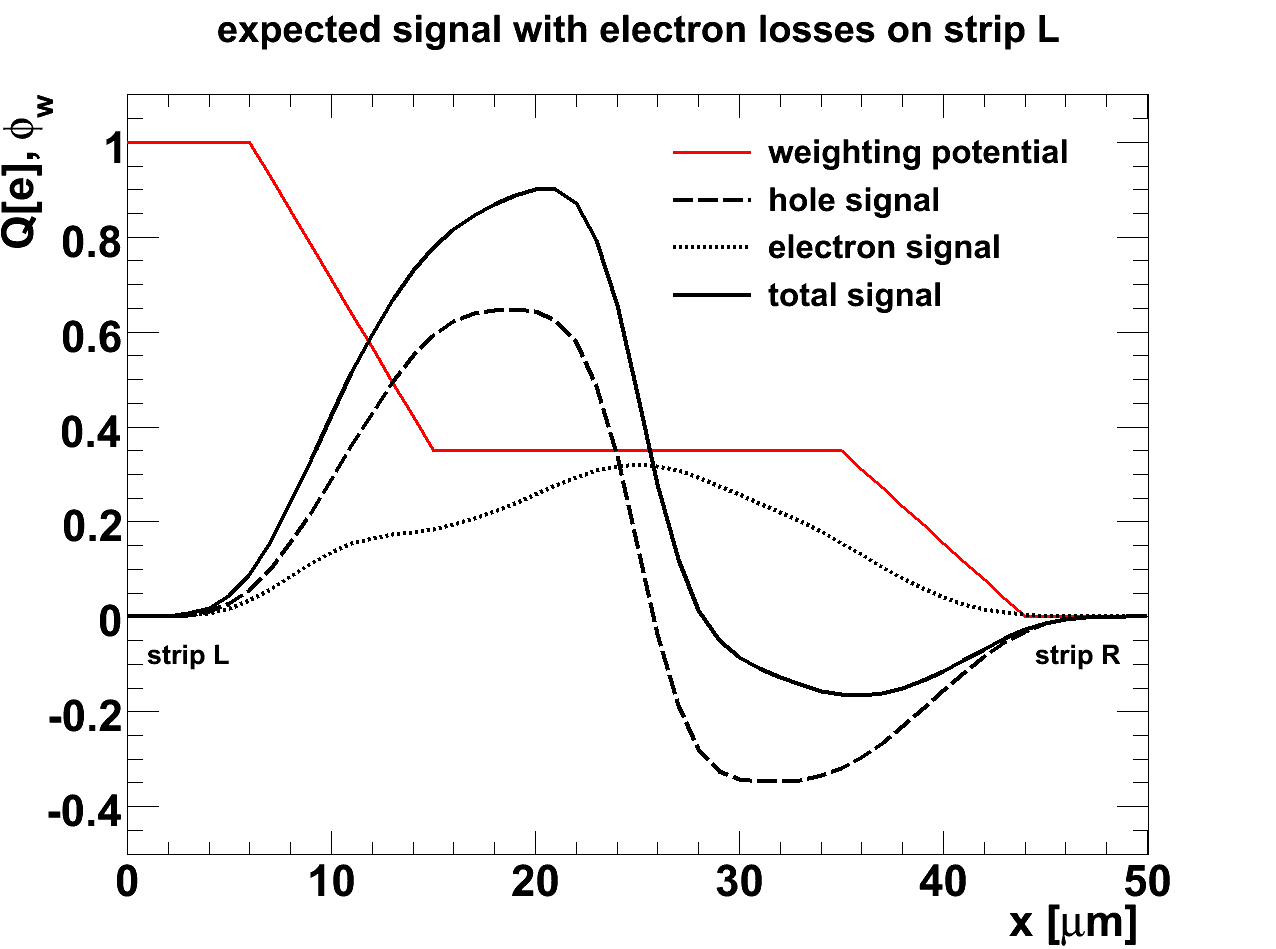}
		\includegraphics[width=7.4cm]{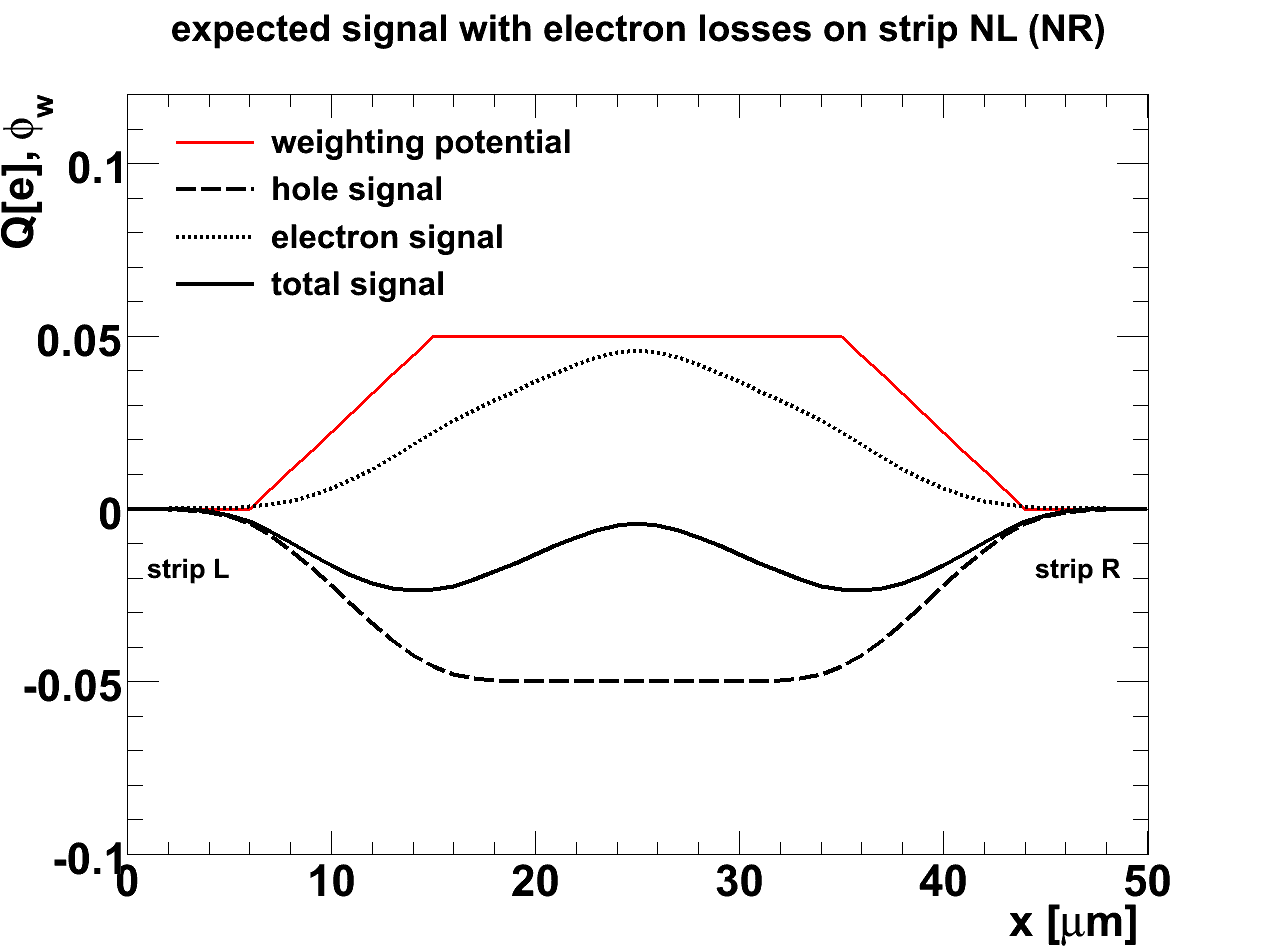}\\
		\includegraphics[width=7.4cm]{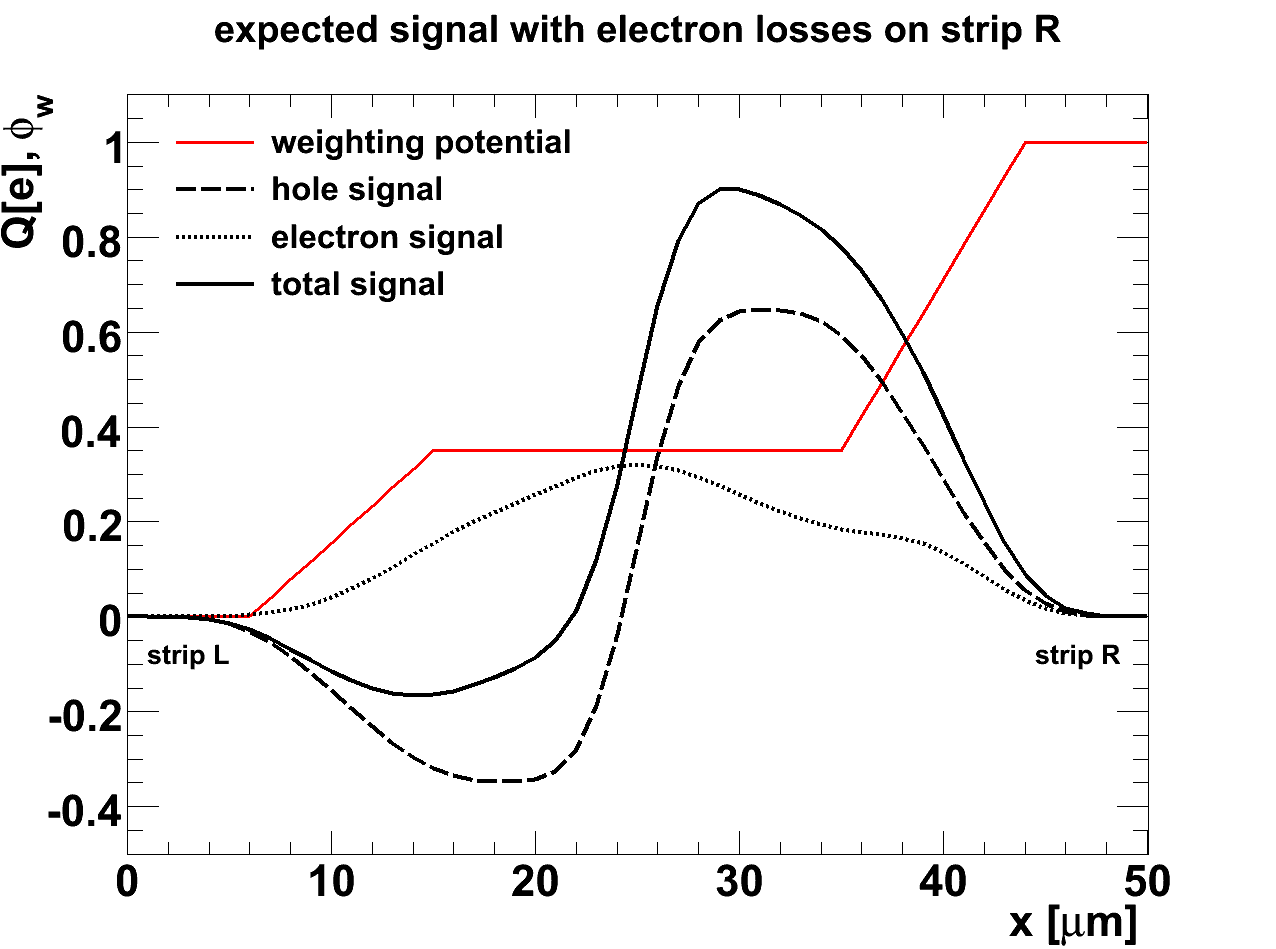}
		\includegraphics[width=7.4cm]{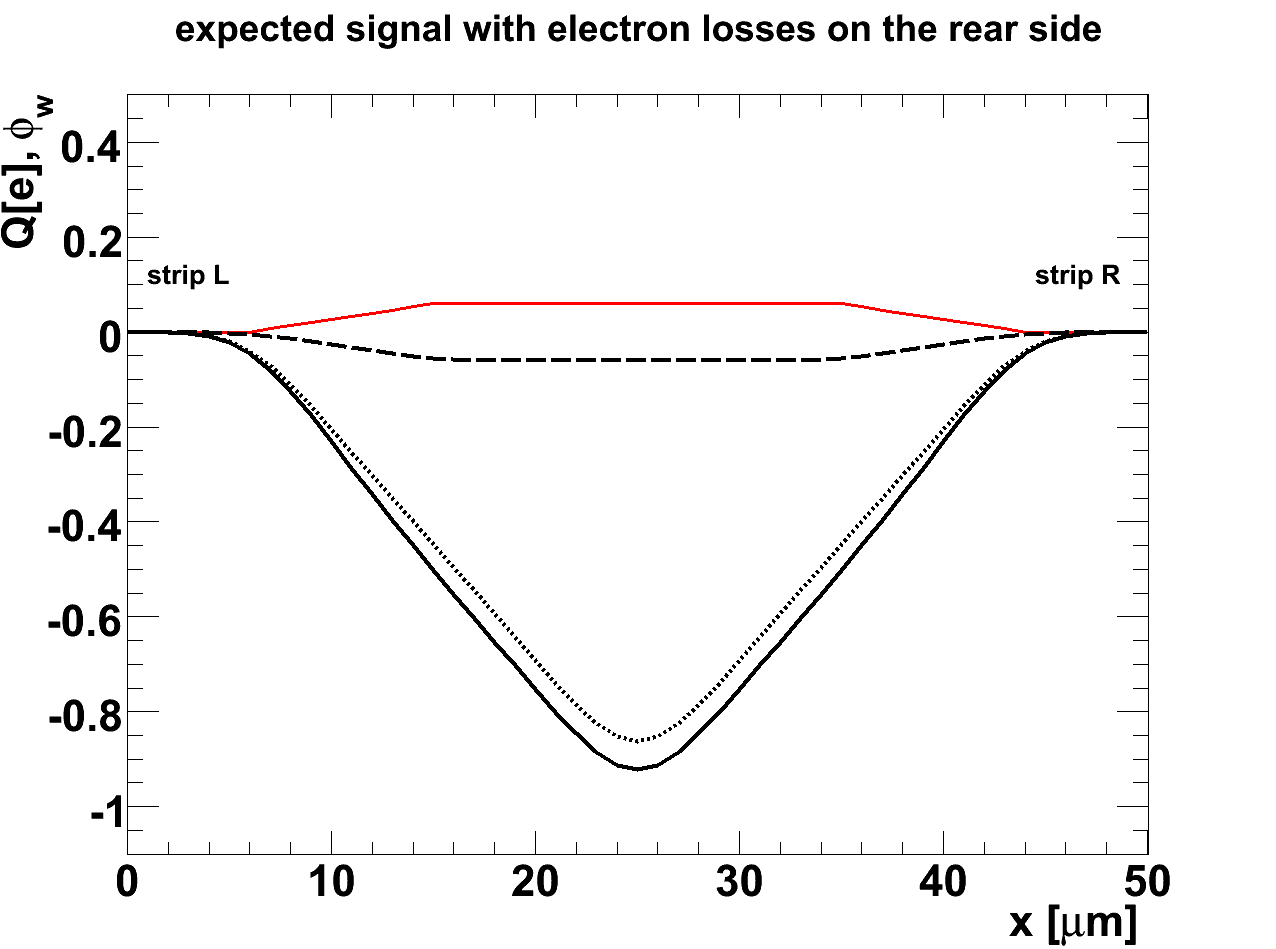}\\
		\includegraphics[width=7.4cm]{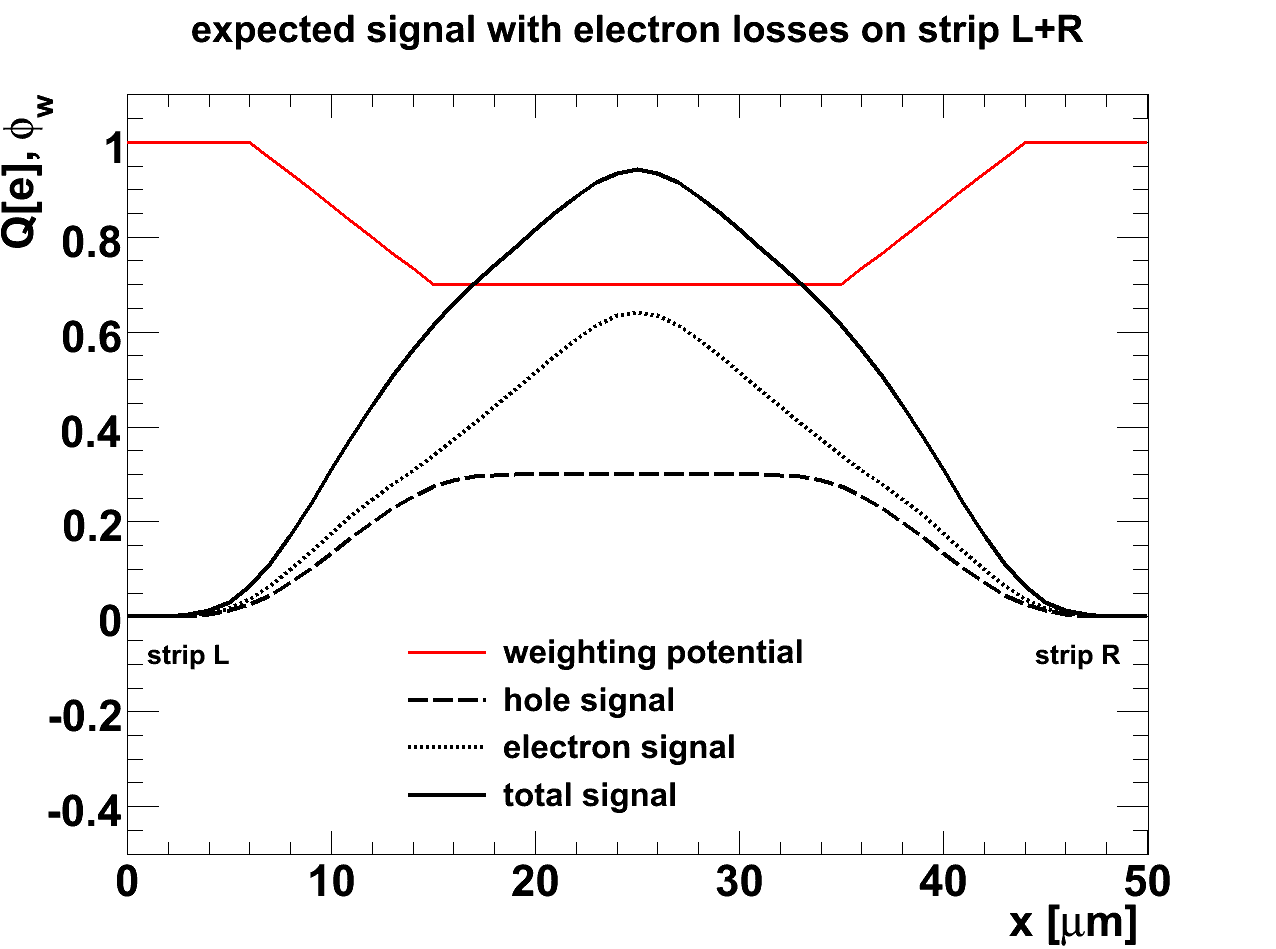}
		\includegraphics[width=7.4cm]{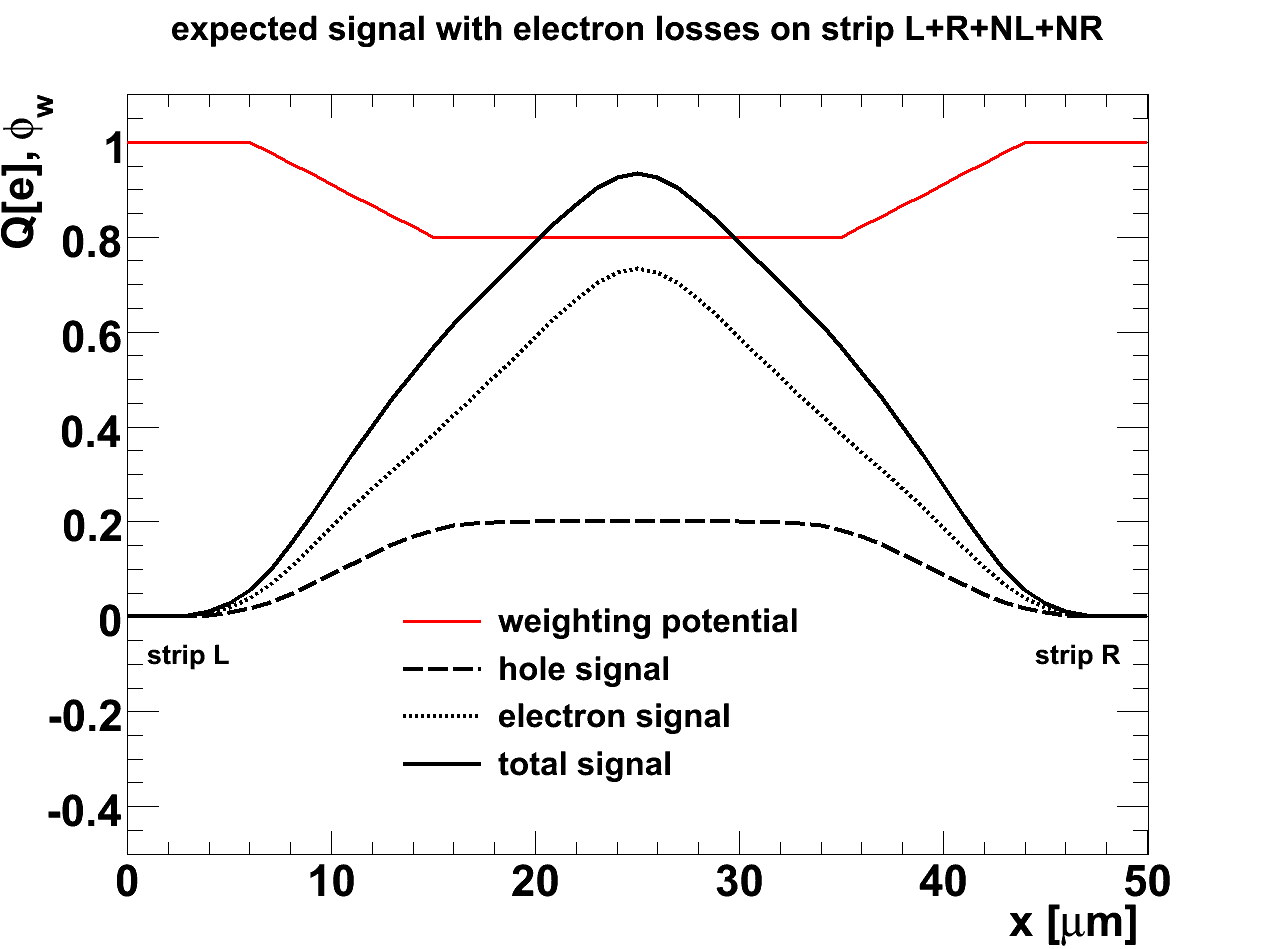}
  \caption{Weighting potential and expected hole and electron signals as a function of the position $x$ of the light spot between the strips L and R for the situation of 100~$\%$ hole collection and position dependent electron losses (100~$\%$ at the strip edge at $x = 6~\upmu$m, and 0~$\%$  at the gap center at $ x = 25~\upmu$m). The signals are normalised to one produced electron-hole pair. For more details see text.}

	\label{fig:expected_charge_positiondependent}
\end{figure}

   The ad-hoc assumption of an $x$-independent loss of electrons at the Si-SiO$_2$ interface has to be replaced by a loss linear in $x$ between the centre of the gap and the strip edge. Introducing position dependent hole losses, however, does not improve the description of the data, and is therefore not used.
 Figure~\ref{fig:expected_charge_positiondependent}
   shows the predicted signal distributions for position dependent electron losses, with losses of 100~$\%$ at the Al-strip edge and 0~$\%$ at the centre between strips.

   In addition we observe in some of the measurements, that even when the light is injected quite some distance from the centre between the strips, a significant number of holes reach the far readout strip. This is taken into account by an additional hole-diffusion term.

\subsection{Model for determining the charge losses from the data}

  This section discusses the parameters of the model which is used in Chapter~3 to fit the charges $Q_i$ collected at the electrodes $i$ as a function of the position $x$ of the laser beam. For the laser beam between strips L and R $(0 \le x \le 50~\upmu$m) it is assumed that no holes are collected by strip NL nor by strip NR. $N_h$ is the total number of holes collected: $f_L(x) \cdot N_h$ by strip L, and $(1 - f_L(x)) \cdot N_h$ by strip R. The number of electrons collected by the rear contact is $N_e$. Using the weighting potentials discussed in the previous section, we obtain for the signal charge induced in strips L, NL and in the rear contact:

% \begin{equation}
  $$ Q_L(x) = N_h(x) \cdot [f_L(x) \cdot (1-\phi_w^L(x)) + (1-f_L(x))\cdot (-\phi_w^L(x))] - N_e(x) \cdot (-\phi_w^L(x)) $$
  $$ Q_{NL}(x) = N_h(x) \cdot (-\phi_w^{NL}(x)) - N_e(x) \cdot (-\phi_w^{NL}(x)) $$
  $$ Q_{rear}(x) = N_h(x) \cdot (-\phi_w^{rear}(x)) - N_e(x) \cdot (-\phi_w^{rear}(x)) $$
% \end{equation}

 Similar formulae apply for strips R and NR.\\

 The individual parameters of the model are:
  \begin{itemize}
    \item
     $x_0$, the position of strip L relative to the zero of the moving stage of the laser (free); ($x$ is replaced by $x - x_0$ in the fit),
    \item
     the width of the Al strips: $d_{Al}= 14~\upmu$m (fixed),which is assumed to have zero light transmission; it should also be noted that the actual width of the $p^+$ implant is 11~$\upmu$m and the metal overhang over the oxide 2~$\upmu$m,
    \item
     $N_e(x) = N_e^c + (N_e^e-N_e^c) \cdot \frac{\mid 0.5\cdot pitch - x \mid}{0.5 \cdot (pitch - d_{Al})}$, with
     $N_e^c $ the number of electrons contributing to the signal in the gap centre, $N_e^e$ the number of electrons contributing to the signal at the Al edge of the strip, and  the strip pitch $pitch$,
    \item
     $N_h$, the number of holes contributing to the signal (free), which is assumed to be independent of $x$,
    \item
     $d_{acc}$, the width of the accumulation layer, which corresponds to the plateau of the weighting potential (free, however, can only be determined if either hole or electron losses occur),
    \item
     the weighting potentials at the accumulation layer: 0.35 for the strips adjacent to the accumulation layer, 0.05 for the next to neighbour strips, and 0.06 for the rear contact (all fixed),
    \item
     three parameters for the shape of the light spot (all fixed): Two Gaussians with widths $\sigma_{light}^{(1)}$ = 3 $\upmu$m and $\sigma_{light}^{(2)}$ = 9 $\upmu$m and relative normalisation $3 : 1$; this shape is derived from the measurements reported in Chapter~3, and is compatible with the laser spot measured using a metal edge on a pad sensor,
    \item
     the parameter of the hole diffusion $\sigma_{diff}$ (free), which is parameterised by
     $$ f_L^{diff}(x) = 0.5\cdot (1 + \text{erf}((0.5 \cdot pitch - x) / \sigma_{diff}));$$ $f_L(x)$ is obtained by convoluting $f_L^{diff}(x)$ with the two Gaussians describing the light spot.
  \end{itemize}

 The free parameters of this model are obtained from a fit to the measured distributions of the signals from strip L and the rear contact when the laser is moved from $x = -100$~$\upmu$m to $x = 100$~$\upmu$m in steps of 2~$\upmu$m.

\section{Results}

 In order to study the charge losses close to the Si-SiO$_2$ interface as a function of the irradiation dose, environmental conditions (humidity) and biasing history, many measurements have been performed for the two sensor types. They will be documented in
  \cite{Poehlsen:Thesis}.
 Here we only show a limited number of results for the DC-coupled Hamamatsu sensor, which are typical for the effects observed. The effects observed for the AC-coupled CiS sensor are similar.

 \subsection{Evidence for charge losses}

  In this section three measurements are presented, which are examples for the situations of no charge losses, of electron losses, and of hole losses. The measurements were performed at room temperature and at a bias voltage of 200~V:

 \begin{itemize}
   \item "humid - 0~Gy":~non irradiated sensor biased to 200~V in a humid atmosphere (relative humidity $>$~60~$\%$),
%C) the DC coupled sensor before irradiation (DC0) in humid atmosphere.\\
   \item "dried at 0~V - 1~MGy":~irradiated sensor (1 MGy) stored at 0~V for a long time, then put into a dry atmosphere for $>$~60 minutes (relative humidity $<$~5~$\%$), and then biased to 200~V for the measurements,
%A) the DC coupled sensor after irradiation (DC1M) in dry atmosphere, dried at 0 V \\
   \item "dried at 500~V - 0~Gy":~non-irradiated sensor kept for a few hours at 500~V in a humid atmosphere (relative humidity $>$~60~$\%$), then dried for $>$~60 minutes and afterwards biased in the dry atmosphere to 200~V for the measurements.
%B) the DC coupled sensor before irradiation (DC0) in dry atmosphere, dried at 500 V\\
 \end{itemize}

 For the values of oxide charge density, $N_{ox}$, interface trap density, $N_{it}$, and surface current density, $I_{surf}$, for the irradiated and non-irradiated sensor we refer to
  Table \ref{tab:irrad}.

 As will be shown later, a sensor is not in a steady state after the bias voltage is changed, and the time constant to reach the steady state depends on the humidity of the surrounding atmosphere: It is of the order days in a dry atmosphere and two orders of magnitude shorter in a humid atmosphere.

 The measurements for "humid" were started $\sim$~20~ minutes after reaching the bias voltage of 200~V, the ones for "dry" after $\sim$~5~ minutes. The focussed light with an attenuation length of $\sim 3.5~\upmu$m in silicon, was scanned in steps of $2~\upmu$m from $x = -100~\upmu$m to $x = 100~\upmu$m and the signals from strips L and R, and from the rear contact were recorded by the digital oscilloscope with 10~Gsamples/s. As shown in
  Figure \ref{fig:sensor},
 $x = 0$ corresponds to the centre of strip L. The repetition rate of the laser was 1~kHz. In order to reduce the fluctuations due to random noise the average of 5000 pulses has been taken. One light scan took $\sim$~90 minutes.

\begin{figure}
	\centering
		\includegraphics[width=7.4cm]{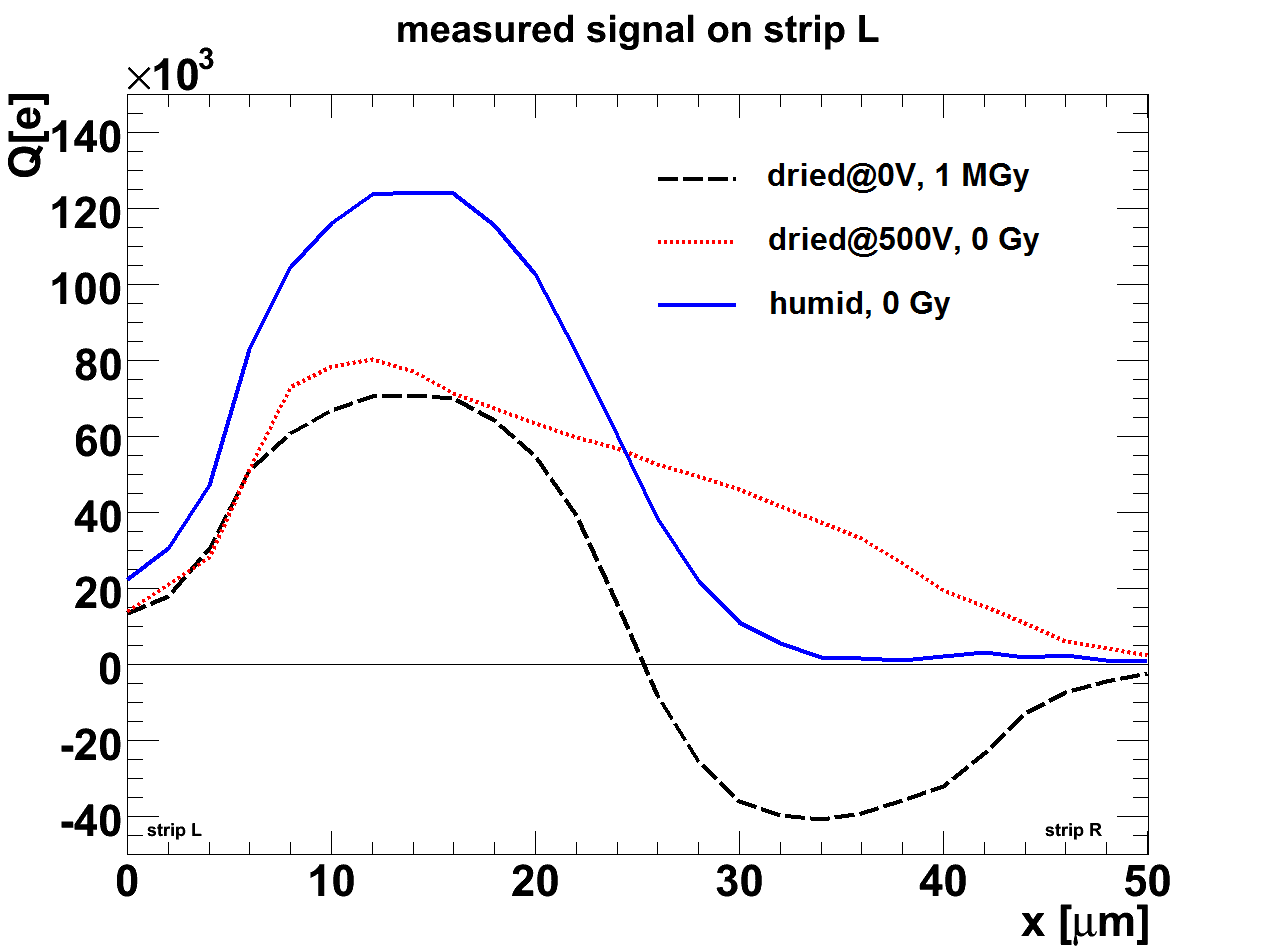}
		\includegraphics[width=7.4cm]{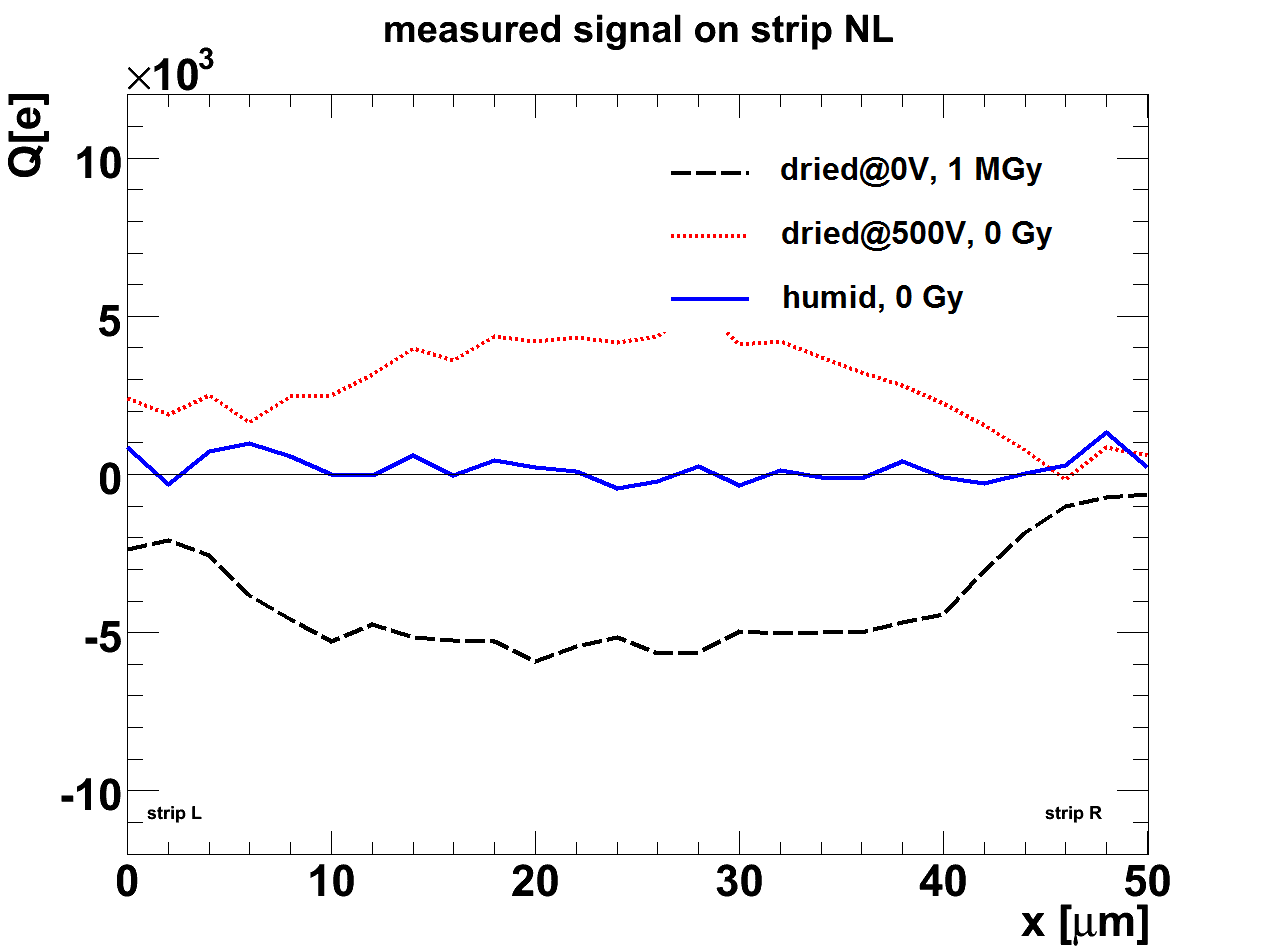}\\
		\includegraphics[width=7.4cm]{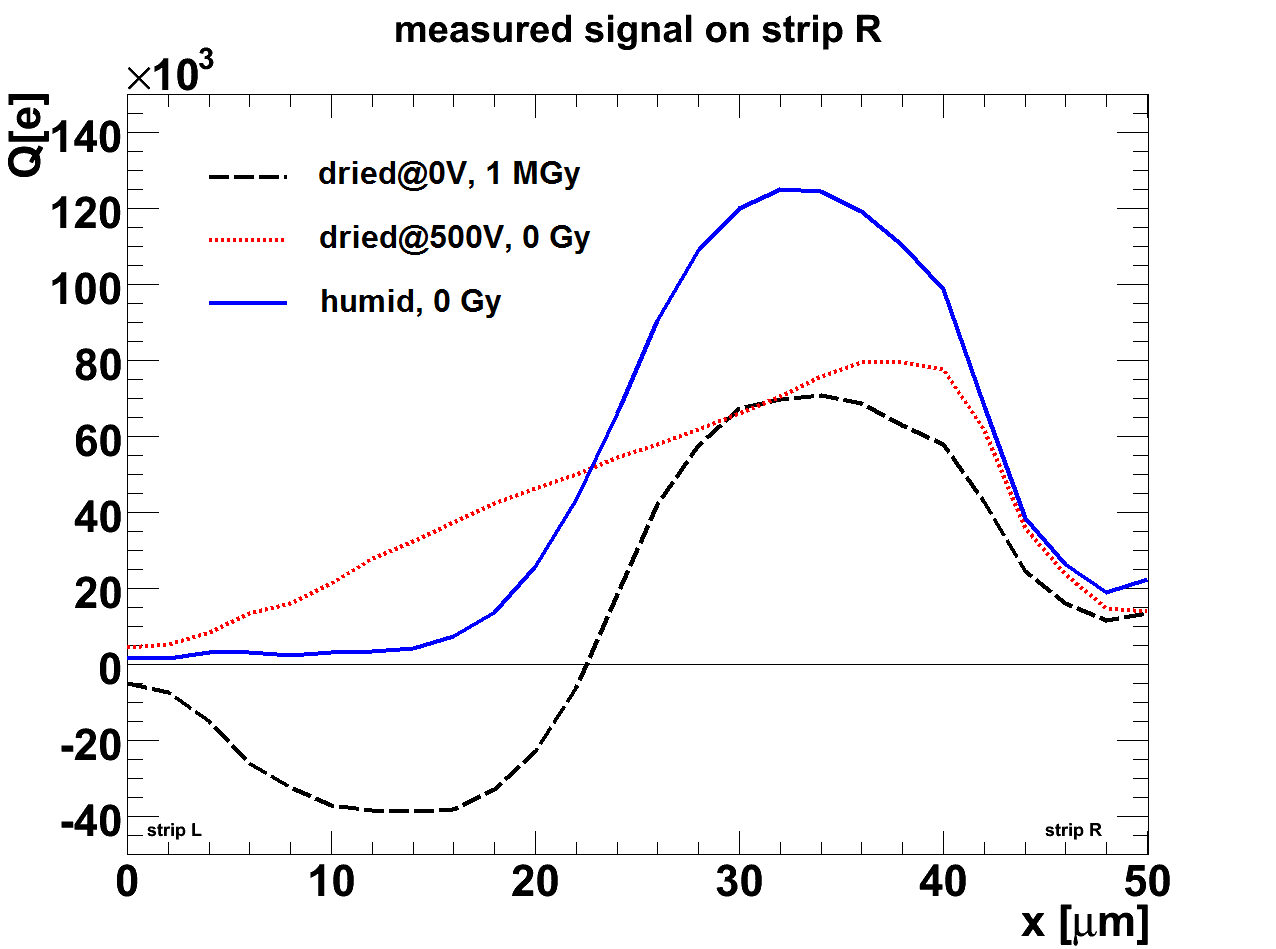}
		\includegraphics[width=7.4cm]{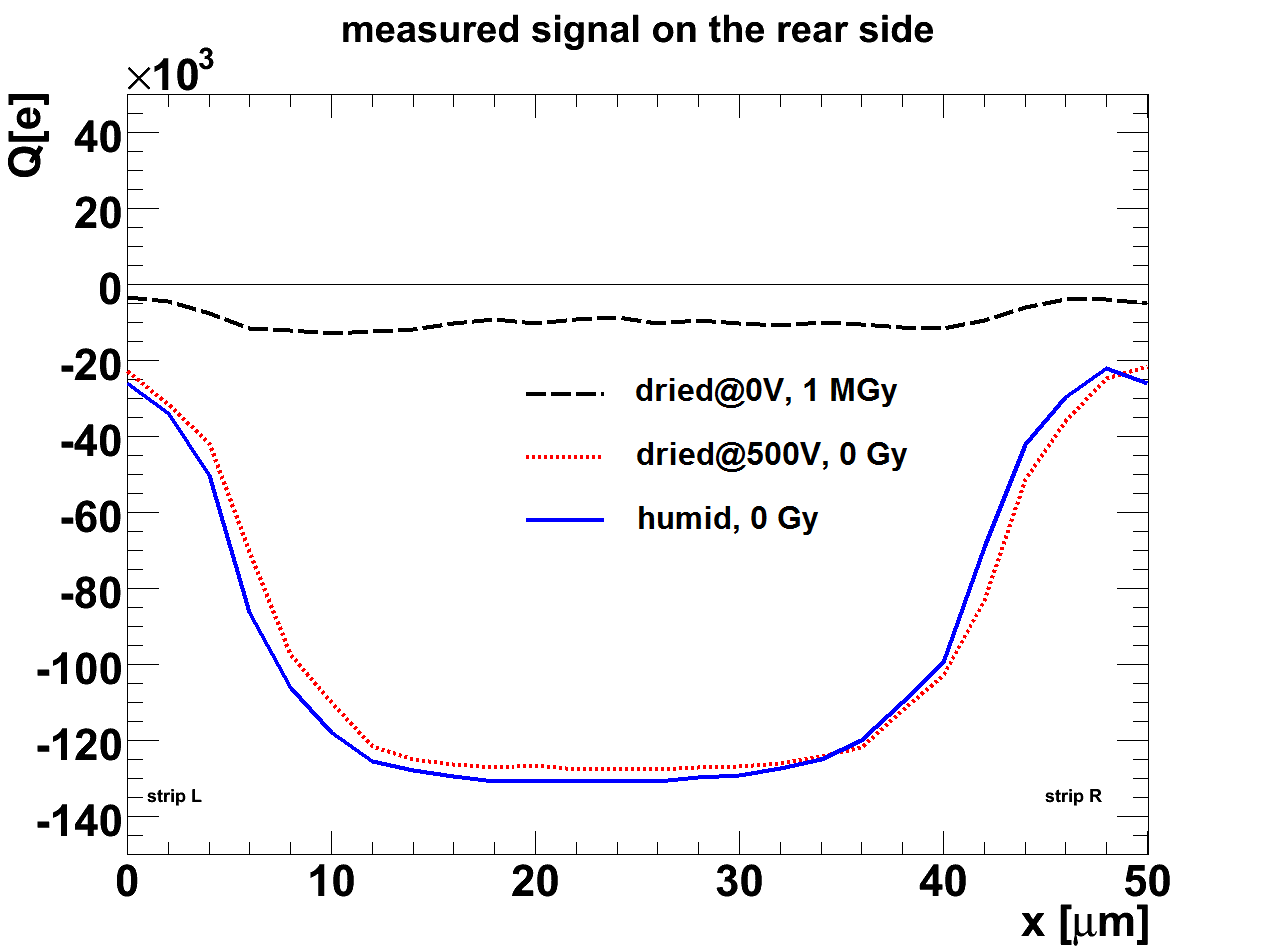}\\
		\includegraphics[width=7.4cm]{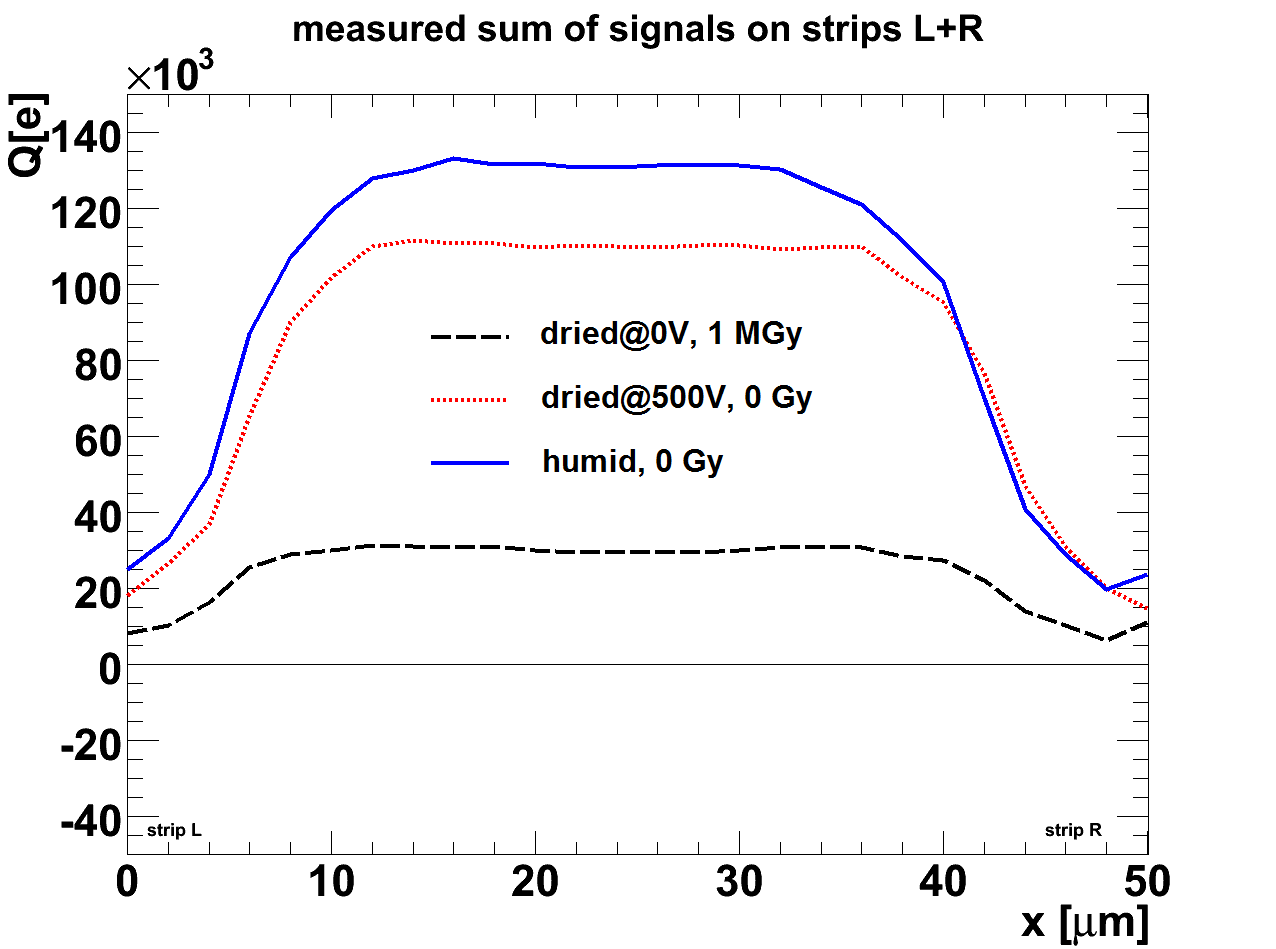}
		\includegraphics[width=7.4cm]{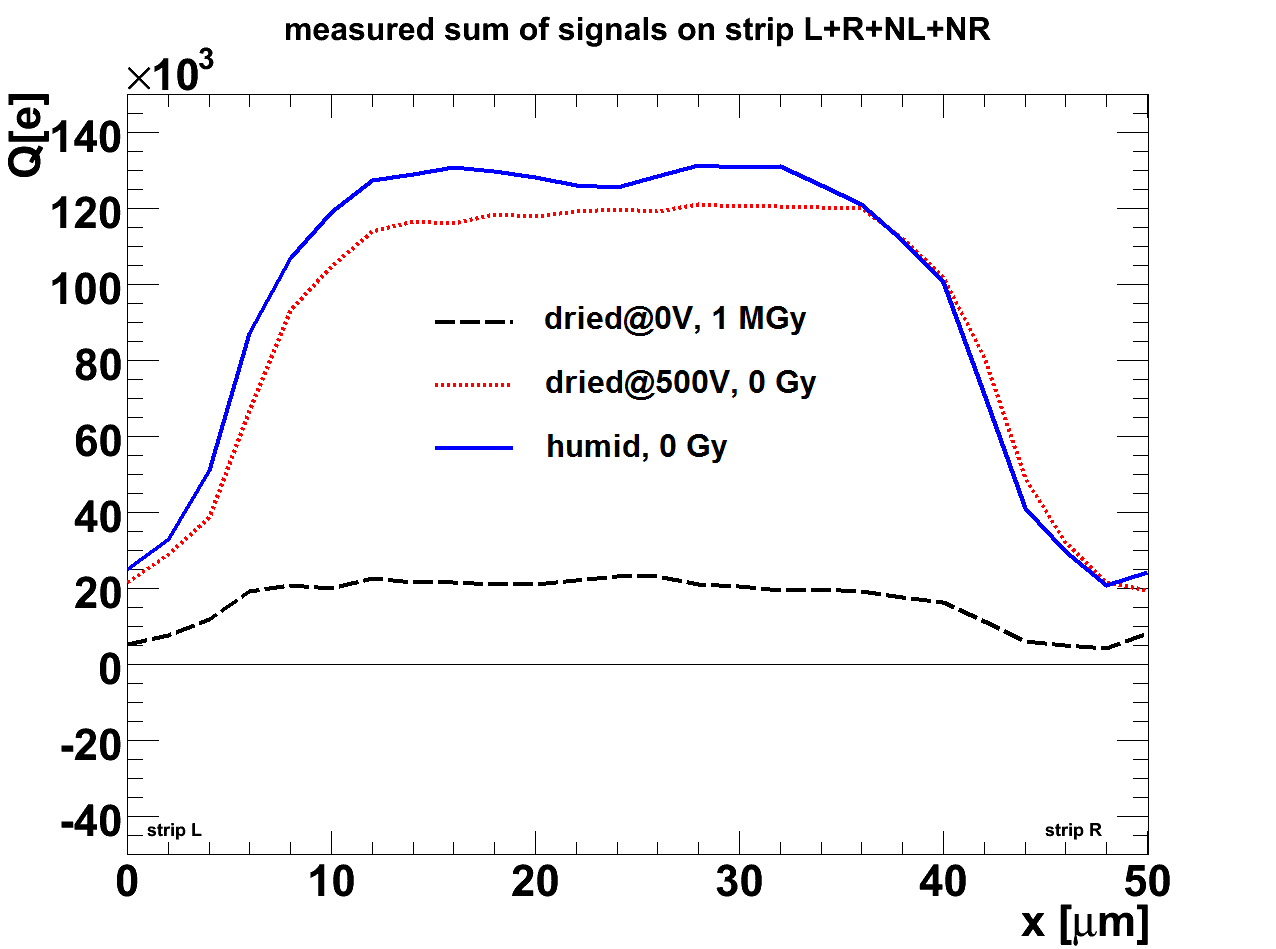}
	\caption{Charge signals measured in strips L, R, NL, NR and the rear contact and summed signals as
 function of the position $x$ of the light spot between strips L and R for the three sets of measurements described in the text. For the definition of $x$, see Figure \ref{fig:sensor}. "humid  - 0~Gy", "dried at 0~V - 1~MGy", and "dried at 500~V - 0~Gy, correspond to the situations "no losses", "electron losses" and "hole losses".}
	\label{fig:charge_profile}
\end{figure}

   Figure~\ref{fig:charge_profile}
 shows, as a function of the position $x$ of the light spot,  the charge collected within 25 ns on strips L, R, NL, on the rear contact, the sums  L~+~R, and the sums for all four strips NL~+~L~+~R~+~NR for the three measurement conditions. The signal NL was not measured simultaneously with strips L and R, but obtained from the signal in strip L, when the laser spot was to the right of strip R. The signal NR was obtained in an analogous way. The comparison with the expectations shown in Figures~\ref{fig:expected_charge} and \ref{fig:expected_charge_electronloss} allows to interpret the data:

\begin{itemize}
  \item
  "humid - 0~Gy" resembles the situation without charge losses shown in
Figure~\ref{fig:expected_charge}:
  The signal from strip L has a broad maximum for $x$ values between 10 and 20 $\upmu$m. However, the transition in between the strips at $x = 25~\upmu$m in the data is much more gradual than expected, an indication that ignoring diffusion might be a crude approximation. The signals in NL and NR are zero, and the rear side signal is the negative of the signal L~+~R, as expected for zero charge losses,
  \item
  "dried at 0~V - 1~MGy" resembles the situation of electron losses shown in
Figure~\ref{fig:expected_charge_electronloss}:
  The signal from strip L shows a broad maximum for $x$ values between 10 and 20 $\upmu$m, followed by a sharp drop at $x = 25~\upmu$m, reaching negative values above 30 $\upmu$m. The signals in NL and NR are negative as expected for electron losses. As the sum signals L~+~R and L~+~R~+~NL~+~NR without charge losses as well as the rear-sides signal are dominated by the electrons, the electron losses result in a strong reduction of these signals,
  \item
  "dried at 500~V - 0~Gy" agrees with the expectations from hole losses: The $x$ dependencies are quite similar to the electron signal shown in
    Figure~\ref{fig:expected_charge},
  and a small positive signal is induced in strips NL and NR. Both the rear-side signal as well as the sum signals are only slightly reduced compared to situation of no charge losses.
\end{itemize}

   Fitting the model discussed in Sect.~2.4 to these measurements gives the values for the free parameters shown in
 Table~\ref{tab:fit_param}.

 For all these measurements the laser has been adjusted to the same intensity, corresponding to approximately 140~000 produced $eh$-pairs with an uncertainty of $\sim$~5\%.

 For "humid - 0~Gy" the fit indicates that  in the centre between the strips there are electron losses of $\sim$~10~\%, no electron losses close to the readout strips, and hole losses of $\sim$~7~\%.
 %These differences are close to the model uncertainties and we conclude that at the level of $\pm$5~\% there are neither electron nor hole losses.
 The  width determined for the accumulation layer of $34~\upmu$m is close to the distance of $39~\upmu$m between the $p^+$-implants. However, given the systematic uncertainties of the model, the results could also be compatible with zero charge losses. In this case the value of $d_{acc}$ cannot be determined by this method.

 For "dried at 500~V - 0~Gy" hole losses of $\sim$~55~\% and a width of the accumulation layer of $31~\upmu$m are found. The data for "dried at 0~V - 1~MGy" show that there is an accumulation layer of $36~\upmu$m covering most of the region between the $p^+$-implants, and practically all electrons are lost. This can be directly seen in Figure~\ref{fig:charge_profile} where the signal measured at the rear contact has practically disappeared. In addition, hole losses of about 12~\% are  observed.

\begin {table}
\centering
\begin{tabular}{|c|c|c|c|c|c|c|}
\hline
    	       & humid	        & dried@500~V 	& dried@0~V    & humid      & dried@0~V     & dried@500~V     \\
		       & 0~Gy	        & 0~Gy	        & 0~Gy          & 1~MGy     & 1~MGy         & 1~MGy          \\
\hline
\hline
\ $N_e^c$ 		& 128 000 		& 129 000		& 53 000        & 109 000      & 0          & 123 000 \\ \hline
\ $N_e^e$ 		& 143 000	 	& 145 000       & 63 000        & 72 500       & 9 400      & 124 000 \\ \hline
\ $N_h$ 		& 133 000		& 64 000        & 133 000       & 121 000      & 126 000    & 112 000 \\ \hline
$x_0$    	    & 0.90 $\upmu $m & 1.83 $\upmu $m& - 0.18 $\upmu $m  & 1.64 $\upmu $m& 0.77 $\upmu$m  & 0.80 $\upmu $m \\ \hline
$d_{acc}$       & (34 $\upmu $m) & 31 $\upmu $m   & 36 $\upmu $m   & 34 $\upmu $m  & 36 $\upmu $m& (6 $\upmu $m) \\ \hline
$\sigma _{diff}$& 2.7 $\upmu $m	& 14.1$\upmu $m  & 3.1 $\upmu $m   & 2.2 $\upmu $m  &  2.1 $\upmu $m  & 5.4 $\upmu $m \\ \hline
\hline
\end{tabular}
\caption{Results of the fits using the model described in Sect. 2.4. Values of $d_{acc}$ in parenthesis indicate that the charge losses are insufficient for a reliable determination of the width of the accumulation layer. The laser is adjusted to generate $\sim$~140~000~$eh$-pairs.}
\label{tab:fit_param}
\end{table}

\begin{figure}
	\centering
		\includegraphics[width=7.4cm]{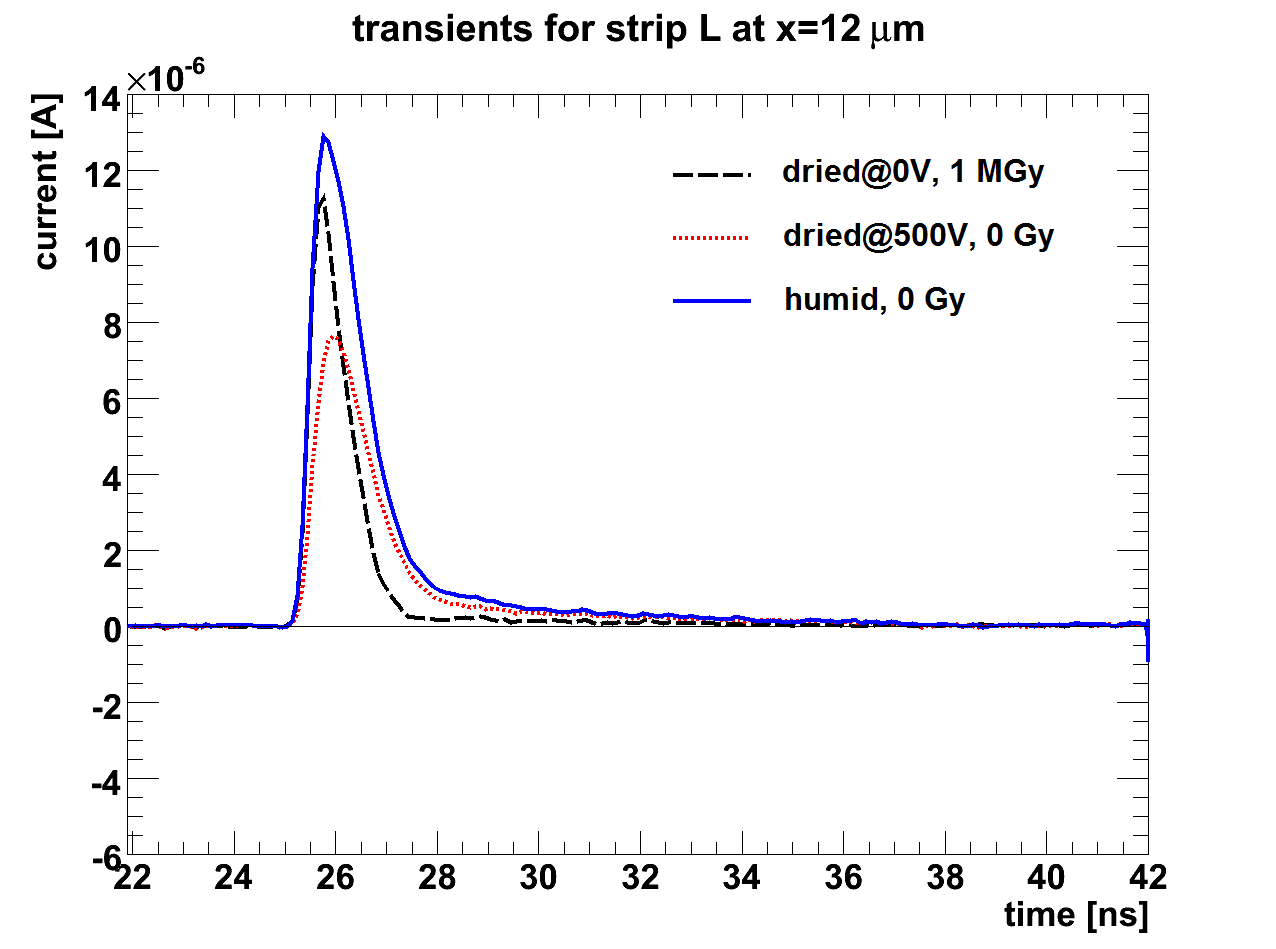}
		\includegraphics[width=7.4cm]{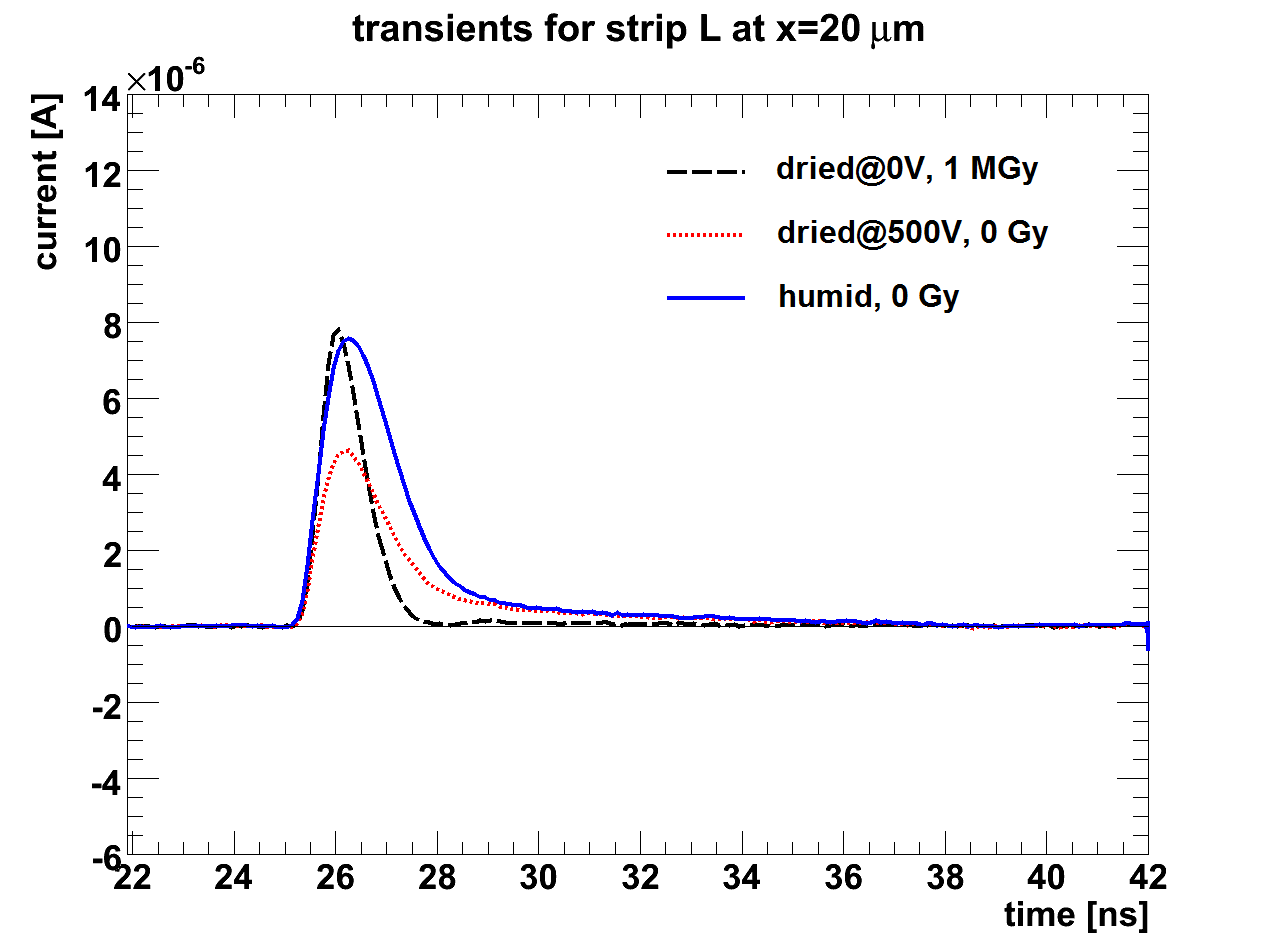}\\
		\includegraphics[width=7.4cm]{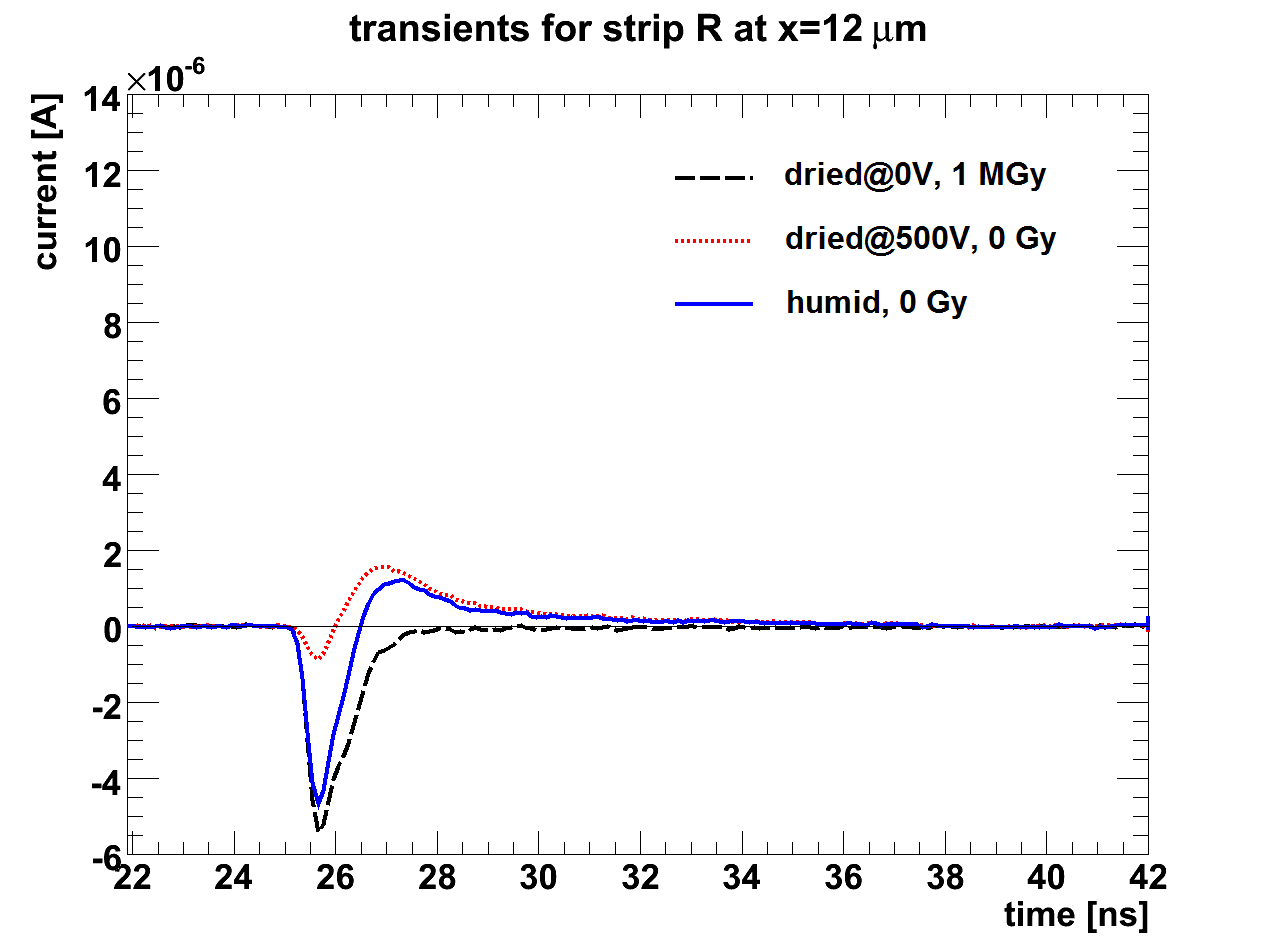}
		\includegraphics[width=7.4cm]{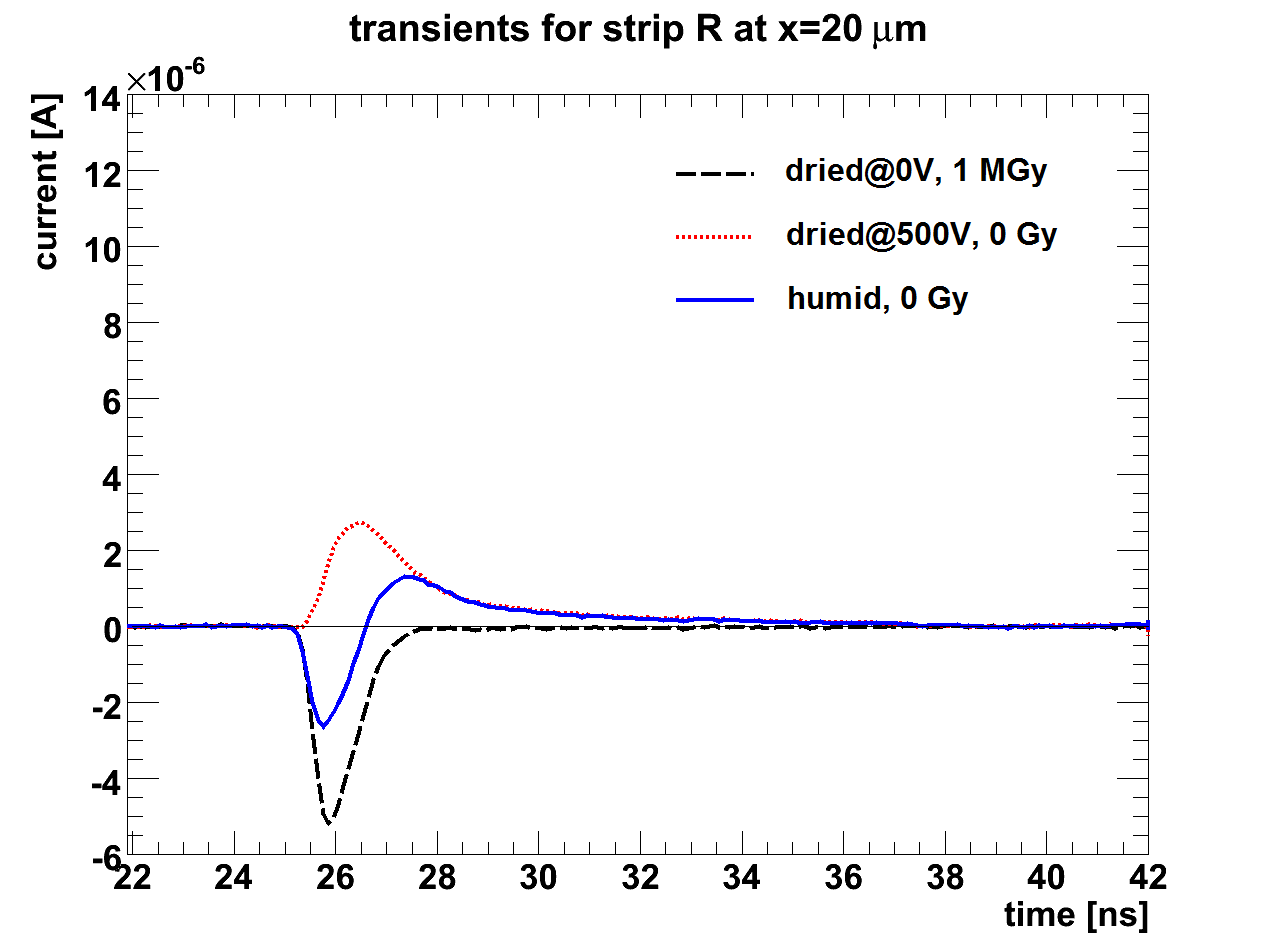}\\
		\includegraphics[width=7.4cm]{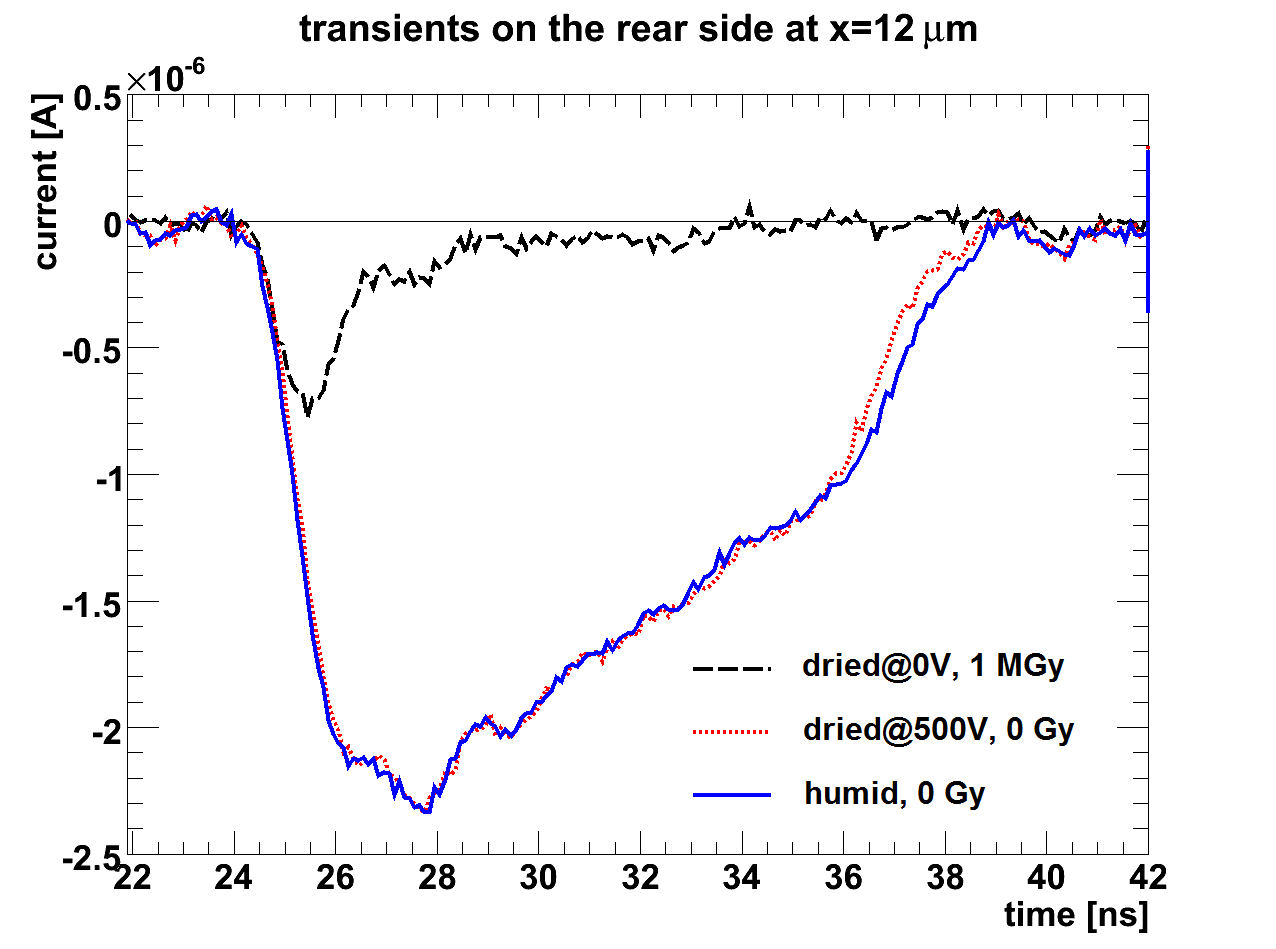}
		\includegraphics[width=7.4cm]{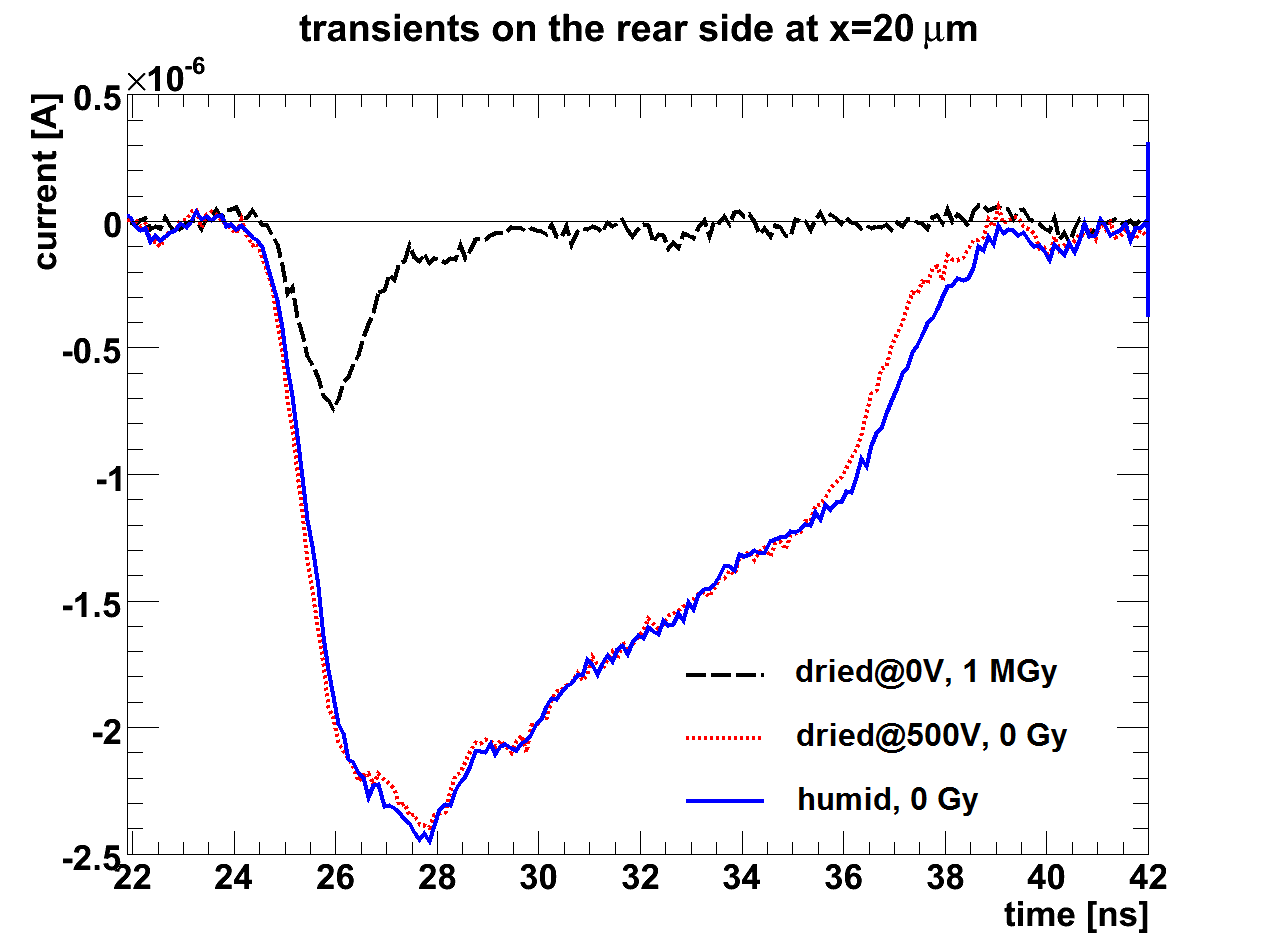}\\
   \caption{Measured transients on strips L and R and the rear contact for light injected at the two positions
   $x$ = 12 $\upmu$m and $x$ = 20 $\upmu$m for the three sets of measurements described in the text. "humid - 0~Gy", "dried at 0~V - 1~MGy", and "dried at 500~V - 0~Gy, correspond to the situations "no losses", "electron losses" and "hole losses".}
	\label{fig:transients}
\end{figure}

  The measured current transients shown in
     Figure~\ref{fig:transients}
  confirm above interpretation of the data, in particular of no or little charge losses for "humid - 0~Gy", electron losses for "dried at 0~V  - 1~MGy", and hole losses for "dried at 500~V - 0~Gy".
  Shown are the transients for the two strips L and R at the positions of the light spot $x = 12~\upmu$m, which is about 5.5~$\upmu$m from the edge of the aluminum of strip L, and at $x = 20~\upmu$m, 5~$\upmu$m away from the centre between the strips towards strip L. \\
 %Physically 8 different positions and only one read out strip is used.

 At $x$ = 12 $\upmu$m:
 \begin{itemize}
   \item
    For strip L in all three cases positive signals are observed. Compared to no charge losses ("humid - 0~Gy"), the signal is shorter for electron losses ("dried at 0~V - 1~MGy"), as the electrons which drift through the entire sensor generate the signal at later times. For hole losses ("dried at 500~V - 0~Gy") the signal is correspondingly smaller at short times.
   \item
    For strip R, for no charge losses, a bipolar signal with an integral of zero is observed. The initial negative signals are due to holes drifting away from strip R minus a positive signal from the electrons moving towards strip R. Once the holes are collected the positive signal from the electrons drifting towards the rear contact remains. For electron losses there is just the negative signal due to the holes drifting away from strip R, and for hole losses there is essentially just the longer positive signal due to the electrons drifting through the entire sensor.
   \item
    For the rear contact the signal is mainly due to electrons, and the holes drifting to the strips contribute only little to the signal. Therefore, the signals with and without hole losses are essentially identical. For electron losses the signal is drastically reduced and a small fast negative signal due to the holes drifting to the readout strip is observed. The signal from the rear contact, which has the biggest capacitance, shows some ringing at a frequency of about 1~GHz.
\end{itemize}

  At $x$ = 20 $\upmu$m:

 \begin{itemize}
   \item
    For strip L, for electron losses and for no charge losses, a shift of the peak to later times is observed compared to $x$ = 12 $\upmu$m, due to the longer distance the holes have to drift.
    For no charge losses the signal at strip L is reduced, which is described in the model by holes diffusing to strip R.
    % In the case of hole losses the signal is compared to $x$ = 12 $\upmu$m, which may be due to a larger impact of hole losses close to the gap centre (lower weighting potential) or due to a larger fraction of hole losses close to the centre (not included in the model).
   \item
    For strip R, corresponding effects are observed: For electron losses a shift of the negative peak to later times, for no charge losses an increased signal (holes diffusing to strip R), and for hole losses the negative part of the signal is absent, and only a positive signal mainly due to  electrons remains.
   \item
    As expected, for the rear side little dependence of the signal on the $x$ position  is observed.
 \end{itemize}

 To summarise: Convincing evidence has been found for situations with little or no charge losses, complete hole collection  and significant electron losses, and complete electron collection and significant hole losses. The signals from the readout strips and the rear contact as a function of the position of the laser spot can be described by the model presented, which allows determining quantitatively the charge losses and the width of the accumulation layer at the Si-SiO$_2$ interface.

\subsection{Time constants to reach stable charge losses}

  In Section~3.1 it has been  shown, that significant hole losses are observed for the non-irradiated sensor biased at 200~V for the scenario "dried at 500~V - 0~Gy", and hardly any charge losses for "humid - 0~Gy". Next we show that, after changing the voltage from 500~V to 200~V, the hole losses decrease with time, with a time constant which depends on the humidity.

\begin{figure}
 \centering
   \includegraphics[width=8cm]{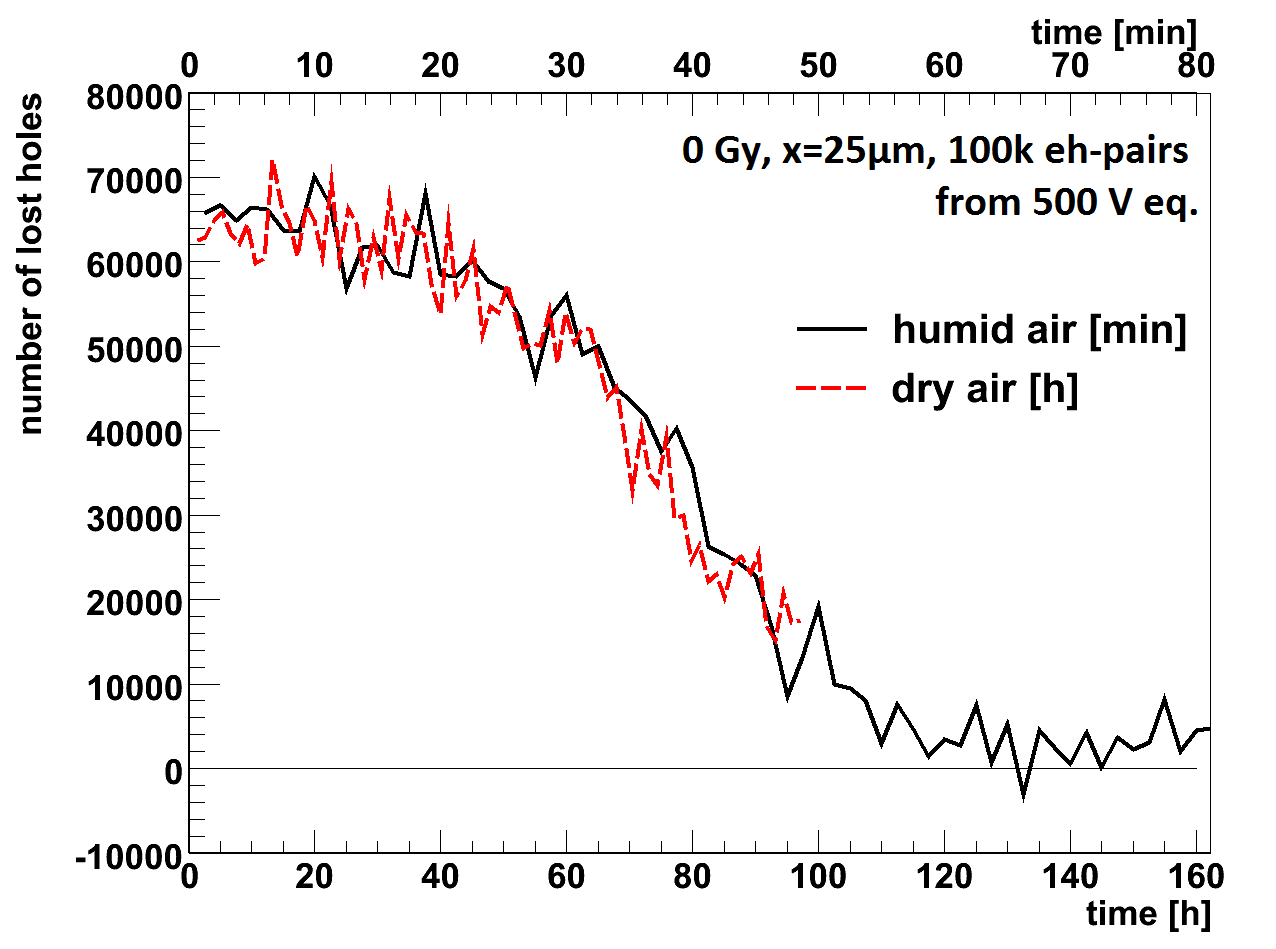}
	\caption{Number of holes lost for light pulses generating $\sim$~100 000 $eh$-pairs as a function of the time after
    the voltage applied to the non-irradiated sensor has been reduced from 500~V in steady-state conditions, to 200~V . The upper scale of the horizontal axis (in minutes) refers to the situation "humid", the lower one (in hours) to "dry". }
	\label{fig:time_dep}
\end{figure}

  For this measurement approximately 100~000 $eh$-pairs were generated by the laser light at $x\sim 25~\upmu$m, close to the centre between the strips L and R. The hole losses are calculated from  $Q_{NL}$, the signal measured on strip NL, using  $Q_{NL}/(q_0 \cdot 0.05)$.  For the weighting potential a value of $\phi_w^{NL}=0.05$ is assumed, and $q_0$ is the elementary charge. The measurements were made at room temperature in a dry (relative humidity $<$~5~$\%$) and a humid ($>$~65~$\%$) atmosphere. The results are shown in
    Figure~\ref{fig:time_dep}.
  At time $t$ = 0, when the voltage was changed from 500 to 200~V, approximately 65~$\%$ of the holes are lost. As a function of time, the hole losses decrease following an S-shaped curve and finally reach a steady state of little hole losses. The shape of the curves are similar for "dry" and "humid", the time constants however, are very different: The reduction to 50~$\%$ of the initial losses is reached after $\sim$~40 minutes for "humid", and after $\sim$~80 hours for "dry". The measurements were made once in the middle of the $\sim$~8~mm long strips, and once $\sim$~1~mm from the end of the strips, where the openings in the passivation layer for the bonding are located. The results are similar. In Sect.~4.1 the explanation for these results will be given: The change of the charge losses with time is due to the change of the charge distribution on the surface of the sensor, which influences both the field close to the Si-SiO$_2$ interface as well as the extension of a  possible accumulation or inversion layer. The time constant for reaching the steady-state conditions depends on the surface resistivity, which is a strong function of humidity. We note that the situation for the irradiated sensor is quite different. In particular, for "humid~-~1~MGy" the steady state situation is reached on a much shorter time scale \cite{Poehlsen:Thesis}.

\subsection{Correlation of the dark current and the accumulation layer width}

 For X-ray irradiated sensors the dark current is expected to be dominated by the surface generation current from the depleted Si-SiO$_2$ interface.
 %From measurements on irradiated gate-controlled diodes
%\cite{Perrey:Thesis, Zhang:Thesis}
% it has been found, that after a dose of 1~MGy and annealing for 60 minutes at 80$^\circ$C, the surface generation current from the depleted Si-SiO$_2$ interface is $\sim$~5~$\upmu$A/cm$^{2}$.
  As the width of the depleted Si-SiO$_2$ interface is $d_{gap} - d_{acc}$, where $d_{gap}$ is the distance between the $p^+$ implants and $d_{acc}$ the width of the accumulation layer, an anti-correlation of the dark current and the width of the accumulation layer is expected.
 Table~\ref{tab:IV}
  shows for the 1~MGy-irradiated sensor biased at 200~V for the three experimental conditions:
  "dried at 0~V - 1~MGy", "humid - 1~MGy", which corresponds to  steady-state conditions on the sensor surface,     and "dried at 500~V - 1~MGy" the measured dark currents for a single strip, $I_{dark}^{meas}$, the widths of the accumulation layer, $d_{acc}^{fit}$, determined from the fits, and the widths of the accumulation layer, $d_{acc}^{calc}$, calculated from the measured dark current and the value of the surface generation current $I_{surf} = 2.2~\upmu $A/cm$^2$ from
 Table~\ref{tab:irrad}.
  For the measurement "dried at 500~V - 1~MGy" the charge losses are small and $d_{acc}$ cannot be determined reliably.
  The agreement is certainly not perfect, however qualitatively the results are similar, and give us some confidence in the method used to determine the widths of accumulation layers from the TCT measurements.
  Also the measured time dependence of the dark current to a steady state (not shown) agrees with the measured time dependence to reach a steady state for the charge losses, which has been discussed in the previous Section.

\begin {table}
\centering
\begin{tabular}{|c|c|c|c|}
\hline
					& dried at 0~V - 1~MGy  & humid - 1~MGy     & dried at 500~V - 1~MGy \\ \hline
 $I_{dark}^{meas}$ 	& 1.2~nA				& 1.8~nA			& 3.3~nA  \\ \hline
 $d_{acc}^{fit}  $  & 36~$\upmu$m			& 34~$\upmu$m		& -  \\ \hline
 $d_{acc}^{calc} $  & 32~$\upmu$m			& 29~$\upmu$m		& 21~$\upmu$m  \\ \hline

% \hline
 \end{tabular}
 \caption{For the sensor irradiated to 1~MGy, biased at 200~V under three different measurement conditions:
   $I_{dark}^{meas}$, the dark current for a single strip, $d_{acc}^{fit}$, the width of the accumulation layer determined from the fit, and the width of the accumulation layer, $d_{acc}^{calc}$, obtained from the measured dark current, the surface generation current from the test structures (Table~\ref{tab:irrad}), and the geometrical parameters of the sensor (Table~\ref{tab:sensors}). The current values refer to a temperature of 22.9$^{\circ}$C}.

\label{tab:IV}
\end{table}

\subsection{Charge losses as a function of voltage for the irradiated sensor}

 As discussed in Section~3.1, electron losses are observed for the irradiated sensor for the measurement conditions "humid~-~1~MGy" and "dried~at~0~V~-~1~MGy". Here we present the number of holes and electrons collected as a function of the applied voltage ramped up from 0 to 500~V in 50~V steps with the sensor in a dry atmosphere (relative humidity $<$~5~$\%$). At each voltage a position scan between $x = -100~\upmu$m and $x = 100~\upmu$m was performed. A position scan took $\sim$~1.5 hours, and there may be some time effects, in particular at longer times and thus at the higher voltages. The number of generated $eh$-pairs was approximately 30~000. The number of collected electrons and holes is determined from the fit of the model described in Sect.~2.4 to the data.

\begin{figure}
	\centering
	  \includegraphics[width=10cm]{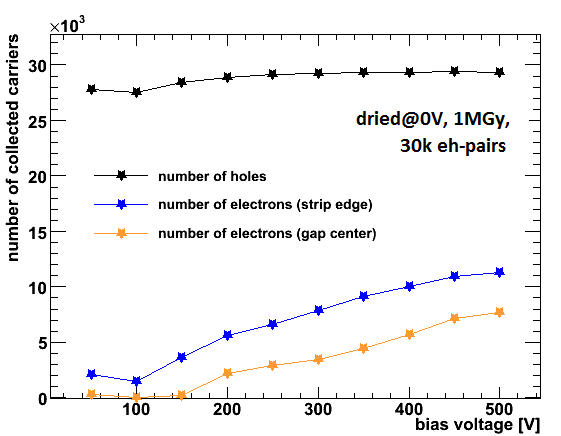}
        \caption{Irradiated sensor operated in a dry atmosphere for voltages between 50 and 500~V: Number of electrons collected for light injected in the middle between and close to the readout strip, and number of holes collected. The number of  $eh$-pairs generated by the light is approximately 30~000.}
	\label{fig:voltage_dependence}
\end{figure}

 The results are shown in Figure~\ref{fig:voltage_dependence}. The number of holes collected shows a small increase between 50 and 150~V, and then remains constant. Up to 150~V practically no electrons are collected at the center between the strips. Above this voltage the fraction of electrons collected increases approximately linearly with voltage, reaching $\sim$~25~\% at 500~V. The fraction of  electrons collected close to the edge of the readout strips, increases from $\sim$~5~\% to $\sim$~40~\%. We note that for all voltages significant electron losses are observed and, that the width of the accumulation layer, $d_{acc}$ (not shown), is between 34 and 36~$\upmu$m.  Thus the accumulation layer covers most of the 39~$\upmu$m wide gap between the $p^+$-implants. We also observe (not shown) that the diffusion term is below  3~$\upmu$m, the width of the light spot.
% For comments on these results we refer to Section~3.2.

\section{Discussion of the results}

 In this Chapter an attempt is made to qualitatively interpret the results and discuss their relevance for the operation of $p^+n$~sensors. We first discuss the cause of the charge losses and their dependence on time and humidity. Our explanations are supported by detailed two-dimensional simulations of the sensor
%  (see Figure~\ref{fig:sensor})
 assuming different boundary conditions on the surface of the passivation and different values of
    $ q_0 \cdot N_{int}^{eff} = q_0 \cdot (N_{ox} + N_{it}^{don} - N_{it}^{acc})$,
 the effective charge density at or close to the Si-SiO$_2$ interface. $q_0$ is the elementary charge, $N_{ox}$ the density of positive oxide charges, and $N_{it}^{don}$ and $N_{it}^{acc}$ the density of filled donor and acceptor states at the interface, integrated over the silicon band gap.

\subsection{Charge losses and their time dependence}

\begin{figure}
	\centering
		\includegraphics[width=7.4cm]{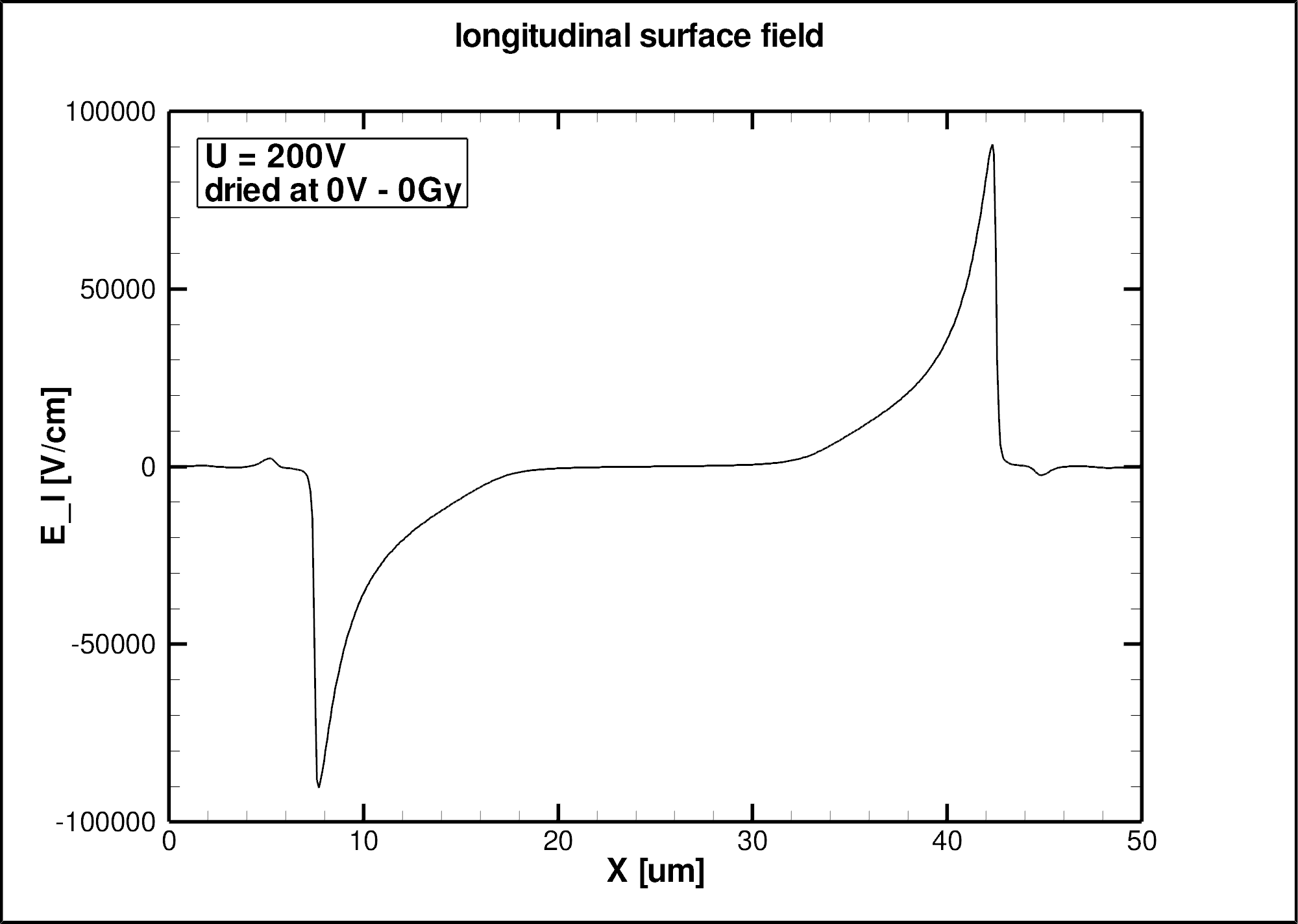}
		\includegraphics[width=7.4cm]{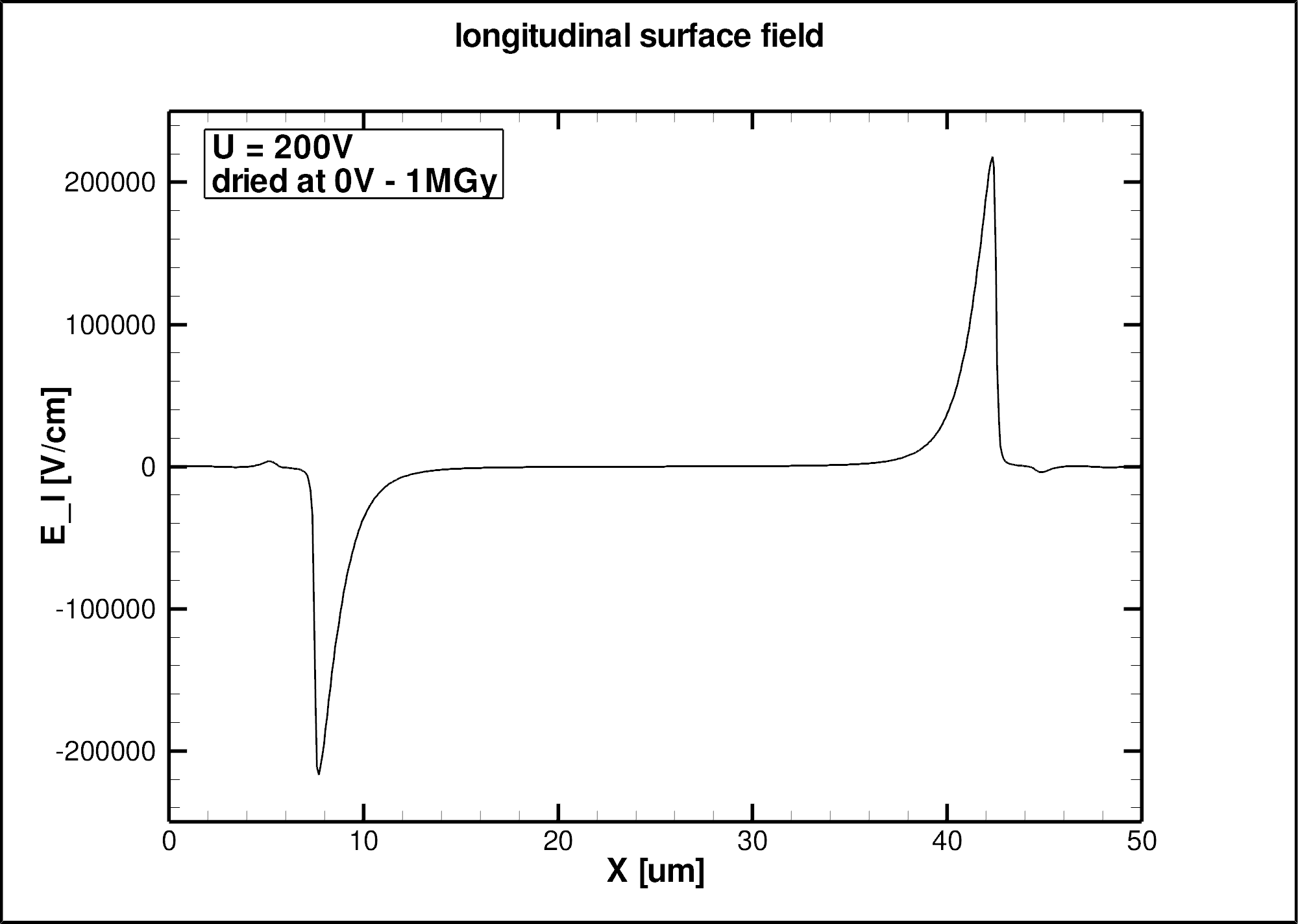}\\
		\includegraphics[width=7.4cm]{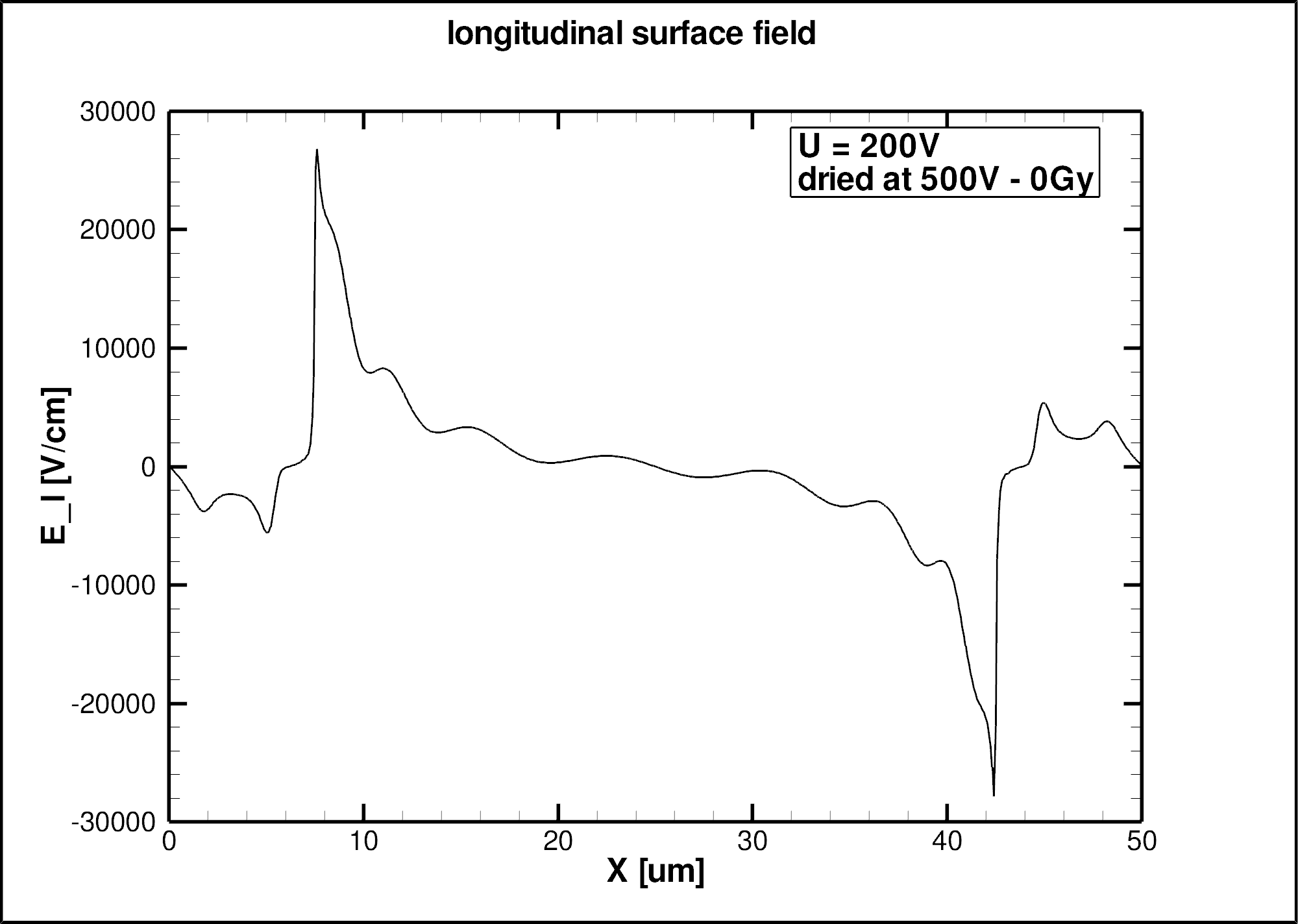}
		\includegraphics[width=7.4cm]{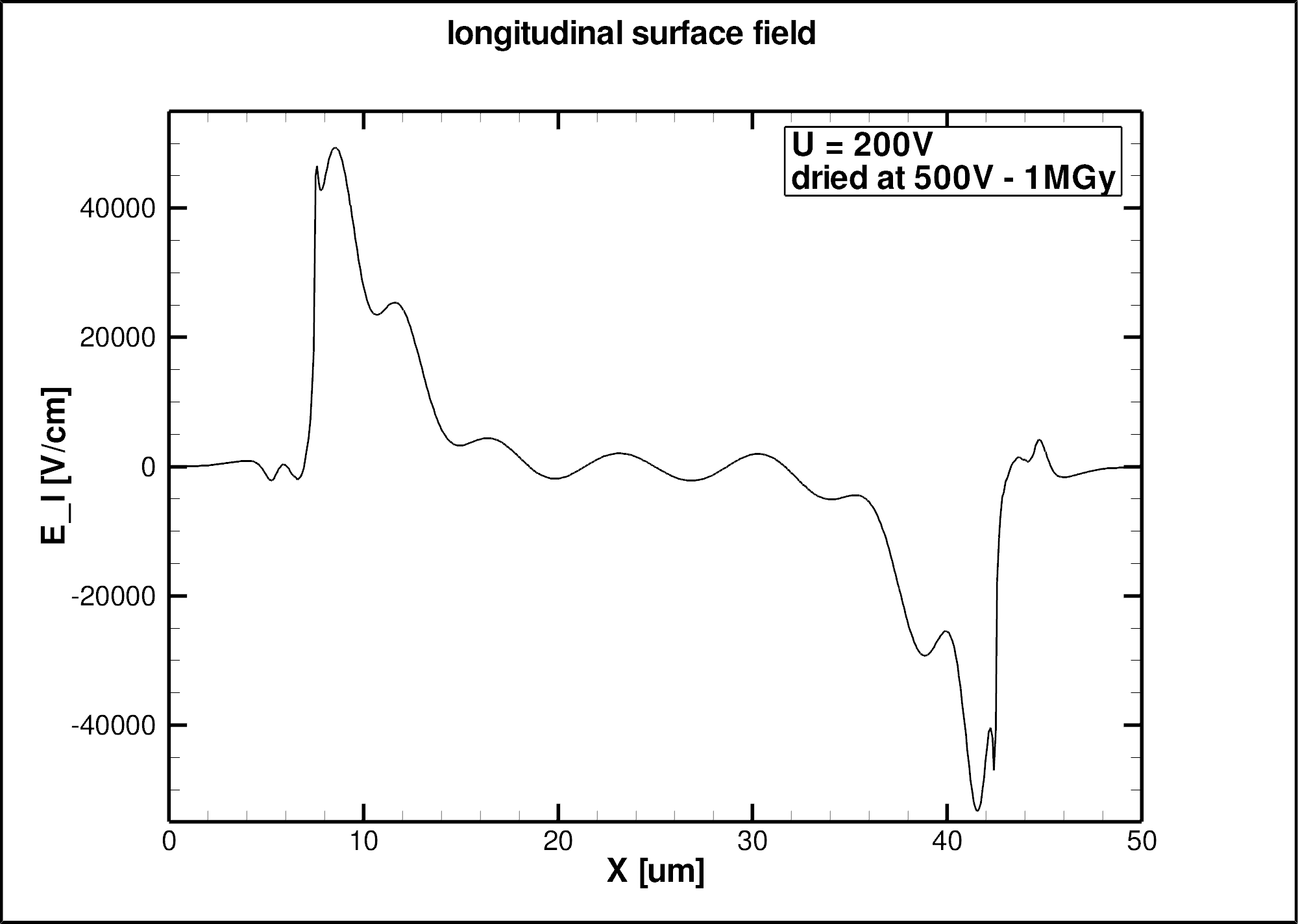}
  \caption{Simulated longitudinal component of the surface field for the sensor biased to 200~V under dry conditions. Left: Non-irradiated sensor ($N_{int}^{eff} = 10^{11}$~cm$^{-2}$); right: Irradiated sensor ($N_{int}^{eff} = 10^{12}$~cm$^{-2}$); top: Steady-state conditions at 0~V, and bottom at 500~V. The strips are centred at $x = 0$ and 50~$\upmu$m.}
%  The field points towards the strips for the top, and away from the strips for the bottom plots.}
  	\label{fig:long_field}
 \end{figure}

\begin{figure}
	\centering
		\includegraphics[width=7cm]{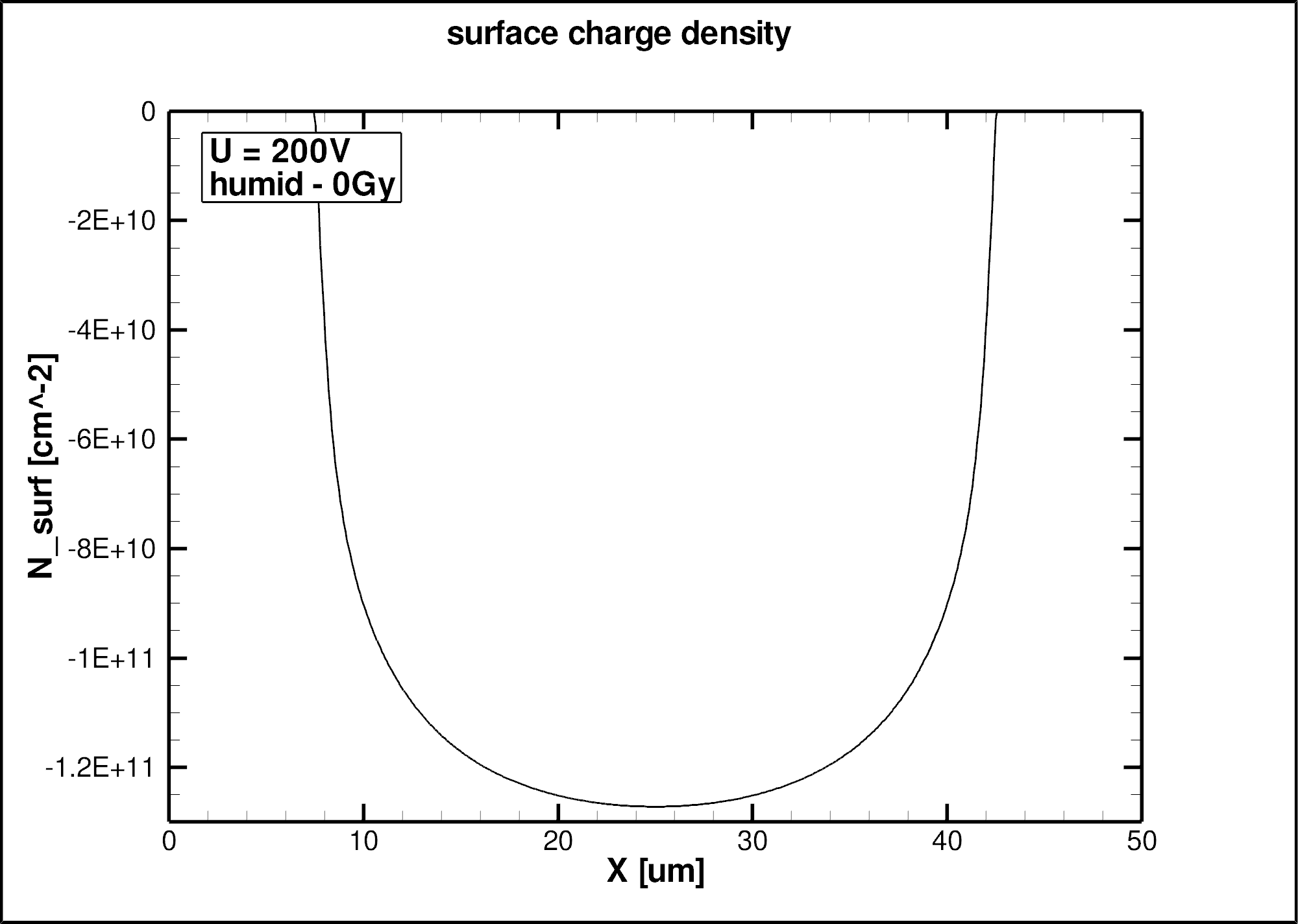}
		\includegraphics[width=7cm]{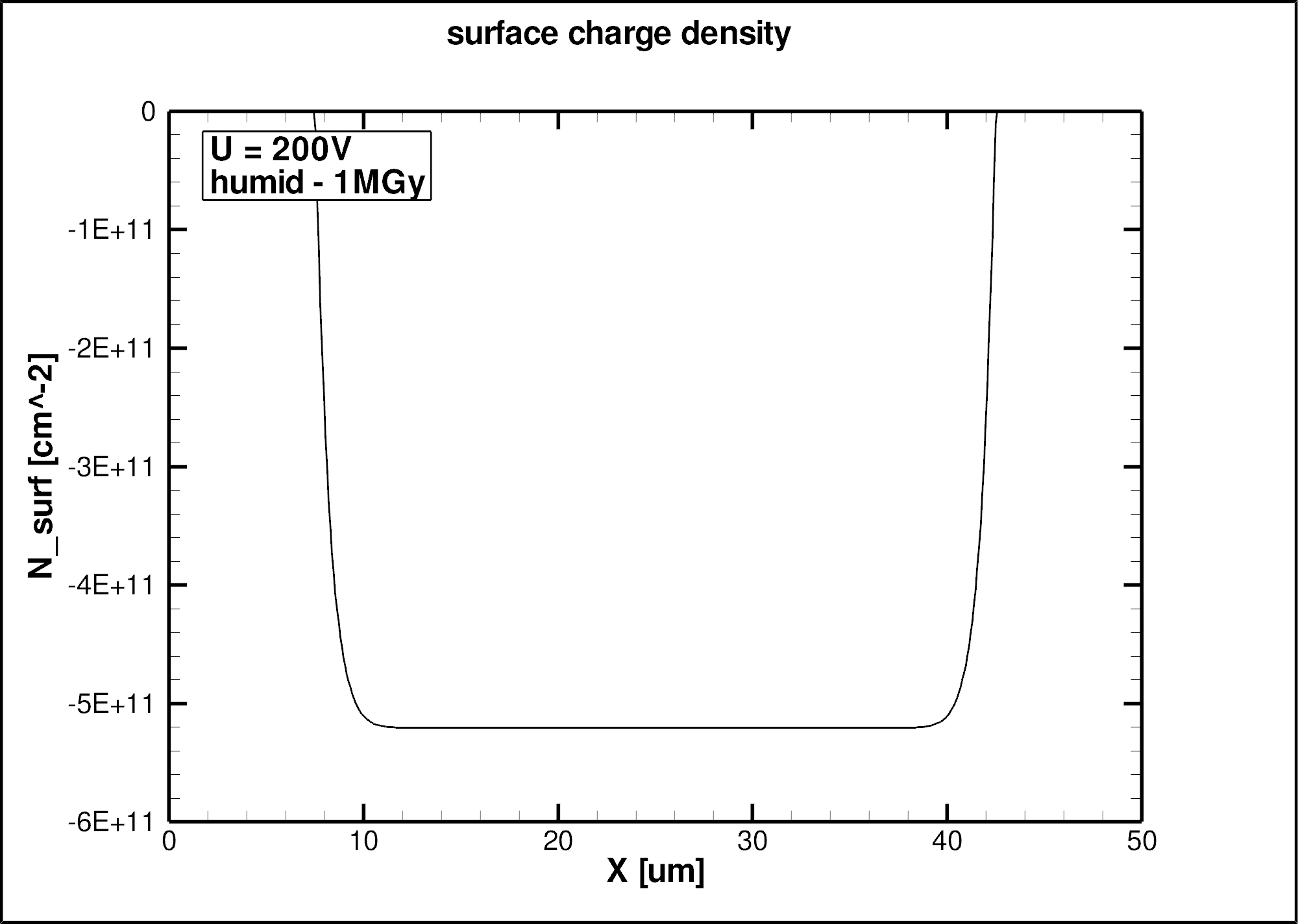}\\
  \caption{Simulated distribution of the charge-carrier density on the surface of the sensor biased to 200~V for steady-state conditions. Left: Non-irradiated sensor ($N_{int}^{eff} = 10^{11}$~cm$^{-2}$); right: Irradiated sensor ($N_{int}^{eff} = 10^{12}$~cm$^{-2}$). }
  	\label{fig:surf_charge}
\end{figure}

% All measurements presented are performed at room temperature ($22.5^{\circ}$C).
 For the non-irradiated strip sensor the most relevant results, which were presented in Chapter~3, are:
  \begin{itemize}
   \item  Electron losses when ramping up the voltage in a dry atmosphere ("dried~at~0~V~-~0~Gy"),
   \item  hole losses when  ramping down the voltage in a dry atmosphere ("dried~at~500~V~-~0~Gy"),
   \item  no or little charge losses in a humid atmosphere ("humid~-~0~Gy"), and
   \item  the time to reach the steady state after a voltage change is about an hour in a humid and $\sim $~100~hours in a dry atmosphere.
  \end{itemize}
   We explain these observations in the following way:  We assume that the sensor is in steady-state conditions at 0~V with zero charge density on its surface. When the sensor is biased, parts of the $p^+$~implants will be depleted, resulting in negative charges at the $p^+n$~junctions of the strips. These negative charges are balanced by the positive charges of the depleted $n$~bulk and of the $n^+$~implant of the rear contact, if the sensor is biased above depletion. These charges produce an electric field at the Si-SiO$_2$ interface and at the sensor surface. If surface charges on top of the passivation do not move, the electric field at the surface will have a longitudinal component which points to the $p^+$~implants and a transverse component at the  Si-SiO$_2$ interface which points into the sensor~\footnote{In the simulation this is realised by defining a boundary at
   $y = - 100$~$\upmu$m where Neumann boundary conditions are applied.  On the surface of the passivation layer, fixed charges, in this case zero, are put. For the simulation "dried at 500~V" the surface-charge distribution obtained for the steady-state conditions at 500~V is used.}. This is seen in the top left plot of
 Figure~\ref{fig:long_field},
   which shows for a sensor biased to 200~V a TCAD simulation of the longitudinal surface field for a density $N_{int}^{eff} = 10^{11}$~cm$^{-2}$ and zero surface-charge density. This longitudinal surface field, which reaches values of 100~kV/cm in the simulation, will cause the redistribution of surface charges until a uniform surface potential is reached. This is the steady-state condition for a given applied voltage. As the effective surface conductivity increases with increasing humidity, the steady state will be reached in a shorter time for humid than for dry conditions. The left plot of
 Figure~\ref{fig:surf_charge}
   shows the simulated surface charge distribution for $N_{int}^{eff} = 10^{11}$~cm$^{-2}$, the $p^+$~implants at 0~V, and the rear contact at $200$~V, which approximately represents the steady-state condition. The potential on the surface is also set to 0~V. In principle the potential on the surface should have been set to a voltage so that the integrated charge on the surface is zero. We however did not manage to perform such a simulation. A crude estimation indicates that the surface potential is  between 1/3 to 1/2 of the potential of the accumulation layer.

   If the sensor is in steady-state conditions under bias and the voltage is ramped down, the field direction will be opposite to the situation discussed above: The transverse component of the electric field will point into the SiO$_2$ and the longitudinal component of the surface field will point away from the $p^+$~implants.
   This can be seen at the bottom left plot of
 Figure~\ref{fig:long_field},
  which shows the simulated longitudinal field distribution for the sensor initially in steady-state conditions at 500~V and then biased to 200~V in a dry atmosphere, i.e. assuming the surface charge distribution from the steady-state simulation at 500~V. We note, that the maximum value of the simulated surface field is only 25~kV/cm, significantly smaller than the value for the case discussed above. The wiggles in the curves are an artifact of the simulation: The surface charge distribution at 500~V has been parameterised by the sum of 10 Gaussians. We stress that, given the assumptions made in the simulations, the results should  be understood as indicative only.

   For completeness we also show on the right sides of
 Figures~\ref{fig:long_field} and \ref{fig:surf_charge}
   the simulated surface fields and surface-charge distributions for a density $N_{int}^{eff} = 10^{12}$~cm$^{-2}$, which corresponds to a sensor with X-ray radiation damage. We note that qualitatively the results are similar to the non-irradiated situation. However, both the surface fields and the surface-charge densities are significantly higher than for the non-irradiated sensor.

  The dependence of the surface sheet resistance, $R_\Box$, on humidity, and the impact on the performance of MOS structures, are well documented
    \cite{Atalla:1960, Shockley:1964, Grove:1967}.
  Under the simplified assumption that the sheet resistance is independent of the electric field, the time dependence of the distribution of the surface charge on the way to the steady state scales with $R_\Box$. This scaling is observed in the measurements shown in
    Figure~\ref{fig:time_dep}.
  The ratio of the time constants of $\sim$~120 is compatible with values from the literature. A crude estimation of the surface resistivity, following the approach presented in
    \cite{Heimann:1982},
  gives values for $R_\Box$ of the order of $10^{17}$~$\Omega $ for the humid, and approximately a factor 120 higher for the dry conditions. Similar values for $R_\Box$ are reported in
    \cite{Grove:1967, Heimann:1982}.

  In Section 3.2 we have mentioned that the time it takes to reach the steady state on the surface of the sensor does not depend on the distance between the injected light and the end of the strips where the openings in the passivation are located. This agrees with the expectation that the steady state on the sensor surface is reached by a local redistribution of the surface charges, and not by charges moving from or to the bond pads.
%  , where the aluminium is not covered by the passivation.

\begin{figure}
	\centering
		\includegraphics[width=7.4cm]{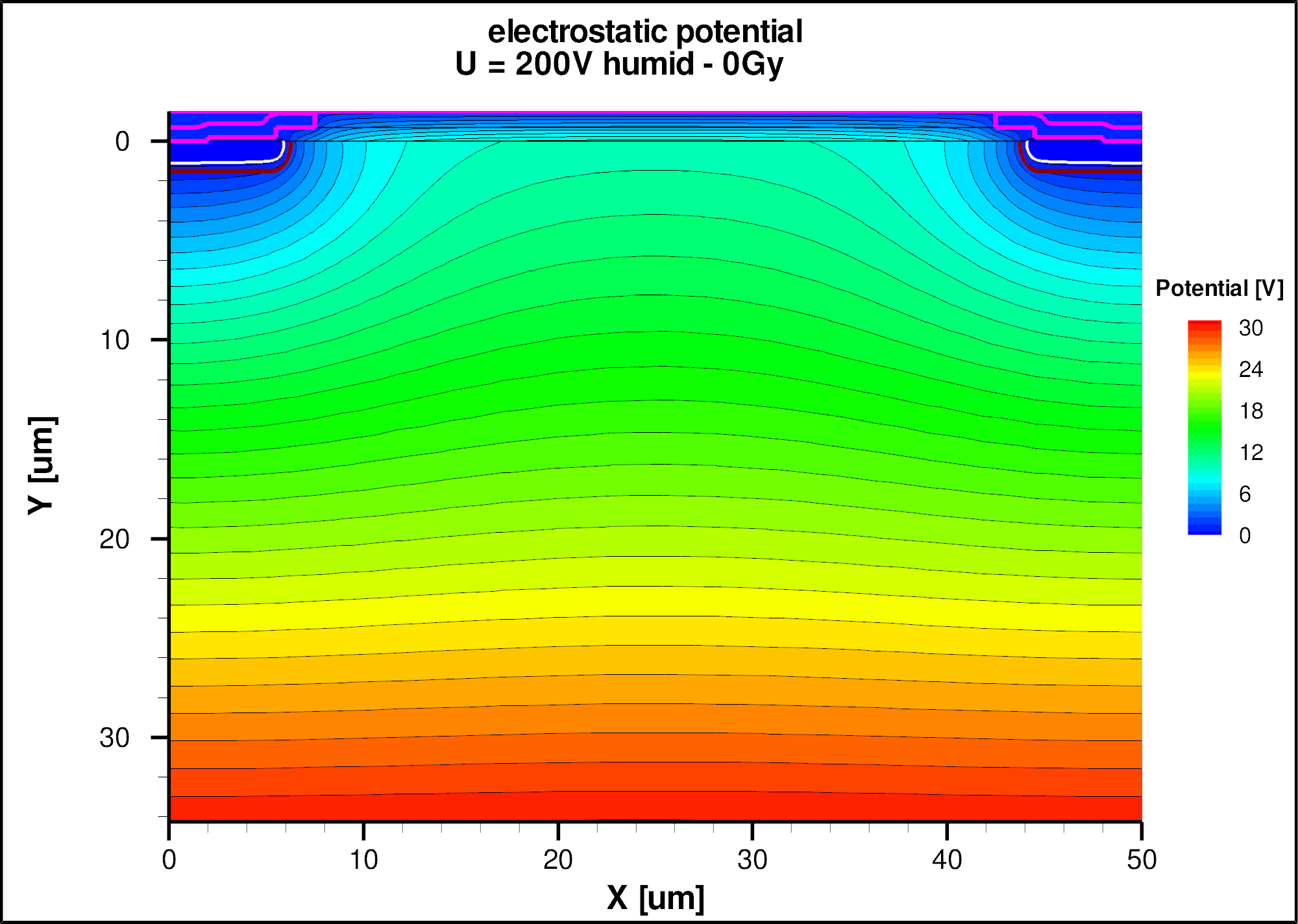}
		\includegraphics[width=7.4cm]{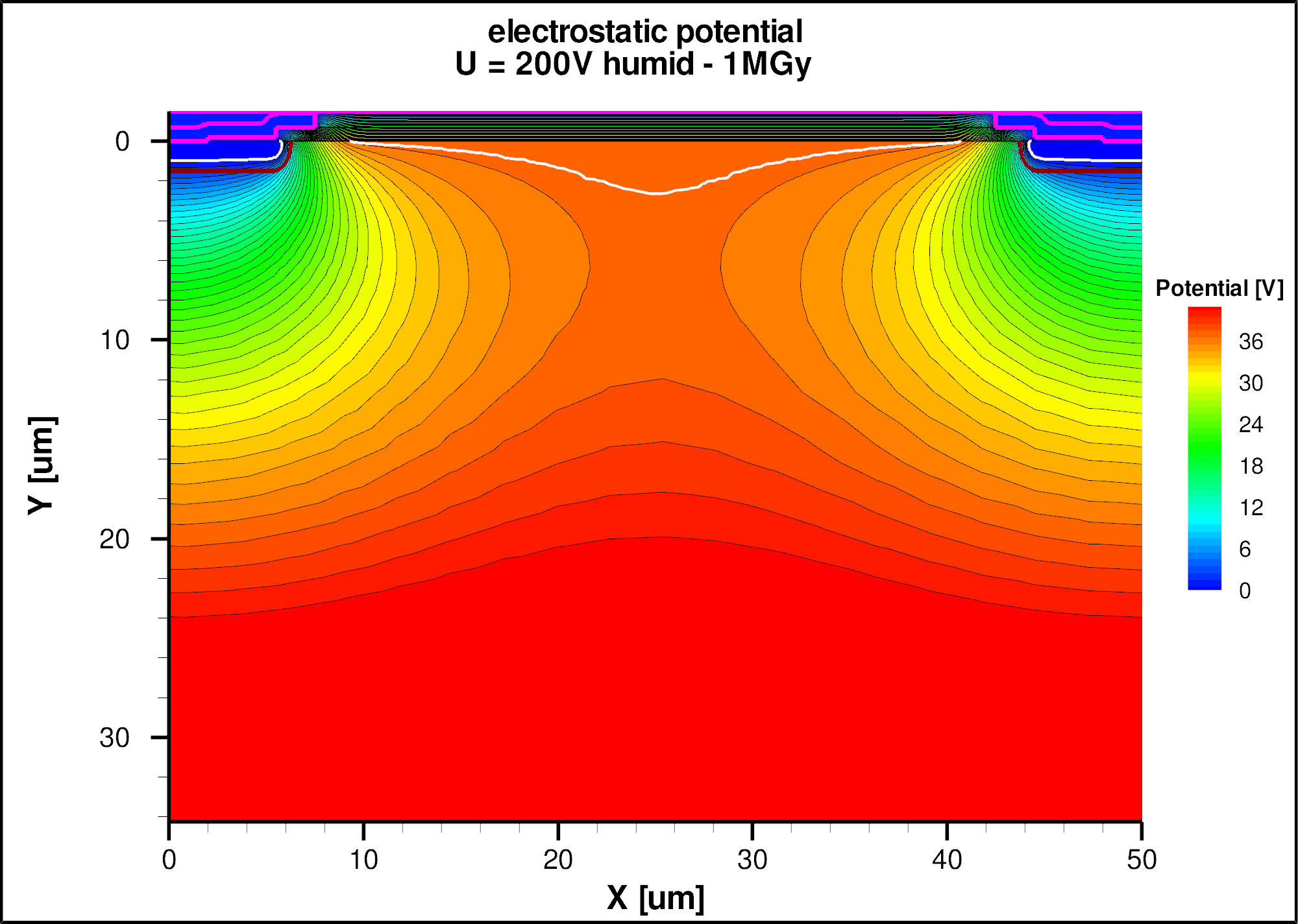}\\
		\includegraphics[width=7.4cm]{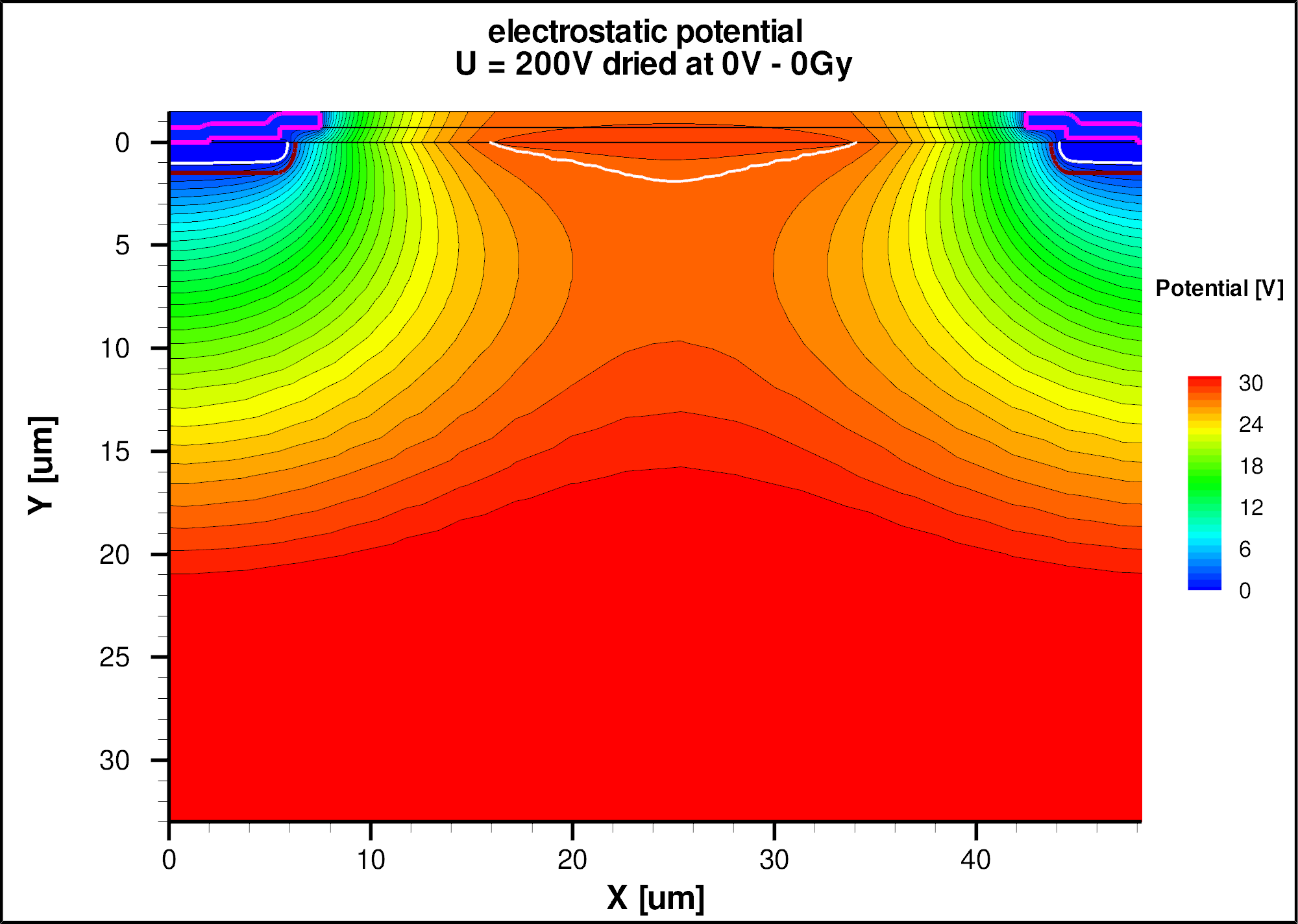}
		\includegraphics[width=7.4cm]{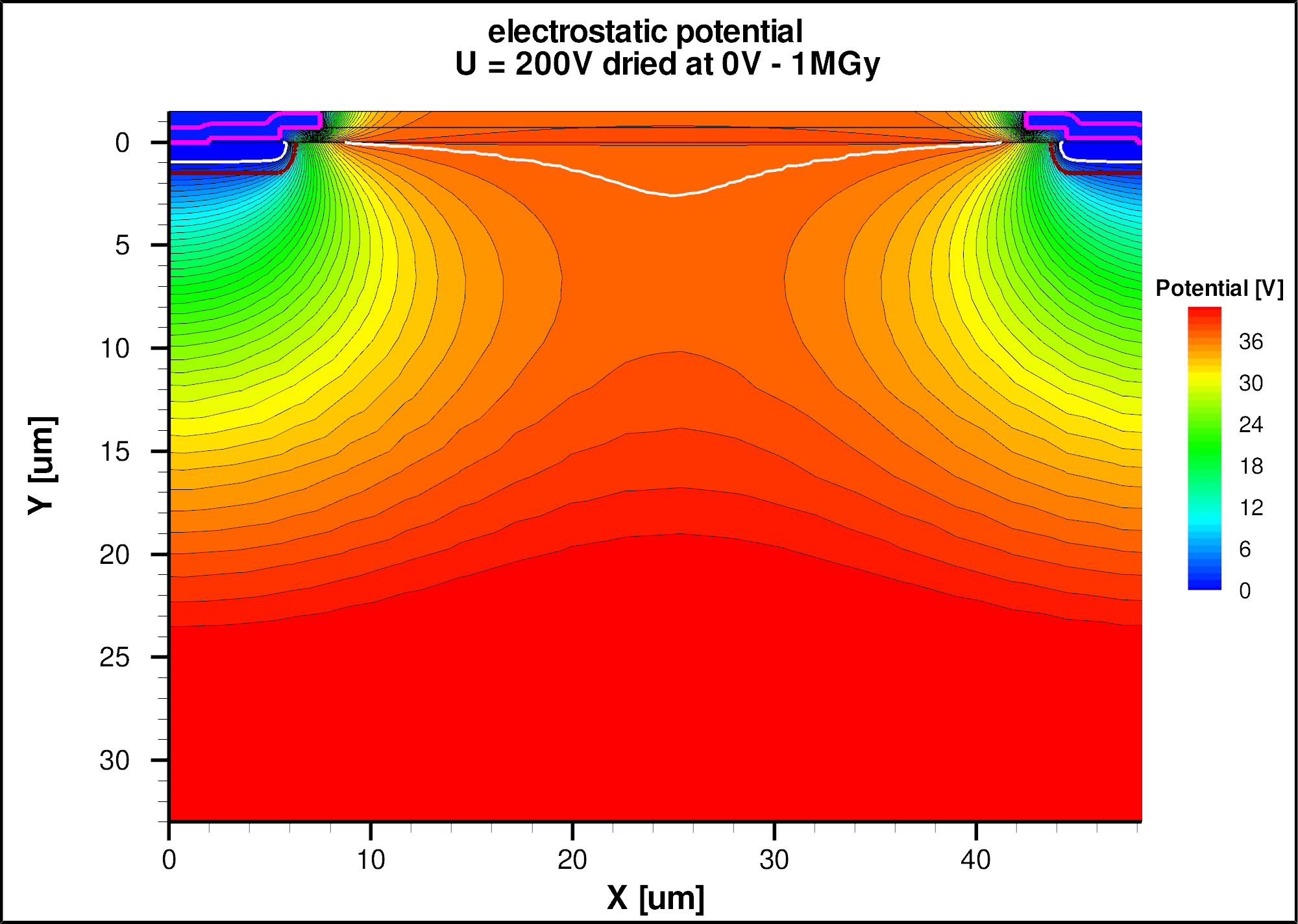}\\
		\includegraphics[width=7.4cm]{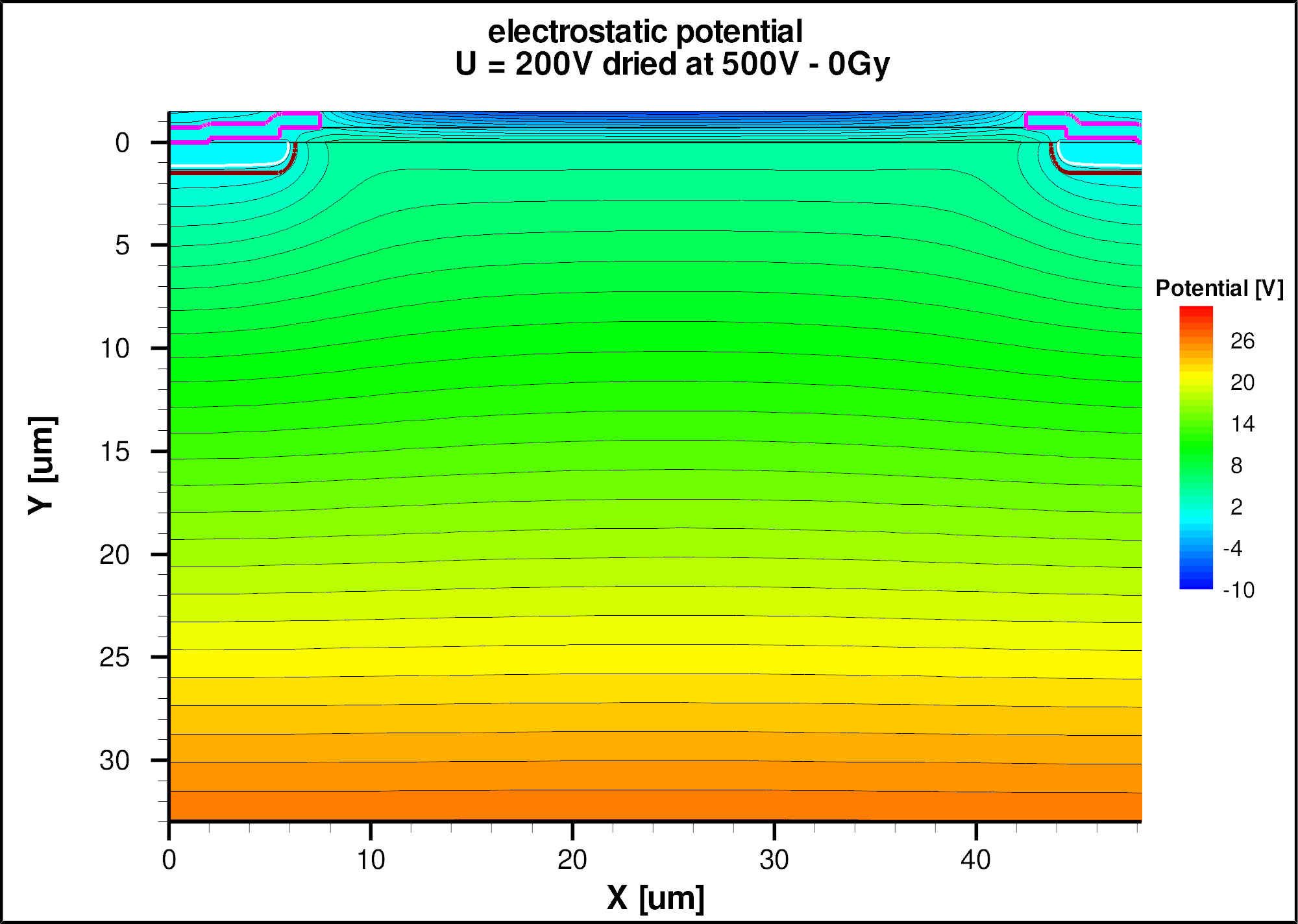}
		\includegraphics[width=7.4cm]{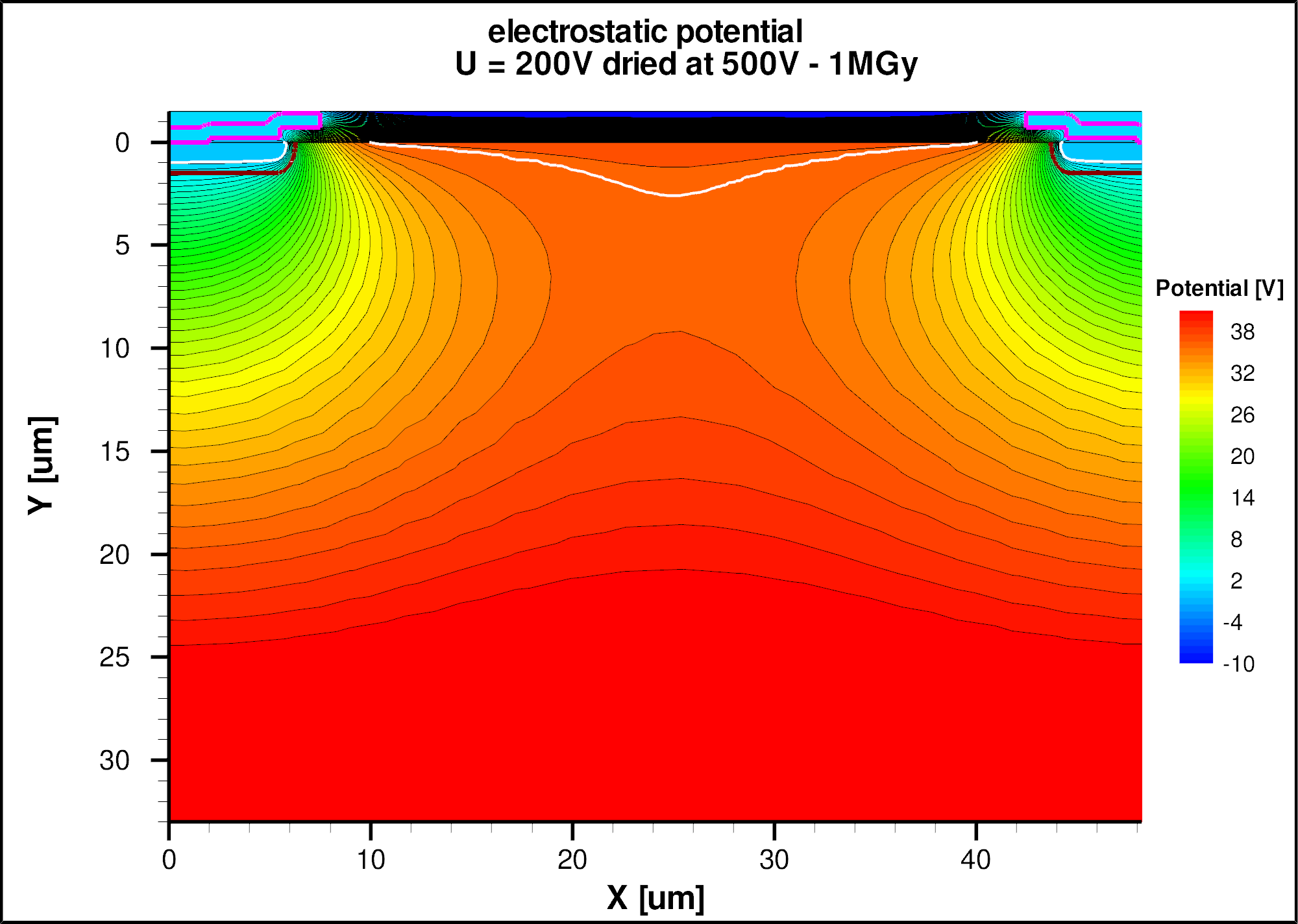}\\
  \caption{Simulated potential distribution for the sensor biased to 200~V: Non-irradiated (left - $N_{int}^{eff} = 10^{11}$~cm$^{-2}$) and  irradiated to 1~MGy  (right - $N_{int}^{eff} = 10^{12}$~cm$^{-2}$) for the conditions "humid" (top), "dried at 0~V (middle)" and "dried at 500~V (bottom)".}
  	\label{fig:potentials}
 \end{figure}

\begin{figure}
	\centering
		\includegraphics[width=7.4cm]{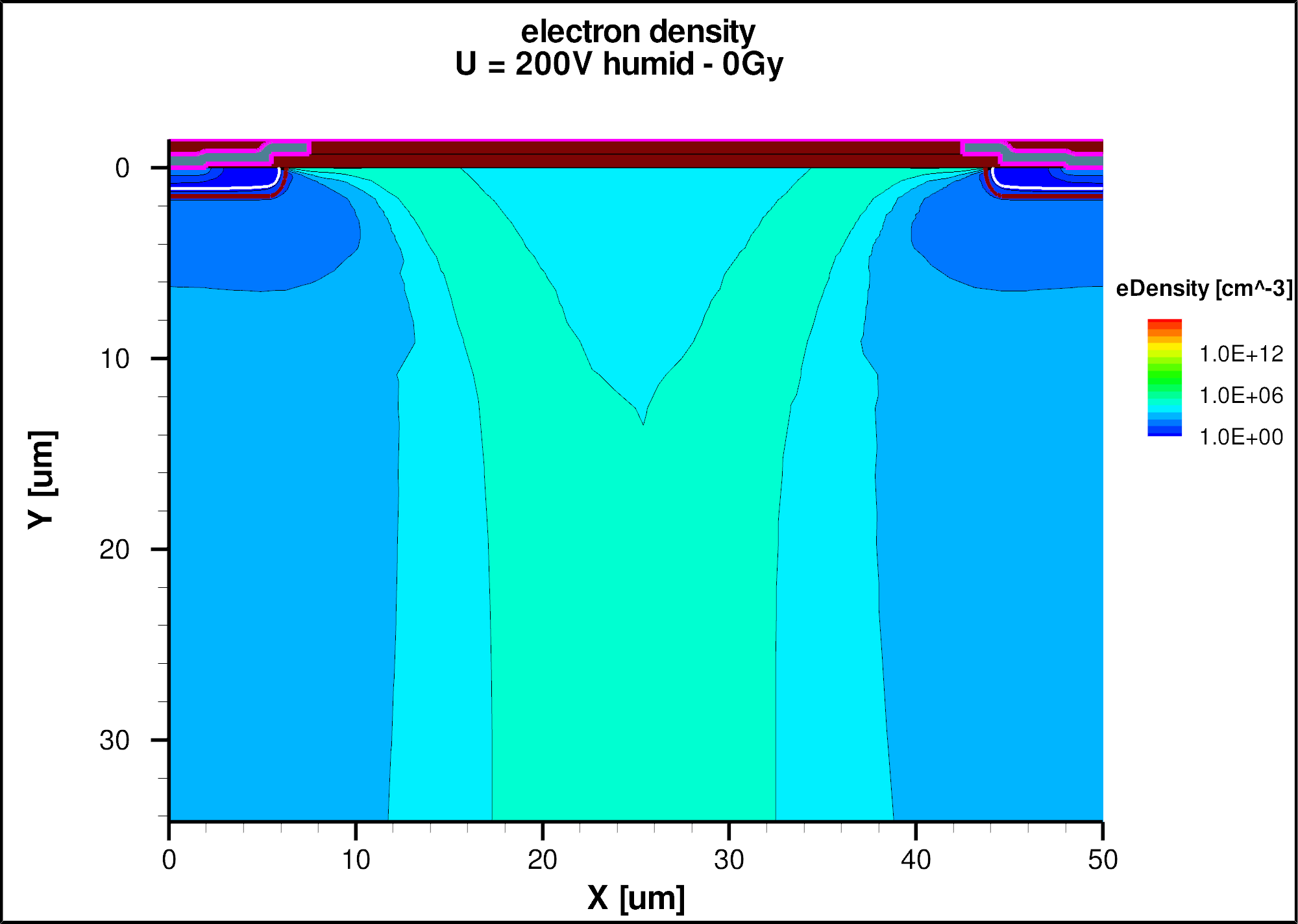}
		\includegraphics[width=7.4cm]{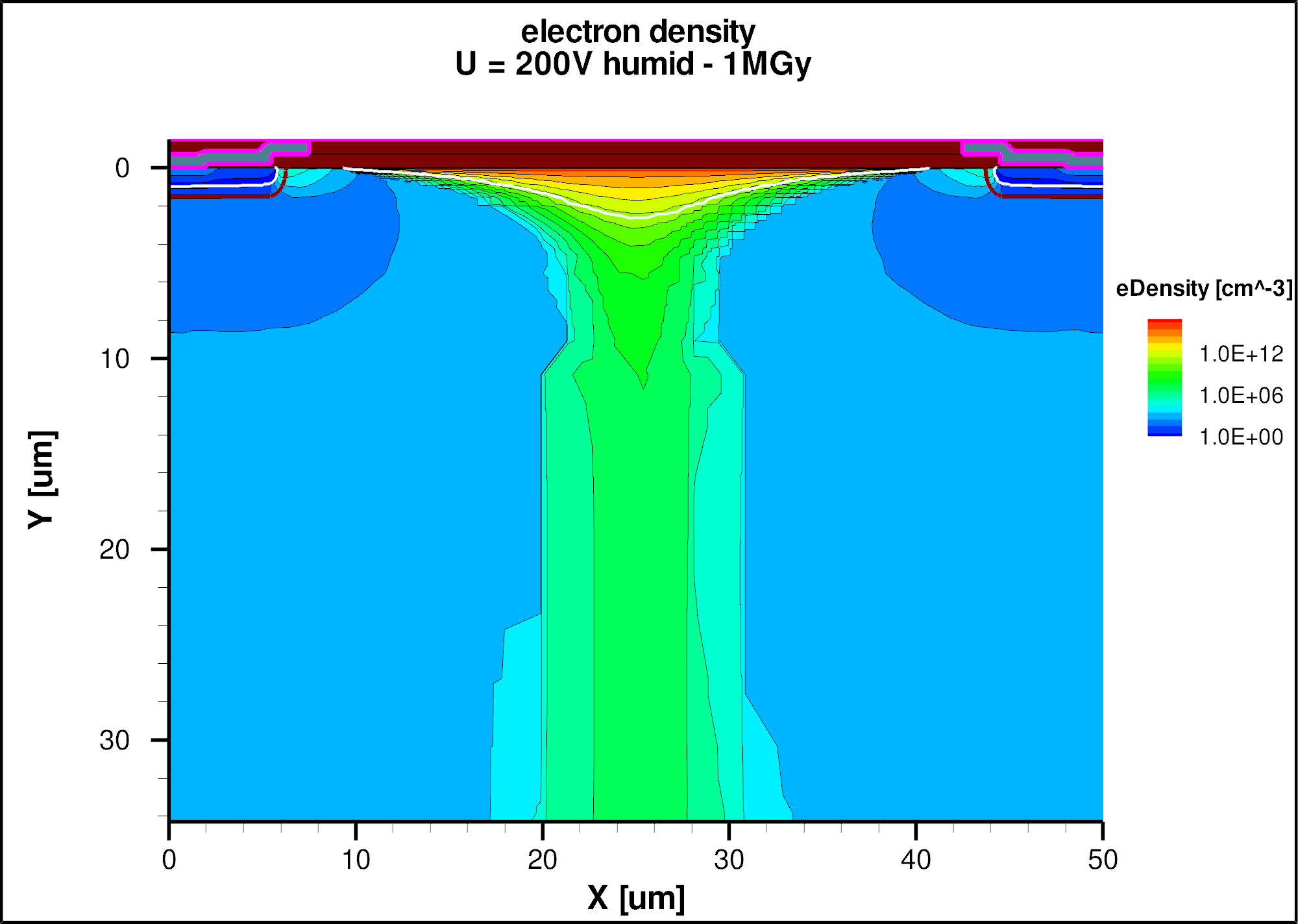}\\
		\includegraphics[width=7.4cm]{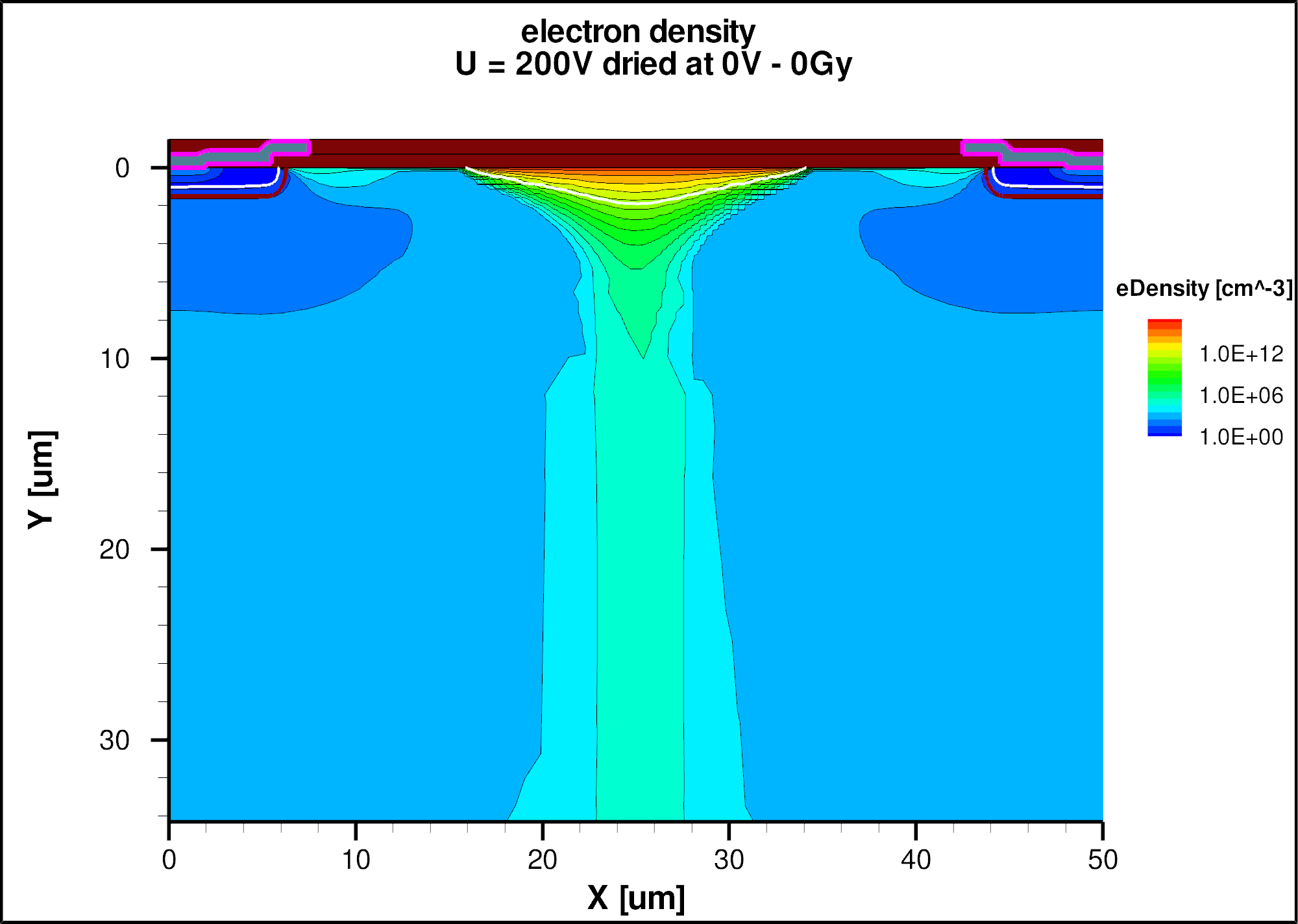}
		\includegraphics[width=7.4cm]{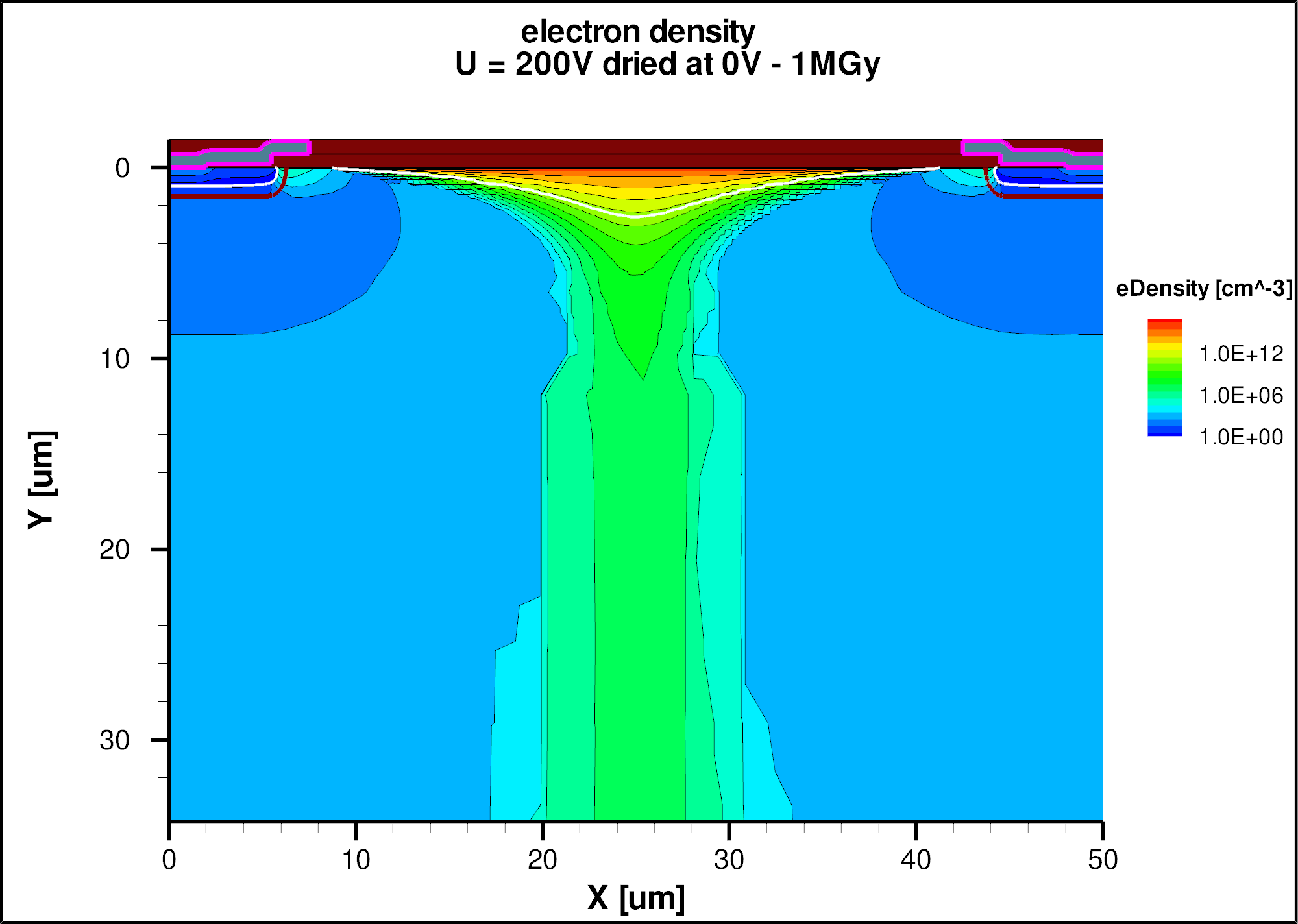}\\
		\includegraphics[width=7.4cm]{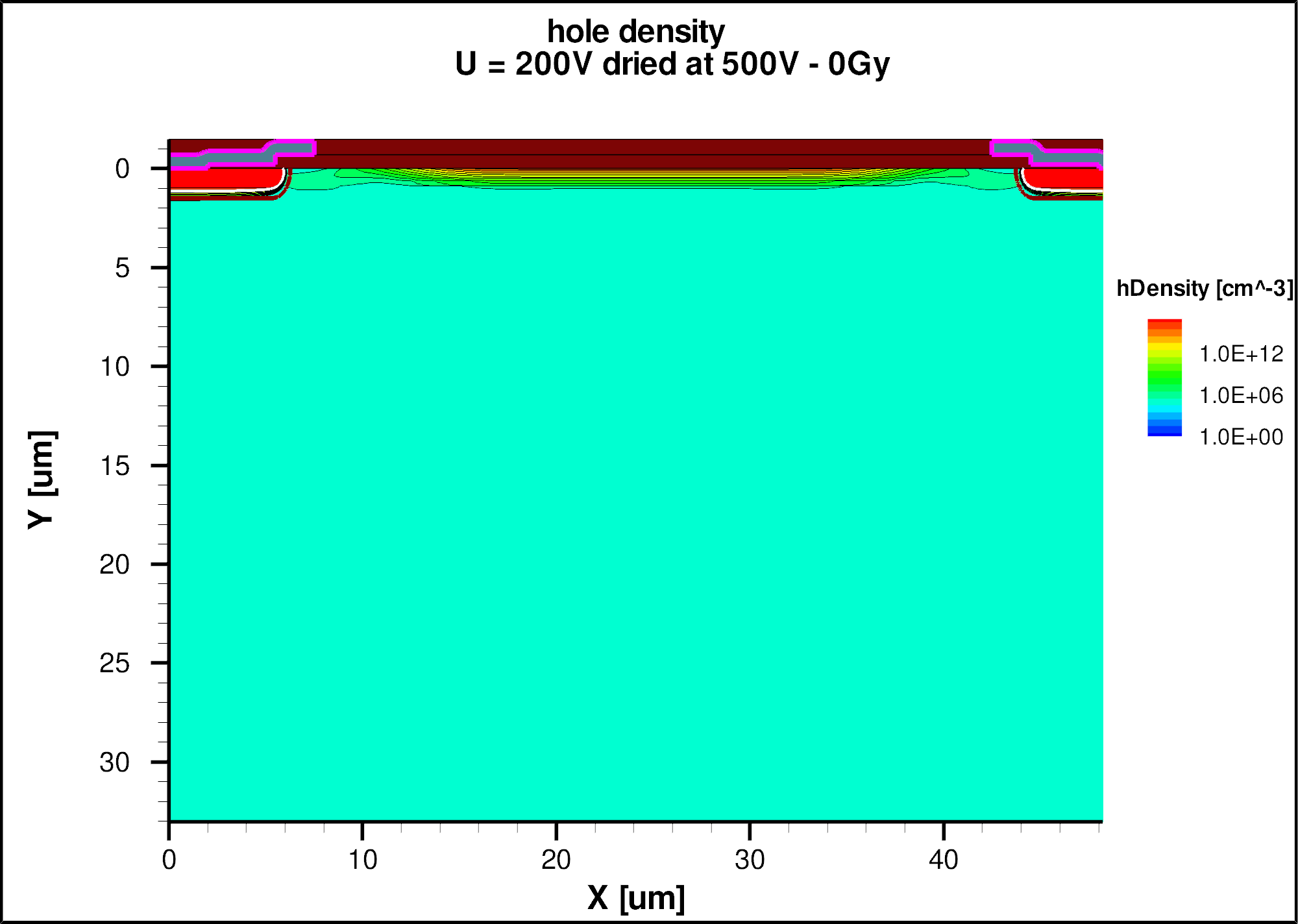}
		\includegraphics[width=7.4cm]{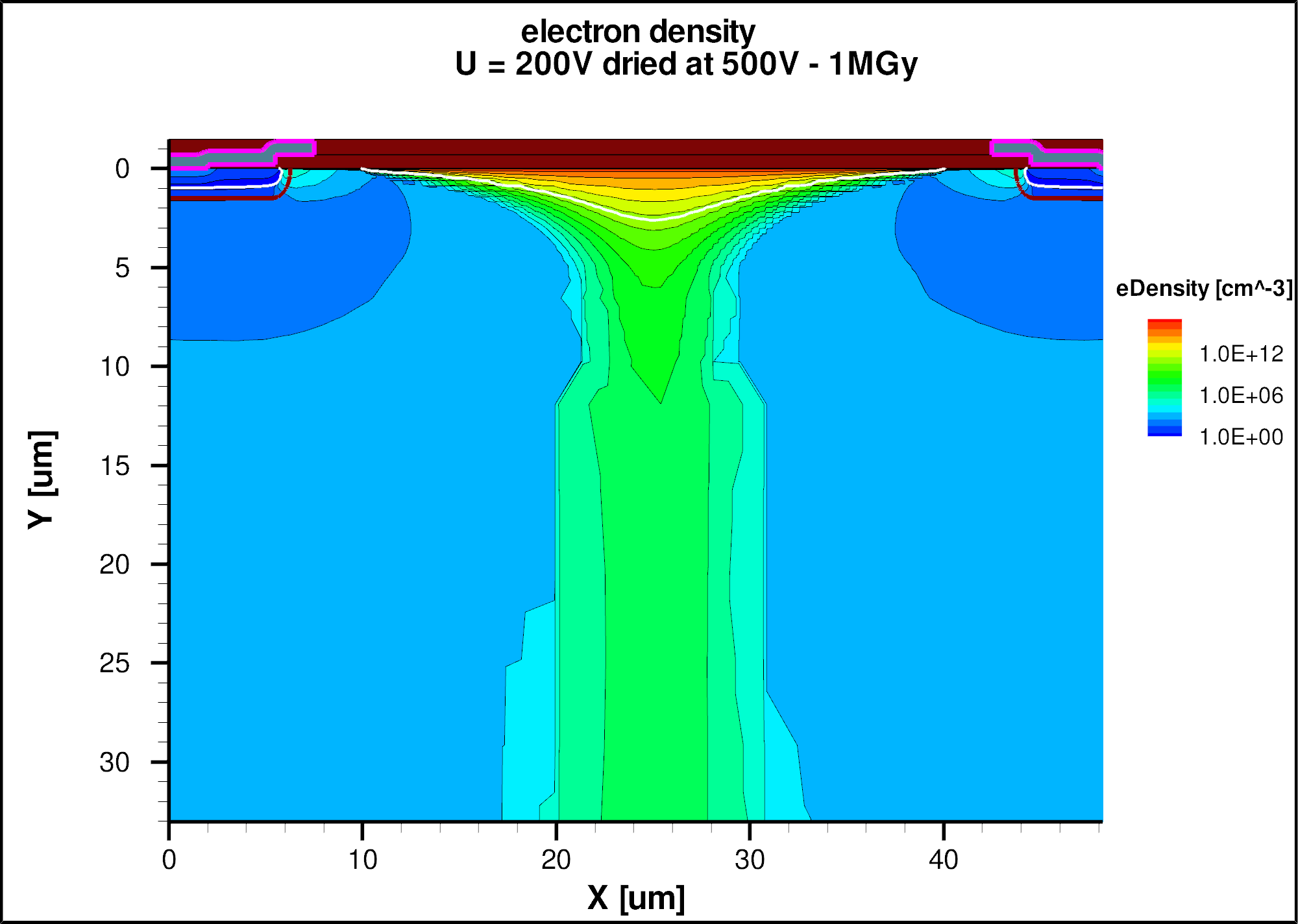}\\
  \caption{ Simulated electron and hole densities close to the Si-SiO$_2$ interface for the sensor biased to 200~V: Non-irradiated (left - $N_{int}^{eff} = 10^{11}$~cm$^{-2}$) and  irradiated to 1~MGy  (right - $N_{int}^{eff} = 10^{12}$~cm$^{-2}$) for the conditions "humid" (top), "dried at 0~V (middle)" and "dried at 500~V (bottom)". Except for the bottom left plot, which shows the holes density, only the electron densities are shown. }
  	\label{fig:acc_layers}
\end{figure}

    Next we explain the reasons for the different type of charge losses for the different measurement conditions with the help of selected TCAD simulations
 \cite{Schwandt:Thesis}.
 Figures~\ref{fig:potentials} and \ref{fig:acc_layers}
    show simulated distributions of the electric potential and of the electron and hole densities in the sensor close to the Si-SiO$_2$~interface for different measurement conditions. We start by discussing the situation of significant electron losses, which are observed for the condition "dried~at~0~V~-~0~Gy". For an explanation of the nomenclature we refer to Section 3.1.

   \begin{figure}
	\centering
		\includegraphics[width=12cm]{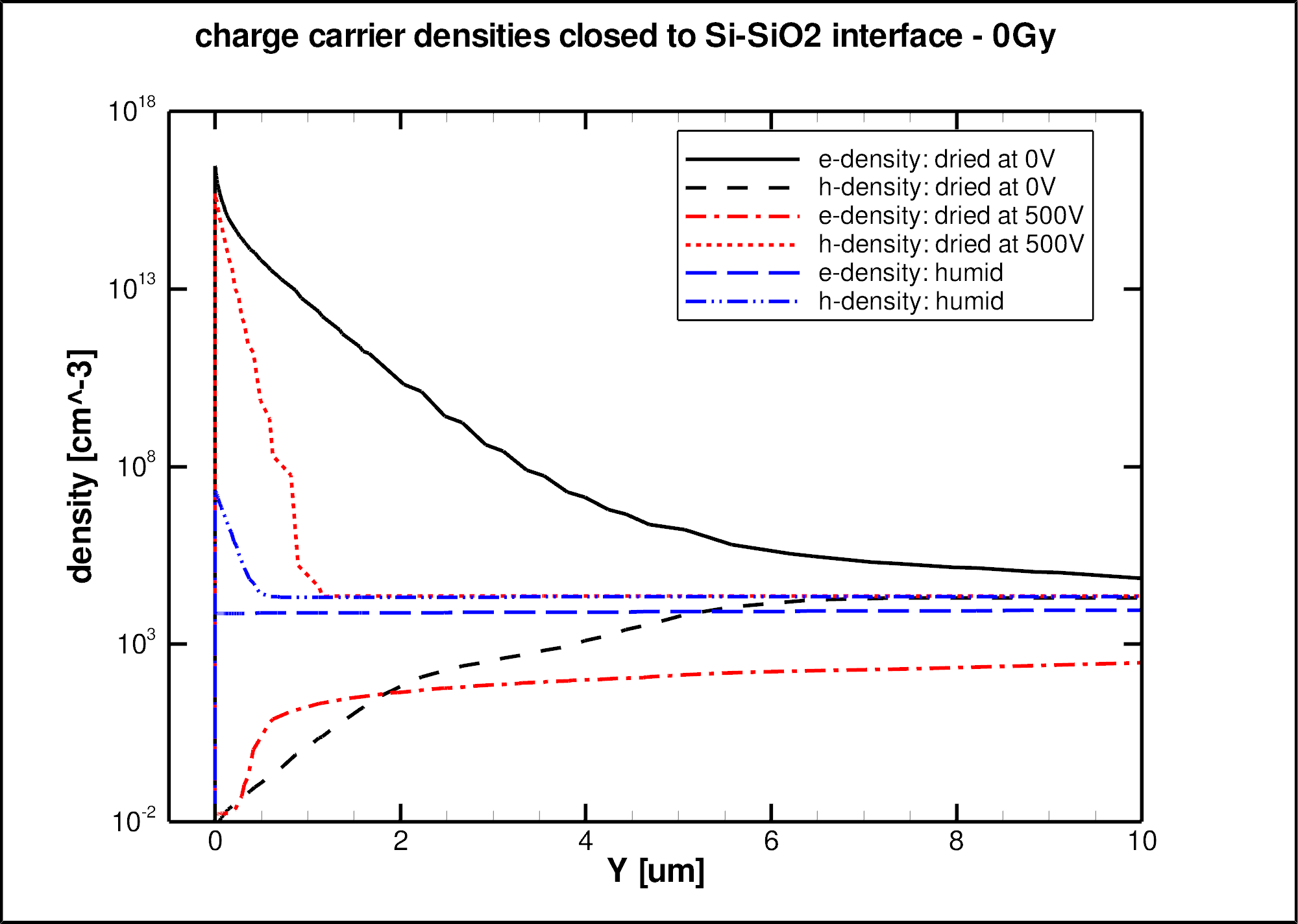}\\
		\includegraphics[width=12cm]{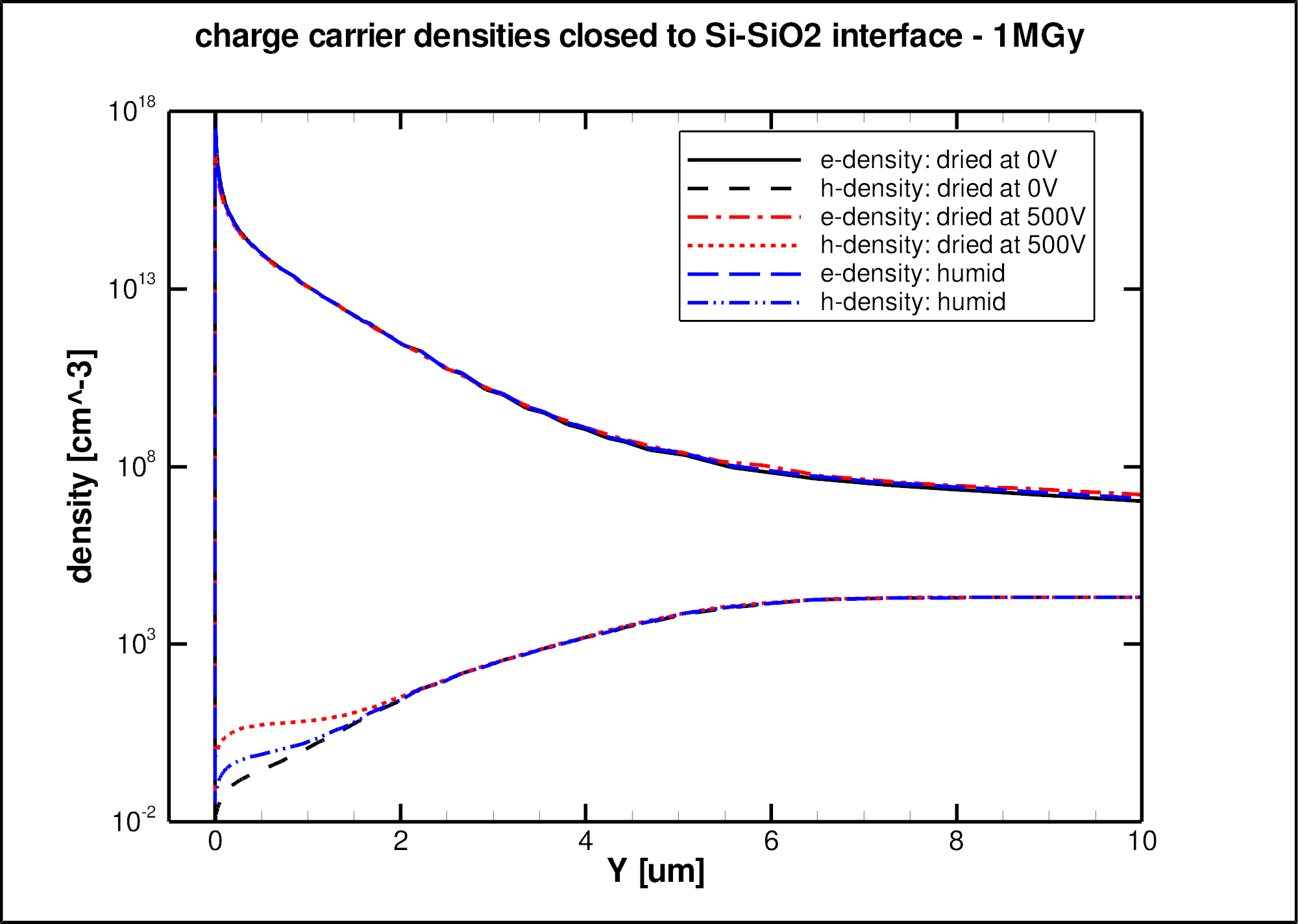}\\
  \caption{ Simulated charge carrier densities close to the Si-SiO$_2$~interface in the symmetry plane between the strips for the sensor biased to 200~V and different measurement conditions. Left for $N_{int}^{eff} = 10^{11}$~cm$^{-2}$, corresponding to a non-irradiated sensor, and right for $N_{int}^{eff} = 10^{12}$~cm$^{-2}$, corresponding to a sensor irradiated to dose of 1~MGy. A high electron density at the interface ($y = 0$) is evidence for an electron-accumulation layer, a high hole density for an inversion layer.}
  	\label{fig:e-h-cut}
\end{figure}

   We assume that in steady-state conditions with the strips and the rear contact at 0~V, the charge density on the surface of the sensor is zero. If the sensor is biased to 200~V in the condition "dried~at~0~V~-~0~Gy" the surface-charge density remains zero, and an electron-accumulation layer forms below the  Si-SiO$_2$~interface as seen by the high electron density visible in the middle left plot of
 Figure~\ref{fig:acc_layers} and the top plot of
 Figure~\ref{fig:e-h-cut}:
   The  electron density reaches a maximum value of $\sim 3 \cdot 10^{16}$~cm$^{-3}$ at the interface for a depth of $\sim$~5~nm in $y$. The white lines in Figures~\ref{fig:potentials} and \ref{fig:acc_layers} indicate the electron density of $\sim 10^{12}$~cm$^{-2}$, which corresponds to the $n$ doping of the sensor. Inspection of the corresponding potential distribution (middle left plot of
 Figure~\ref{fig:potentials})
   shows, that the potential has a saddle point $\sim 5$~$\upmu$m below the Si-SiO$_2$~interface, and that the electric field points from the interface into the sensor. Thus holes produced close to the interface will drift in a short time along the field lines to the readout strips where they are collected. Hole losses due to recombination in the accumulation layer are estimated to be negligible. However, a fraction of the electrons produced close to the accumulation will reach the accumulation layer and, like those produced in this layer, will spread over the layer with a time constant given by the dielectric relaxation time
 $\tau_R = \varepsilon_{Si}/(q_0\cdot \mu_e\cdot n)$.
   $\varepsilon_{Si}$ denotes the dielectric constant of silicon, $\mu_e$ the electron mobility and $n$ the position dependent electron density. For $n = 10^{15}$~cm$^{-3}$ the value is $\tau_R = 5$~ps
 \cite{Grove:1967},
    which is short compared to the charge collection time of a few nanoseconds. These electrons are "lost", as they do not induce a significant signal within the integration time of the measurements. As they spread at least over the entire length of the accumulation layer along the sensor strips, the resulting increase in electron density and change of the local electric field will  be quite small. A study of the impact of the number of electrons "lost" on the charge collection, and the time required to return to the pre-light injection state, is the topic of a separate publication
 \cite{Poehlsen:Thesis, Chargelosses:2012}.
    We also note from
  Figure~\ref{fig:acc_layers},
    that the depths of the electron-accumulation layers have their maxima in the symmetry plane between the readout strips and decrease towards the readout strips.
% This explains, why a triangular distribution for the electron losses had to be assumed in the model described in Section~2.4.

   Next we discuss the situation of hole losses which occur for the condition "dried~at~500~V~-~0~Gy". In this case the negative charges on the surface overcompensate the positive charges at the interface, and a hole-inversion layer forms below the Si-SiO$_2$~interface. This can be seen in the bottom left of
 Figure~\ref{fig:acc_layers} and in Figure~\ref{fig:e-h-cut}
   as a high density of holes at the interface.
   The value found for the maximum hole density is $\sim 4 \cdot 10^{15}$~cm$^{-3}$. We note that the depth of the inversion layer is much smaller than for the accumulation layer and essentially independent of position $x$.
% which explains why position independent hole losses could be assumed in the model described in Section~3.2.
    The potential distribution presented in the bottom left plot of
 Figure~\ref{fig:potentials}
   shows that the electric field distribution resembles the situation of a pad sensor: The electric field in the sensor points towards the Si-SiO$_2$~interface and the transverse field component is small. This explains the large value of the hole diffusion term $\sigma _{diff}$ observed for this condition and reported in
 Table~\ref{tab:fit_param}.

   Finally we discuss the situation of no or little losses, which is observed for the condition "humid~-~0~Gy". In this case the potential on the surface of the sensor is uniform, the redistributed surface charges compensate the positive interface charges of density $N_{int}^{eff}$, and neither an electron-accumulation nor a hole-inversion layer forms at the Si-SiO$_2$~interface. As shown in
 Figure~\ref{fig:e-h-cut} top,
   both electron and  hole densities are below $\sim 10^{8}$~cm$^{-3}$ at the Si-SiO$_2$~interface.  From the corresponding potential distribution shown on the top left plot of
 Figure~\ref{fig:potentials}
    we conclude, that the electric field close to the Si-SiO$_2$~interface is weak and points to the interface. Nevertheless, most holes generated in this region will diffuse until they reach a region of higher field and then drift to a readout strip.
% This explains why, as discussed in Chapter 2.4, an increased diffusion term is required to describe the measurements.

   The results of the simulations for the irradiated sensor for the three measurement conditions are shown on the right sides of
 Figures~\ref{fig:potentials},  \ref{fig:acc_layers} and the bottom of Figure~\ref{fig:e-h-cut}.
   Given the high positive charge density at the interface of $N_{int}^{eff} = 10^{12}$~cm$^{-2}$ electron-accumulation layers are present under all three conditions. Given that the charge densities change by 20 orders of magnitudes, the curves for the electron densities appear indistinguishable on top of each other. Nevertheless, the values at $y = 0$ are quite different: $2.5\cdot 10^{18}$, $6\cdot 10^{17}$ and $8 \cdot 10^{16}$~cm$^{-3}$ for "dried~at~0~V~-~1~MGy" "humid~-~1~MGy" and "dried~at~500~V~-~1~MGy", respectively. Qualitatively this dependence is expected from the differences in surface-charge distributions and also fits the measured numbers of the electrons and holes collected as presented in
 Table~\ref{tab:fit_param}:
   Electron losses $\sim 90\%$ for "dried~at~0~V~-~1~MGy" and $\sim 30 \%$ for "humid~-~1~MGy". For the situation "dried~at~500~V~-~1~MGy", different to the expectation from the simulation, no or only minor electron or hole losses are observed. Given all the uncertainties in the assumptions made in the simulations, we are not worried by this difference. We note, that by changing the value of $N_{int}^{eff}$ in the simulation, the accumulation layer also changes.

   We summarise: The observed losses of electrons (holes) close to the Si-SiO$_2$~interface can be qualitatively understood by the shape of the electric field in the sensor close to the interface and by the formation of an electron-accumulation (hole-inversion) layer at the interface. Both are influenced by the distribution of charges on the surface of the sensor and by the density of charged states in the region of the Si-SiO$_2$ interface. The latter is a strong function of X-ray radiation damage. After changing the sensor voltage it takes some time until the steady-state conditions of the surface-charge distribution are reached. The time constant, which can be as long as several days, depends on the humidity.

\subsection{Relevance of the results}

 This chapter discusses the relevance of the measurements for the understanding, the design and the operation of segmented $p^+n$ sensors.

 First of all, the measurements give some insight into the complexities of the Si-SiO$_2$ interface and of possible effects of the properties of the passivation on sensors with and without surface radiation damage. They also show the difficulties in defining the correct boundary conditions on the surface of sensors, required for a realistic sensor simulation. Some insight has also been gained into the well known effect
   \cite{Longoni:1990, Richter:1996}
 that, after changing the voltage it takes some time until steady-state conditions are reached and that the corresponding time constant is a strong function of the humidity
   \cite{Chilingarov:2006}.
 Time constants can be as long as several days. Many users of silicon sensors have actually observed such "long-time run-aways". Depending on the design of the sensor, in particular if the metal of the strips does not overlap the SiO$_2$, breakdown at low bias voltages has been observed when the sensors are ramped up in a dry atmosphere
   \cite{Hartjes:2005}.
 We conclude that for sensors exposed to high doses of X-rays the oxide area not covered by metal should be minimized. However, the increase in capacitance and the resulting  increase in readout noise has to be taken into account. For a hybrid pixel sensor, like the AGIPD sensor with a pixel size of 200~$\upmu$m~$\times~200~\upmu$m, a small inter-pixel gap of order 20~$\upmu$m appears as a reasonable compromise.

 Concerning the relevance for the efficient detection of radiation, we remind that light of 660~nm wavelength has an absorption length in silicon at room temperature of 3.5~$\upmu$m and that the observed charge losses only occur close to the interface.
 Therefore the charge losses are relevant for short range particles only, like low-energy heavy ions entering the $p^+$~strip side of the sensor. It is clear from the studies presented, that the precise  measurements of their energy is difficult and both positive and negative signals will appear on the readout strips. Such effects have actually been observed already quite some time ago
   \cite{Yorkston:1987}
 and a simplified explanation has been given.

 For the detection of minimum-ionizing particles, the effects can certainly be ignored, as long as the thickness of the sensor is large compared to $\sim 50~\upmu$m. To the authors' knowledge, no effects of charge losses due to surface effects have been reported in such applications. The same is true for the detection of X-rays, in particular when they enter through the rear $n^+$ contact. As an example, for 12 keV X-rays the attenuation length is 250~$\upmu$m and only 0.3~\% of the photons convert in the last 5~$\upmu$m of a 500~$\upmu$m thick sensor. Measurements with light with 1052~nm wavelength, which also has an attenuation length of 250~$\upmu$m, show no measurable charge losses
  \cite{Poehlsen:Thesis}
 and confirm this estimate.

  One worry, which actually has triggered the study, is the situation which occurs at the European X-ray Free-Electron Laser, XFEL
    \cite{XFEL}:
  $\sim 10^5$ photons of 12~keV  will be deposited within tens of femtoseconds in a pixel of size 200~$\upmu$m~$\times~200~\upmu$m followed by the next XFEL pulse after 220~ns. The question is: How long is the "memory" of the pixel for such large pulses, and does the memory change with  surface damage? The studies presented show, that as far as surface effects are concerned, there are no problems: The charge lost locally in the accumulation layer spreads in picoseconds over the accumulation layer of the entire sensor. Plasma effects however, due to the high charge density generated by the high density of interacting photons, may very well result in memory times beyond the XFEL pulse spacing of 220~ns as shown in
    \cite{Becker:Thesis, Becker:2010b, Becker:2010a}.

\section{Summary}

 Using the Transient Current Technique, TCT, with sub-nanosecond focussed light pulses of 660~nm wavelength, the charge collection of $p^+n$ strip sensors before and after 1~MGy of X-ray irradiation for charges produced close to the Si-SiO$_2$~interface has been investigated. The absorption length of 660~nm light in silicon at room temperature is 3.5~$\upmu$m.
Depending on applied bias voltage, bias history, humidity and X-ray irradiation, incomplete collection of either electrons or holes has been observed. The charge losses are due to electron-accumulation or hole-inversion layers at the Si-SiO$_2$ interface and the close-by electric field. The data can be described by a model which allows a quantitative determination of the losses of holes and electrons, and the widths of the accumulation and inversion layers below the Si-SiO$_2$~interface.

 After changing the bias voltage, the charge distribution on the sensor surface is not in a steady state. The time it takes to reach steady-state conditions depends on humidity: Several days in a dry, and two orders of magnitude faster in a humid atmosphere. This difference is related to the decrease of surface resistivity with humidity.

 For non-irradiated sensors little or no charge losses are observed when the surface of the sensor is in steady-state conditions. However, immediately after applying the bias voltage, electron losses occur when the voltage is increased, and hole losses when the voltage is decreased from a steady-state condition.  For irradiated sensors electron losses are observed in steady state. The fraction of electrons lost increases when the voltage is ramped up after exposure to the initial test conditions, and decreases when ramped down.

 The observations for sensors from two vendors built from silicon with different crystal orientations and different coupling of the aluminium strips to the $p^+$~implants are  similar.

 The results provide information on the surface-charge distribution and its impact on the accumulation or inversion layer, as well as on the electric field close to the interface of segmented sensors with and without X-ray-radiation damage. They also demonstrate the difficulties in defining the boundary conditions on the sensor surface for a realistic sensor simulation.

 The observed charge losses are relevant for the detection of radiation with short attenuation length, like low-energy heavy ions, but have a minor effect for the detection of X-rays and minimum-ionizing particles for silicon sensors with thicknesses above 50~$\upmu$m.

\section*{Acknowledgements}

  This work was performed within the AGIPD Project which is partially supported by the XFEL-Company. We would like to thank the AGIPD colleagues for the excellent collaboration. Support was also provided by the Helmholtz Alliance "Physics at the Terascale" and the German Ministry of Science, BMBF, through the Forschungsschwerpunkt "Particle Physics with the CMS-Experiment". J.~Zhang is supported by the Marie Curie Initial Training Network "MC-PAD". We also are thankful to H.~Spieler for essential advice and enlightening discussions.

%\section*{References}

%\include{chapters/bibliography}


\begin{thebibliography}{9}

\bibitem{Kemmer:1980}
 J.~Kemmer,
 \emph{Fabrication of low noise silicon radiation detectors by the planar process},
   Nucl. Instr. and Meth.~169~(1980)~499.

\bibitem{Kemmer:1984}
 J.~Kemmer,
 \emph{Improvement of detector fabrication by the planar process},
   Nucl. Instr. and Meth.~226~(1984)~89.

\bibitem{Nicollian:1982}
 E.H.~Nicollian and J.R.~Brews,
 \emph{MOS (Metal Oxide Semiconductor) Physics and Technology},
    New York, Wiley-Interscience, 1982.

\bibitem{Longoni:1990}
 A.~Longoni, M.~Sampietro and L.~Strüder,
 \emph{Instability of the behaviour of high resistivity silicon detectors due to the presence of oxide charges},
    Nucl. Instr. and Meth. A~288~(1990)~35.

\bibitem{Richter:1996}
 R.H.~Richter et al.,
 \emph{Strip detector design for ATLAS and HERA-B using two-dimensional device simulation},
    Nucl. Instr. and Meth. A~377~(1996)~412.

\bibitem{Eremin:2003}
 V.~Eremin et al.,
 \emph{The charge collection in single side silicon microstrip detectors},
    Nucl. Instr. and Meth. A~500~(2003)~121.

\bibitem{Verbitskaya:2003}
 E.~Verbitskaya et al.,
  \emph{Effect of SiO2 Passivating Layer in Segmented Silicon Planar Detectors on the Detector Response},
    IEEE TRANSACTIONS ON NUCLEAR SCIENCE, VOL. 52, NO. 5, OCTOBER 2005.

\bibitem{AGIPD}
 B.~Henrich et al.,
  \emph{The adaptive gain integrating pixel detector AGIPD, a detector for the European XFEL},
     Nucl. Instr. and Meth. A~500~Suppl.~1(2011)~S11, and      \url{http://hasylab.desy.de/instrumentation/detectors/projects/agipd/index_eng.html}.

\bibitem{XFEL}
 M.~Altarelli, et al. (Eds.),
  \emph{XFEL: The European X-Ray Free-Electron Laser, Technical Design Report},
    Preprint DESY 2006-097, DESY, Hamburg 2006, and \url{http://www.xfel.eu/de/}.

\bibitem{Hamamatsu}
  \url{http://www.hamamatsu.com/}.

\bibitem{CIS}
   \url{http://www.cismst.org/}.

\bibitem{Zhang:2011a}
 J.~Zhang et al.,
 \emph{Study of radiation damage induced by 12~keV X-rays in MOS structures built on high resistivity n-type silicon}, Journal of Synchrotron Radiation, 19~(2012)~340.%, and arXiv 1107.5949.

\bibitem{Zhang:2011b}
 J.~Zhang et al.,
 \emph{Study of X-ray Radiation Damage in Silicon Sensors},
 Journal of Instrumantation~6~C11013~(2011).

\bibitem{Perrey:Thesis}
  H.~Perrey,
   \emph{Jets at Low Q2 at HERA and Radiation Damage Studies for Silicon Sensors for the XFEL},
    Ph.D. Thesis, Universität Hamburg, DESY-THESIS-2011-021 (2011).

\bibitem{Zhang:Thesis}
  J.~Zhang,
   \emph{X-ray Radiation Damage Studies and Design of a Silicon Pixel Sensor for Science at the XFEL}, Ph.D. Thesis, Universität Hamburg, in preparation.

\bibitem{Kraner:1993}
 H.W.~Kraner, Z.~Li and E.~Fretwurst,
  \emph{The Use of the Signal Current Pulse Shape to Study the Internal Electric Field Profile and Trapping Effects in Neutron Damaged Silicon Detectors}, Nucl. Instr. and Meth. A~326~(1993)~350.

\bibitem{Becker:2011}
 J.~Becker, D.~Eckstein, R.~Klanner and G.~Steinbrück,
  \emph{Measurements of charge carrier mobilities and drift velocity saturation in bulk silicon of
  $\langle 111 \rangle $ and
  $\langle 100 \rangle $ crystal orientation at high electric fields}, Solid State Electronics, 56~(2011)~104.

\bibitem{Becker:Thesis}
  J.~Becker,
   \emph{Signal development in silicon sensors used for radiation detection},
   Ph.D. Thesis, Universität Hamburg, DESY-THESIS-2010-33 (2010).

\bibitem{Synopsys}
 Synopsys TCAD webpage:
  \url{http://www.synopsys.com}.

\bibitem{Schwandt:Thesis}
  J.~Schwandt,
   \emph{Design of a radiation hard silicon pixel sensor for X-ray science}, Ph.D. Thesis, Universität Hamburg, in preparation.

\bibitem{Shockley:1938}
   W.~Shockley,
	 \emph{Currents to Conductors Induced by a Moving Point Charge}
	J. Appl. Phys. 9~(1938)~635.

\bibitem{Ramo:1939}
	S.~Ramo, 
	\emph{Currents Induced by Electron Motion}
  IRE~27~(1939)~636.

\bibitem{Hamel:2008}
  L.-A.~Hamel and M.~Julien,
   \emph{Generalized demonstration of Ramo's theorem with space charge and polarization effects},
   Nucl. Instr. and Meth. A~597~(2008)~207.

\bibitem{Poehlsen:Thesis}
  T.~Poehlsen,
   \emph{Charge collection in irradiated silicon sensors}, Ph.D. Thesis, Universität Hamburg, in preparation.


%\bibitem{Sommerfeld:1977}
% Arnold Sommerfeld,
% \emph{Elektrodynamik, Vorlesungen über Theoretische Physik Band III},
% Nachdruck der 4. durchgesehenen Auflage, Verlag Harry Deutsch (1977) p.55.

% \cite{Atalla:1960, Shockley:1964, Grove:1967

 \bibitem{Atalla:1960}
  M.M.~Atalla, A.R.~Bray and R.~Lindner,
   \emph{Stability of thermally oxidized silicon junctions in wet atmospheres},
    Proc.~IEEE 106 B (1960)~1130.

 \bibitem{Shockley:1964}
  W.~Shockley et al.,
   \emph{Mobile electric charges on insulating oxides with applications to oxide covered p-n junctions},
    Surface Science 2~(1964)~277.

 \bibitem{Grove:1967}
  A.S.~Grove,
   \emph{Physics and Technology of Semiconductor Devices},
    John Wiley \& Sons (1967).

 \bibitem{Chargelosses:2012}
  T.~Poehlsen, E.~Fretwurst, R.~Klanner, J.~Schwandt and J.~Zhang,
   \emph{Study of the accumulation layer and charge losses at the Si-SiO$_2$ interface in p+n-silicon
strip sensors}, to be published in Nucl. Instr. and Meth. A.

 \bibitem{Heimann:1982}
  P.P.~Heimann and J.E.~Olsen,
   \emph{A sensitive method for measuring surface conductivity of insulators},
     J. Appl. Phys. 53~(1982)~546.

 \bibitem{Chilingarov:2006}
  A.~Chilingarov, D.~Campbell and G.~Huges,
   \emph{Interstrip capacitance stabilization at low humidity},
   Nucl. Instr. and Meth. A~560~(2006)~118.

 \bibitem{Hartjes:2005}
  F.G.~Hartjes,
   \emph{Moisture sensitivity of AC-coupled silicon strip sensors},
   Nucl. Instr. and Meth. A~5552~(2005)~168.

 \bibitem{Yorkston:1987}
  J.~Yorkston, A.C.~Shotter, D.~Syme and G.~Huxtable,
   \emph{Interstrip surface effects in oxide passivated ion-implanted silicon strip detectors},
    Nucl. Instr. and Meth. A~262~(1987)~353.

\bibitem{Becker:2010b}
  J.~Becker, K.~Gärtner, R.~Klanner and R.H.~Richter,
  \emph{Simulation and experimental study of plasma effects in planar silicon sensors},
   Nucl. Instr. and Meth. A~624~(2010)~715.

\bibitem{Becker:2010a}
  J.~Becker, D.~Eckstein, R.~Klanner and G.~Steinbrück,
  \emph{Impact of plasma effects on the performance of silicon sensors at an X-ray FEL},
   Nucl. Instr. and Meth. A~615~(2010)~230.

\end{thebibliography}
\end{document}